\documentclass[journal]{IEEEtran}
\ifCLASSINFOpdf
\else
\fi

\normalsize

\usepackage{booktabs} % For formal tables
\usepackage{bm}
\newcommand{\tabincell}[2]{\begin{tabular}{@{}#1@{}}#2\end{tabular}}
\usepackage{multirow}
\usepackage{subfigure}
\usepackage{graphicx}
\usepackage{hyperref}
\usepackage{bm} %bold
\usepackage{cite}  % \cite{bibtex1��bibtex2��bibtex3} --> [1]-[3]
\usepackage{enumerate}

\usepackage{color}
\usepackage{xcolor}

\newcommand{\GCL}{\text{GCL}}
\newcommand{\TAS}{\text{TAS}}
\newcommand{\SP}{\text{SP}}
\newcommand{\ATS}{\text{ATS}}
\newcommand{\CBS}{\text{CBS}}
\newcommand{\TT}{\text{TT}}
\newcommand{\M}{\text{M}}

\newcommand{\GB}{\text{GB}}
\newcommand{\F}{\text{F}}
\newcommand{\NF}{\text{NF}}
\newcommand{\link}{\text{link}}
\newcommand{\TDMA}{\text{TDMA}}
\newcommand{\BE}{\text{BE}}
\newcommand{\LP}{\text{LP}}

\newtheorem{corollary}{Corollary}

\usepackage{amsmath}
\usepackage{mathtools}
\usepackage{amssymb}

\usepackage{url}

\usepackage{indentfirst} 
\usepackage[shortlabels]{enumitem}

\newcommand{\subparagraph}{}
\usepackage{lipsum}
\usepackage{titlesec}
\titlespacing\subsubsection{0pt}{12pt plus 4pt minus 2pt}{0pt plus 2pt minus 2pt}
\makeatletter
\renewcommand{\p@subsection}{\arabic{section}.\arabic{subsection}\expandafter\@gobble}
\renewcommand{\p@subsubsection}{\thesubsectiondis\arabic{subsubsection}\expandafter\@gobble}
\makeatother

\usepackage{geometry}
\begin{document}
\title{Quantitative Performance Comparison of Various \\Traffic Shapers in Time-Sensitive Networking}

%\author[a]{Luxi~Zhao\textsuperscript{*}}
%\author[b]{Paul~Pop}
%\author[a]{Sebastian~Steinhorst}
%\thanks{E-mail: luxi.zhao@tum.de; paupo@dtu.dk; sebastian.steinhorst@tum.de.}}
%\affil[a]{Department of Electrical and Computer Engineering, Technical University of Munich, Germany}
%\affil[b]{Department of Applied Mathematics and Computer Science, Technical University of Denmark, Denmark}

\author{Luxi~Zhao, %~\IEEEmembership{xx}
        Paul~Pop, %~\IEEEmembership{xx}
        Sebastian~Steinhorst
	
	\thanks{Manuscript received March 25, 2021; revised November 15, 2021, March
		AQ2 24, 2022, and May 27, 2022; accepted May 29, 2022, IEEE TNSM. DOI: 10.1109/TNSM.2022.3180160}
	\thanks{L.~Zhao is with the Department of Electronic and Information Engineering, Beihang University, Beijing, China. She was with the Department of Electrical and Computer Engineering, Technical University of Munich, Germany (e-mail: zhaoluxi@buaa.edu.cn)}% <-this % stops a space
	\thanks{P.~Pop is with the Department of Applied Mathematics and Computer Science, Technical University of Denmark, Denmark, 2800 DK (e-mail: paupo@dtu.dk)}% <-this % stops a space
	\thanks{S.~Steinhorst is with the Department of Electrical and Computer Engineering, Technical University of Munich, Germany, D-80333 (e-mail: sebastian.steinhorst@tum.de)}% <-this % stops a space
	%\thanks{ Manuscript received April 19, 2005; revised August 26, 2015.}
}


\markboth{IEEE TRANSACTIONS ON NETWORK AND SERVICE MANAGEMENT}%
{Shell \MakeLowercase{\textit{et al.}}: Bare Demo of IEEEtran.cls for IEEE Journals}



\maketitle

\begin{abstract}
Owning to the sub-standards being developed by IEEE Time-Sensitive Networking (TSN) Task Group, the traditional IEEE 802.1 Ethernet is enhanced to support real-time dependable communications for future time- and safety-critical applications. Several sub-standards have been recently proposed that introduce various traffic shapers (e.g., Time-Aware Shaper (TAS), Asynchronous Traffic Shaper (ATS), Credit-Based Shaper (CBS), Strict Priority (SP)) for flow control mechanisms of queuing and scheduling, targeting different application requirements. These shapers can be used in isolation or combination and there is limited work that analyzes, evaluates, and compares their performance, which makes it challenging for end-users to choose the right combination for their applications. This paper aims at (i) quantitatively comparing various traffic shapers and their combinations, (ii) summarizing, classifying, and extending the architectures of individual and combined traffic shapers and their Network calculus (NC)-based performance analysis methods, and (iii) filling the gap in the timing analysis research on handling ATS and CBS used for different priority queues, and two novel hybrid architectures of combined traffic shapers, i.e., TAS+ATS+SP and TAS+ATS+CBS when ATS and CBS used at the same queue. A large number of experiments, using both synthetic and realistic test cases, are carried out for quantitative performance comparisons of various individual and combined traffic shapers, from the perspective of upper bounds of delay, backlog, and jitter. To the best of our knowledge, we are the first to quantitatively compare the performance of the main traffic shapers in TSN. The paper aims at supporting the researchers and practitioners in the selection of suitable TSN sub-protocols for their use cases.
\end{abstract}

\begin{IEEEkeywords}
TSN, traffic shapers, combinations, real-time performance, worst-case, comparison.
\end{IEEEkeywords}

% \fi
\IEEEpeerreviewmaketitle

\section{Introduction}
\IEEEPARstart{N}{owadays}, modern cyber-physical and embedded systems, including systems in the automotive, industrial automation, avionics and aerospace domain increasingly depend on the real-time capabilities of their communication networks. Time-Sensitive Networking (TSN)~\cite{802.1Q} enhances standard Ethernet~\cite{802.3Ethernet}, aiming at providing deterministic communication for real-time traffic. Over the recent years, TSN has become a high-profile and active standardization effort with a strong research community both in academia and in the industry. Several companies, such as Belden, Cisco Systems, Intel Corporation, NXP Semiconductors, Siemens, TTTech Computertechnik and Huawei Technologies are developing TSN switches with various capabilities. The Avnu Alliance consortium has been established to evaluate the interoperability and conformance of such products to the TSN standards. 
TSN integrates multiple traffic types implemented by different scheduling mechanisms (traffic shapers), such as the Time-Aware Shaper (TAS) standardized by IEEE 802.1Qbv~\cite{802.1Qbv}, the Asynchronous Traffic Shaper (ATS) standardized by IEEE 802.1Qcr~\cite{802.1Qcr}, the Credit-Based Shaper (CBS) standardized by IEEE 802.1Qav~\cite{802.1Qav}. These shapers can be used separately or in several combinations. TAS is based on a global clock synchronization (via IEEE 802.1AS~\cite{802.1ASRev}) implementing the time-triggered traffic to guarantee deterministic transmission. ATS avoids using the global clock synchronization, but it is still able to provide real-time guarantees by reshaping traffic flows per hop to reduce the burstiness of traffic. CBS is an asynchronous traffic shaper that implements a bandwidth reservation mechanism.

Many related works have already been proposed for the schedulability analysis and configuration for different traffic shapers. 
%overview: TAS
For TAS, which relies on global clock synchronization, the scheduling synthesis for time-triggered (TT) traffic, which is also called scheduled traffic (ST), has been studied in~\cite{Durr16,Craciunas16:RTNS,Pop16,Raagaard17,Vlk20} using different implementation methods to synthesize Gate Control Lists (GCLs). Vlk et al.~\cite{Vlk20} increase the schedulability and throughput of TT by proposing a simple hardware enhancement of a switch. Ramon et al.~\cite{Oliver18} relax the constraints of the scheduling model to increase the solution space at the expense of the deterministic scheduling of TAS. A more flexible class-based (i.e., window-based) TAS model is proposed in~\cite{Zhao18:Access, Zhao20:ITJ}, which does not require strict flow isolation in queues and supports unscheduled end systems. Reusch et al.~\cite{Reusch20} propose the class-based schedule synthesis for 802.1Qbv. Craciunas et al.~\cite{CraciunasOverview17} give an overview of the comparison of scheduling mechanisms for TAS in TSN networks and time-triggered scheduling in TTEthernet. In~\cite{Mahfouzi18}, researchers solved the stability-aware integrated scheduling and routing problem for networked cyber-physical systems based on the 802.1Qbv TSN standard. 
ATS is developed from the urgency-based scheduler (UBS) proposed by Specht et al.~\cite{Soheil16} and aims at achieving low latency without designing time schedules harmonized among all end systems and switches based on global time synchronization. The same authors~\cite{Soheil17} propose the synthesis of queues and priority assignment for ATS. Zhou et al.~\cite{Zhou18, Zhou19} present the simulation model of ATS implemented in the Riverbed simulator. \cite{LeBoudec18} proves that ATS will not introduce extra overheads to the worst-case delay of the FIFO system. For CBS, several methods related to performance and schedulability analysis have been proposed in~\cite{Diemer12,Bordoloi14,Azua14,ZhaoLin18,Cao16}.

%overview: combination
The above studies all assume the use of a single traffic shaper. There are also some limited studies on the combination of different traffic shapers. An overview of the combined usage of TAS and CBS in controlling flows in in-vehicle networks was presented in~\cite{Bello14}. A simulation study of the coexistence of TAS and CBS is presented in~\cite{Meyer13}. Zhao et al.~\cite{Zhao18} propose the performance analysis of Audio Video Bridging (AVB) traffic under the coexistence of CBS and TAS. The same authors~\cite{Zhao20} extend the timing analysis for the arbitrary number of AVB classes under the same architecture of TAS+CBS, considering both standard credit behavior and more generally assumed credit behavior but deviating from the standard 802.1Q~\cite{802.1Q}. Mohammadpour et al.~\cite{Mohammadpour18} consider the combination of non-time-triggered control-data traffic (CDT), CBS and ATS, and give the latency and backlog bounds for the traffic of CBS affected by ATS. However, the CDT model is not a standard model required by the TSN standard. 
In~\cite{Fang20}, researchers present a simulation model of combined CBS and ATS within the OMNeT++ simulator. 

With the increasing number of sub-standards for TSN networks, there have been several literature reviews related to TSN networks. Researchers~\cite{Nasrallah18} have given a comprehensive survey on TSN networks, from TSN sub-standards to the existing research of TSN before 2018. Maile et al.~\cite{Maile20} provide an overview of the existing publications that use a Network Calculus approach in the timing analysis for TSN networks. 
Researchers in \cite{Hellmanns20} make a comparison between flow-based (i.e., frame-based) and class-based TAS, which concludes that class-based scheduling is easy to plan but loses the advantages of extremely low latency and jitter compared with the flow-based TAS.
Nasrallah et al.~\cite{Nasrallah19} presented the performance comparison of class-based TAS and ATS based on simulations.
Nevertheless, the ATS architecture they considered does not exactly match the general model of ATS~\cite{802.1Qcr, Soheil16}. They apply the ATS shaper at the ingress port of the switch instead, and consider another extra urgent queue with the highest priority before ST traffic.
Thus, as stated above, there are currently no comprehensive and systematic guidelines on the quantitative performance comparison of different traffic shapers, and their further coexistence possibilities and interactions in TSN networks. 

This paper aims at (i) quantitatively comparing various traffic shapers, i.e., TAS, ATS, CBS, strict priority (SP) scheduling and their combinations; (ii) summarizing, classifying and extending the architectures of individual and combined traffic shapers and their performance analysis methods; and (iii) filling the gap in the timing analysis research handling on these novel combinations. We consider the coexistence between time-triggered shapers (TAS) and various event-triggered shapers (ATS, CBS, SP). Our findings will support researchers and practitioners in understanding the performance characteristics and mutual effects of different traffic shapers. The main contributions of the paper are as follows,
\begin{itemize}
	\item[$\bullet$] We summarize the architectures of the main traffic shapers and their combinations in TSN. In order to perform a fair comparison, we use the same method (Network Calculus, NC) to evaluate the performance of each shaper. Based on our and other researchers’ existing NC-based analysis work for different traffic shapers in TSN, we summarize and classify them. The existing work for the NC-based analysis includes: ATS, CBS, SP individually used, and TAS+SP, TAS+CBS used in combination. We complete the general uniform formula for timing analysis of the arbitrary number of AVB classes when CBS is used individually.
	\item[$\bullet$] The NC-based performance analysis approach is extended to the combination used of ATS and CBS for different priority queues, including two cases of ATS+CBS (ATS for high priority queues and CBS for low priority queues) and CBS+ATS (CBS for high priority queues and ATS for low priority queues). Two novel hybrid architectures of traffic shapers, i.e., TAS+ATS+SP (compared with TAS+SP) and TAS+ATS+CBS (compared with TAS+CBS) are proposed to understand the impact of ATS reshaping on the combined architectures, where SP and CBS are used in the same queue. Even though ATS+CBS on the same queue is not supported by the standards, their combination is still worthwhile to be investigated from a research perspective. The NC-based timing analysis method is extended to analyze the real-time performance of traffic in these combinations. The combinations have been selected to provide comprehensive coverage of possible combined traffic shapers in TSN networks, supported by their corresponding NC-based performance analysis.
	\item[$\bullet$] A large number of experiments, using both synthetic and realistic test cases, for quantitative performance comparisons of various individual and combined traffic shapers are carried out, from the perspective of upper bounds of delay, backlog and jitter. Especially with ATS shaping, we highlight interesting results that do not always show the superiority of ATS compared with other shapers, in isolation or combination. Moreover, we compare the NC-based analysis with the closed-form formula proposed in~\cite{Soheil16} for the ATS shaper used individually. We also show the positive function of ATS on the cyclic dependencies. We aim at providing a basic reference for the selection of suitable TSN sub-protocols for researchers and practitioners.
\end{itemize}

The remainder of the paper is organized as follows. Sect.~\ref{sec:PerformanceMetrics} gives the overview of performance metrics for the TSN traffic evaluation. Sect.~\ref{sec:PerformanceCmpIndi} summarizes and supplements the worst-case performance analysis for traffic transmission with individual TAS, ATS and CBS shapers. Sect.~\ref{sec:PerformanceCmpComb} presents novel combined architectures of shapers, and extends the NC-based analysis. The evaluation of our performance comparison of individual traffic shapers and their combinations is provided in Sect.~\ref{sec:Experiment}. Sect.~\ref{sec:conclusion} concludes of the paper. The background of the NC method used is briefly introduced in Appendix~\ref{sec:NC}.

\begin{table}
	\caption{Summary of notation.}
	\label{tab:notation}
	\centering
	\setlength{\tabcolsep}{3pt}
	\begin{tabular}{|p{35pt}|p{210pt}|}
		\hline
		Symbol & Meaning\\
		\hline
		$f$ &A flow\\
		$l_f$ &Frame size of the flow $f$\\
		$T_f$ &Period for periodic flow $f$\\
		$P_f$ &Priority of the flow $f$\\
		$b_f$ &Burst of the leaky bucket model of the flow $f$\\
		$r_f$ &Rate of the leaky bucket model of the flow $f$\\
		$R_f$ &Route of the flow $f$\\
		$D_{\text{pro}}$ &Propagation delay\\
		$D_{\text{fwd}}$ &Forwarding delay in the switch\\
		$d_{Q}(t)$ &Queuing delay of frames in the queue $Q$\\
		$D_Q$ &Latency upper bound of frames waiting in the queue $Q$\\
		$D_{Q,f}$ &Latency upper bound of flow $f$ waiting in the queue $Q$\\
		$D_{\text{E2E},f}$ &Upper bound of end-to-end latency of the flow $f$\\
		$d_{\text{E2E},f}$ &Lower bound of end-to-end latency of the flow $f$\\
		$B_{Q}$ &Backlog upper bound of the queue $Q$\\
		$J_{\text{E2E},f}$ &Upper bound of end-to-end jitter of the flow $f$\\
		$h$ &Output port of a node (link)\\
		$h^-$ &Output port of a preceding node connected to $h$\\
		$C$ &Physical link rate\\
		$\phi_f^h$ &Start transmission time (offset) of the flow $f$ on the link $h$\\
		$n_{SP}$ &Priority number of SP traffic\\
		$n_{CBS}$ &Class number of AVB traffic\\
		$\alpha_Q(t)$ &Input arrival curve of flows arriving before the queue $Q$\\
		$\beta_Q(t)$ &Service curve supplied for flows waiting in the queue $Q$\\
		$\alpha^*_Q(t)$ &Output arrival curve of flows departing from the queue $Q$\\
		$\sigma^{\link}(t)$ &Shaping curve of the physical link\\
		$\sigma^{\CBS}(t)$ &Shaping curve of CBS\\
		$\delta_{D}^q(t)$ &Pure-delay function\\
		$Q_i$&Queue of traffic with priority $i$ in the current node port\\
		$Q_i^-$&Queue of traffic with priority $i$ in the preceding node port\\
		$l^{\max}_{>i}$&Maximum frame size in traffic with priority lower than the priority $i$\\
		$l_Q^{\max}$ &Maximum frame size in the queue $Q$\\
		$l_Q^{\min}$ &Minimum frame size in the queue $Q$\\
		$idSl_i$ &Idle slope for AVB traffic of Class $M_i$\\
		$sdSl_i$ &Send slope for AVB traffic of Class $M_i$\\
		$c^{\max}_i$ &Credit upper bound for AVB Class $M_i$ (no GB / credit frozen during GB)\\
		$\overline{c}^{\max}_i$ &Credit upper bound for AVB Class $M_i$ (credit non-frozen during GB)\\
		$c^{\min}_i$ &Credit lower bound for AVB Class $M_i$\\
		$L^{\GB}$ &Maximum guard band duration ($\mu s$)\\
		$\wedge$ &$x\wedge y = \min\{x,y\}$\\
		$[x]^+$ &$\max\{0,x\}$\\
		$[f(t)]_\uparrow^{+}$ &$\max \limits_{0 \leq s \leq t} \{f(s),0\}$\\
		\hline
	\end{tabular}
\end{table}

\begin{table}
	\caption{Summary of acronym.}
	\label{tab:acronym}
	\centering
	\setlength{\tabcolsep}{3pt}
	\begin{tabular}{|p{55pt}|p{190pt}|}
		\hline
		Acronym & Full Expression\\
		\hline
		TSN &Time-Sensitive Networking\\
		TAS &Time-Aware Shaper\\
		ATS &Asynchronous Traffic Shaper\\
		CBS &Credit-Based Shaper\\
		SP &Strict Priority\\
		TT &Time-Triggered\\
		ST &Scheduled Traffic\\
		AVB &Audio Video Bridging\\
		BE &Best-Effort\\
		GCL &Gate Control List\\
		UBS &Urgency-Based Scheduler\\
		CDT &Control-Data Traffic\\
		RC &Rate-Constrained\\
		ES &end system\\
		SW &switch\\
		TCF &traffic class filtering\\
		TC & test case\\
		SRM &small ring \& mesh\\
		MR &medium ring\\
		MM &medium mesh\\
		ST &small tree\\
		MT &medium tree\\
		NC &network calculus\\
		WCD &upper bound of worst-case end-to-end latency\\
		WCB &upper bound of worst-case backlog\\
		WCJ &upper bound of worst-case jitter\\
		\hline
	\end{tabular}
\end{table}

\section{Overview of Performance Metrics in TSN Evaluation}
\label{sec:PerformanceMetrics}
In this paper, we will compare the quality of service for each individual and combined traffic shapers from the perspective of upper bounds of end-to-end latency, backlog and end-to-end jitter. The end-to-end latency is the time a frame uses to traverse the network from the sending node to the receiving node along its route. The latency upper bound is a significant QoS metric for real-time applications, which is used to check if a message meets its deadline. The backlog is defined as the number of bits waiting in the queue to be served at any time, and the backlog upper bound can be used to determine the buffer size needed to avoid frame loss. The jitter represents the variation in the latency of a flow. High amounts of jitter indicate poor network performance.

In this paper, flows manipulated by time-triggered shapers (TAS) can only be periodic flows, and flows handled by the event-triggered shapers (ATS, CBS, SP) can be periodic or aperiodic flows. For a periodic flow, we know its frame size, period and priority, i.e., $\langle l_f, T_f, P_f \rangle$. For a sporadic flow, we assume as the related work that the flow is regulated by a leaky bucket model $(b_f, r_f)$ before entering the network~\cite{Soheil16}, where $b_f$ and $r_f$ are the burst and rate of the leaky bucket, respectively. Thus, for a sporadic flow we know its $\langle b_f, r_f, l_f, P_f \rangle$. 
In this paper, the traffic class (priority) $P_f$ for the flow $f$ remains the same on all nodes along its path. TSN supports at most eight different priorities (0 of lowest - 7 of highest priority).
Tables~\ref{tab:notation} and \ref{tab:acronym} respectively summarize the notations and acronyms used in this paper.

\subsection{End-to-End Latency Upper Bounds}
Considering a flow $f$, its source of delay consists of: (i) Propagation delay $D_{\text{pro}}$, which is tightly related to the physical medium, and considered constant in this paper; (ii) Forwarding delay $D_{\text{fwd}}$ on the switch, which is the time interval from the time after the frame being fully received, to the time it arrives at the buffer located after the switching fabric. It is also generally considered constant; (iii) Queuing delay $d_{Q}(t)$ in the FIFO queue $Q$ for the egress port, which is a time-variant depending on the flows' contention on the port. The upper bound of queuing delay $D_{Q}$ can be calculated based on the Network Calculus theory~\cite{LeBoudec01}, see Appendix~\ref{sec:NC}. By constructing the input arrival curve $\alpha_{Q}(t)$ of aggregate flows before $Q$, which represents the upper envelope of flows arrival in any time interval, and the service curve $\beta_{Q}(t)$, which represents the service guarantee for these flows, the upper bound of queuing latency of any flows in the queue $Q$ can be calculated by the maximum horizontal deviation of $\alpha_{Q}(t)$ and $\beta_{Q}(t)$,
\begin{equation}\label{maxDQ}
	D_{Q,f}=D_Q=\text{h}\left(\alpha_Q(t),\beta_Q(t)\right),
\end{equation}
which is also the upper bound of delay for each flow $f$ in $Q$.
Then, the upper bound of end-to-end delay of the flow $f$ is obtained by the sum of delays from the source ES to the destination ES along its route $R_f$,
\begin{equation}\label{maxDe2e}
	D_{\text{E2E},f}=\sum_{Q\in R_f}\left(D^Q_f+D_{\text{pro}}+D_{\text{fwd}}\right)-D_{\text{fwd}}.
\end{equation}

\subsection{Backlog Upper Bounds}
\label{sec:Backlog}
According to the Network Calculus theory, the upper bound of backlog in a queue $Q$ is given by the maximum vertical deviation between the arrival curve $\alpha_Q(t)$ of aggregate flows before the queue $Q$ and the service curve $\beta_Q(t)$ offered to flows waiting in the queue $Q$,
\begin{equation}\label{maxBQ}
	B_{Q}=\text{v}\left(\alpha_Q(t),\beta_Q(t)\right).
\end{equation}

\subsection{End-to-End Jitter Upper Bounds}
\label{sec:jitter_ATS}
Jitter refers to the delay variation, i.e., the difference in end-to-end latency between any selected frames in a flow transmitting over a network.
Then, the upper bound of jitter of a flow $f$ is calculated by the difference between the maximum and minimum bounds of end-to-end latency of the flow $f$. The upper bound of end-to-end latency $D_{\text{E2E},f}$ has been discussed previously in Eq.~(\ref{maxDe2e}). The lower bound of end-to-end latency $d_{\text{E2E},f}$ of $f$ is the sum of transmission delays along its route without the interference from other flows, which can be given as follows,
\begin{equation}\label{minDe2e}
	d_{\text{E2E},f}=\sum_{Q\in R_f}\left(l_f/C+D_{\text{pro}}+D_{\text{fwd}}\right)-D_{\text{fwd}}.
\end{equation}
Thus the upper bound of jitter for the flow $f$ is,
\begin{equation}\label{maxJe2e_ATS}
	J_{\text{E2E},f}=D_{\text{E2E},f}-d_{\text{E2E},f}.
\end{equation}

As shown above, in order to obtain the performance metrics for different traffic shapers in TSN, the main objective is to construct the arrival curve $\alpha_Q(t)$ and service curve $\beta_Q(t)$ for the corresponding traffic shapers.

\section{Performance Analysis of Individual Traffic Shapers}
\label{sec:PerformanceCmpIndi}
In the following, the performance analyses for each individual traffic shaper, including Time-Aware Shaper (TAS), Asynchronous Traffic Shaper (ATS) and Credit-Based Shaper (CBS), are summarized and extended from the state-of-the-art, and specified using a citation after the subtitle of each performance analysis section to indicate where the relevant previous work.
Their quantitative performance comparison in Sect.~\ref{sec:CmpIndi} is based on these analyses.
When discussing a certain traffic shaper, it is assumed that all nodes, including end systems (ESes) and switches (SWs), in the network support this traffic shaper.

\subsection{Time-Aware-Shaper (TAS)}
\label{sec:TAS}
Relying on the global network clock (IEEE 802.1ASrev ~\cite{802.1ASRev}), IEEE 802.1Qbv~\cite{802.1Qbv} defines the Time-Aware Shaper (TAS) used to control a gate for each queue of the output port to enable time-triggered communication, enabling the deterministic transmission of extremely low latency and jitter using Gate Control Lists (GCLs). In this paper, we consider the flow-based TAS~\cite{Craciunas16:RTNS, Pop16, Raagaard17, Vlk20}, which is a widely used model compared to class-based TAS~\cite{Zhao18:Access, Zhao20:ITJ}. Researchers in \cite{Hellmanns20} have concluded that it has much better performance in terms of latency and jitter compared with the class-based TAS. Moreover, we consider the case that both ESes and SWs can be scheduled~\cite{Craciunas16:RTNS, Pop16}. With a scheduled ES, the task sending the message and the communication schedule on the ES egress port are synchronized.

\begin{figure}[!t]
	\centering
	\includegraphics[width=0.4\textwidth]{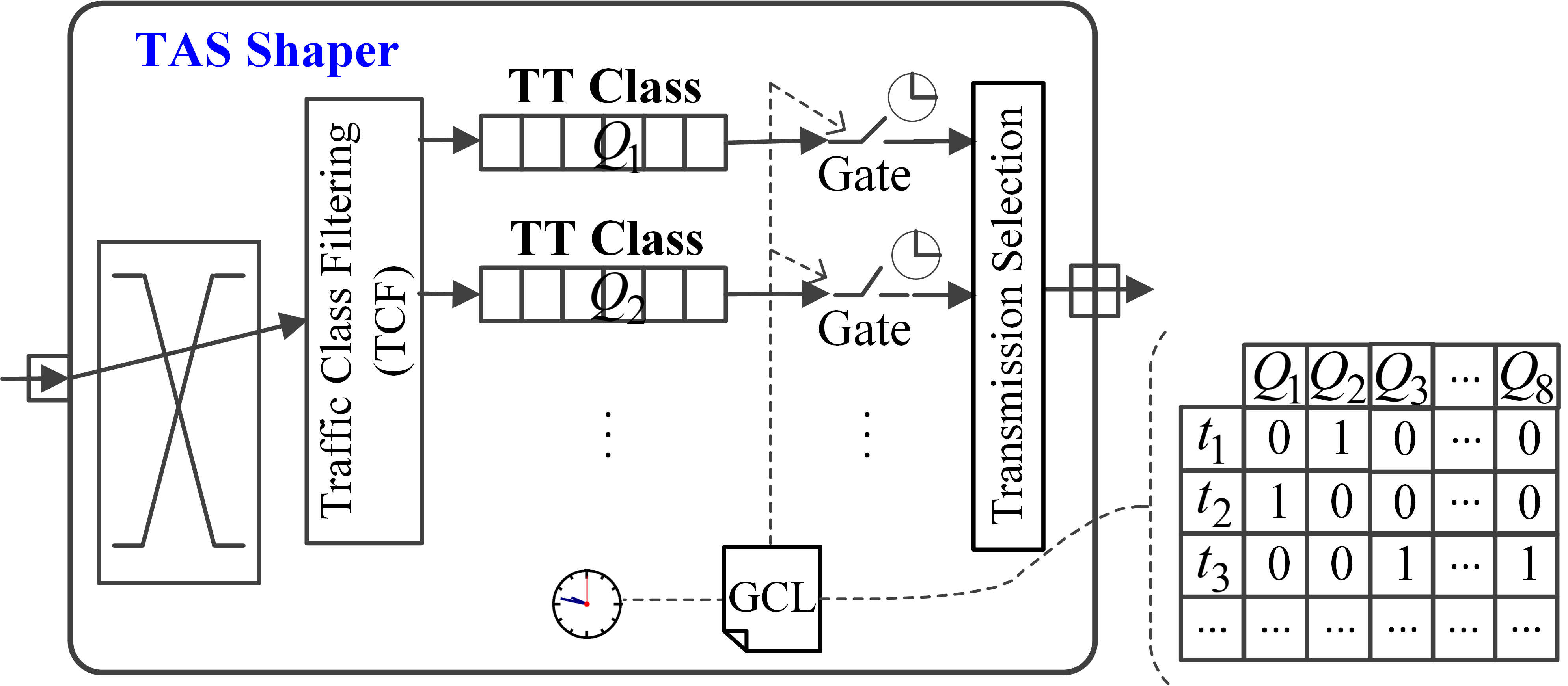}
	\caption{\label{fig:TASArchitecture} TAS Architecture}
\end{figure}

Fig.~\ref{fig:TASArchitecture} depicts a TAS architecture in an egress port of a node supporting 802.1Qbv. The switching fabric forwards input flows to the corresponding output port according to their routing information. The traffic class filtering (TCF) dispatches input frames to the corresponding queue of the output port according to their traffic class. 
For each egress port, there are eight queues, where there may be multiple queues used for TT traffic to achieve completely deterministic transmission, depending on the TT traffic load and construction of GCLs.
Frames waiting in a queue are eligible for transmission only when the corresponding queue gate is open. The TAS control is implemented based on GCLs which dictate the state of the gates. The open and closed states are represented by 1 and 0 respectively in GCLs, as shown with the GCL table beside the TAS architecture in Fig.~\ref{fig:TASArchitecture}. For example, at time $t_1$, the gate for the queue $Q_2$ is open (1) while all the rest are closed (0). Full control of frames can be implemented by mutually exclusive opening queues. 

\begin{figure}[!t]
	\centering
	\includegraphics[width=0.48\textwidth]{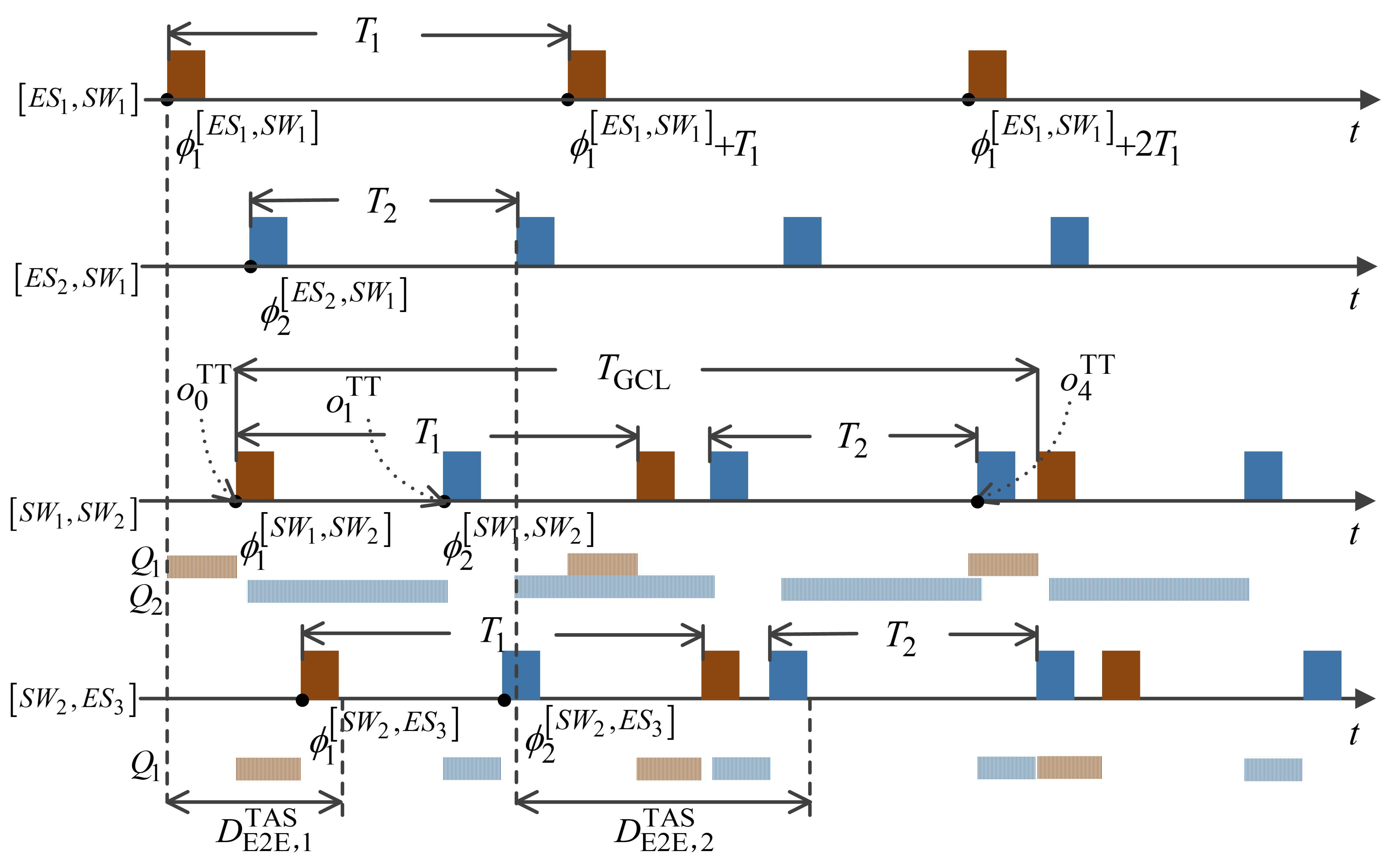}
	\caption{\label{fig:GCL-GanttChart} Example GCL synthesis for ST}
\end{figure}

Currently, the flow-based TAS can only support periodic traffic scheduling. The problem of GCL synthesis is to find feasible and optimized offsets and queue allocations for periodic traffic.
A frame of a TT flow $f$ on a link (egress port) $h=[v_a, v_b]$ is defined by the tuple $\langle \phi_f^h, l_f/C \rangle$, of which $\phi_f^h$ and $l_f/C$ respectively denote the start time (i.e., offset) of transmission and transmission duration of the frame on the respective link. 
Flow $f$ repeatedly sends frames at times $\phi_f^h$, $\phi_f^h+T_f$, $\phi_f^h+2\cdot T_f$, $\phi_f^h+3\cdot T_f$, .... Fig.~\ref{fig:GCL-GanttChart} shows an example of a GCL using a Gantt chart, describing the transmission of two TT flows $f_1$ and $f_2$, with the routes $r_1=[[ES_1,SW_1],[SW_1,SW_2],[SW_2,ES_3]]$ and $r_2=[[ES_2,SW_1],[SW_1,SW_2],[SW_2,ES_3]]$, respectively. The x-axis represents the time dimension, and the y-axis lists the output ports. The rectangles represent the TT frames' transmission, whose length equals to $l_f/C$. The left side of the rectangle is the start time of the transmitted frame, which equals to $\phi^h_f$. The thin shaded row labeled $Q_i$ below the link represents the waiting time of the frame in the corresponding queue $Q_i$.

It can be seen that the transmission time of TT traffic is scheduled in advance. Thus, the performance metrics can be obtained together with the GCLs synthesis without the need for complex performance analysis methods. The performance analysis for TAS was first discussed in \cite{Craciunas16:RTNS}, and we conclude in the following subsection.

\subsubsection{Performance Analysis -- TAS~\cite{Craciunas16:RTNS}}
\textbf{\underline{End-to-End Latency Bounds - TAS}.}
A flow $f$ using flow-based TAS has a completely deterministic end-to-end latency. When the GCL is constructed, its end-to-end delay is known by the time duration between the sending time $\phi^{h_0}_f$ on the source ES $h_0$ and the reception time $\phi^{h_n^-}_f+l_f/C$ on the destination ES $h_n$, where $h_n^-$ is the node before $h_n$. During the design phase of determining the offset $\phi_f^{h_j}$ ($h_j\in[h_0,h_n^-]$), the following inherent delays are considered: propagation delay $D_{\text{pro}}$, forwarding delay $D_{\text{fwd}}$, network precision $\delta$ due to the time-synchronization, and store-and-forward (transmission) delay $l_f/C$, which enforces that a frame is forwarded by a node only after it has been fully received at the node.
Then, the end-to-end latency for the TT flow $f$ is given by,
\begin{equation}\label{EtoETAS}
	D^{\TAS}_{\text{E2E},f}=\phi^{h_n^-}_f+l_f/C-\phi^{h_0}_f.
\end{equation}

\textbf{\underline{Backlog Bounds - TAS}.}
To fully control each frame transmission, researchers have proposed~\cite{Craciunas16:RTNS, Pop16} to isolate the frames in queues, i.e., at most one flow occupies a queue at a time, preventing the frame transmission ordering from being disrupted. We depict the queue occupancy with thin shaded rows in Fig.~\ref{fig:GCL-GanttChart}. Then, the backlog bounds in the queue $Q$ is the maximum frame assigned to such a queue,
\begin{equation}\label{maxBQ_TAS}
	B^{\TAS}_{Q}=\max_{f\in Q} \{l_f\}.
\end{equation}

\textbf{\underline{Jitter bounds - TAS}.}
Real-time communications are typically sensitive to jitter. The flow-based TAS model~\cite{Craciunas16:RTNS, Pop16, Raagaard17, Vlk20} that implements a completely deterministic transmission leads to zero jitter, i.e.,
\begin{equation}\label{maxJe2e}
	J^{\TAS}_{\text{E2E},f}=0.
\end{equation}

\subsection{Asynchronous Traffic Shaper (ATS)}
\label{sec:ATS}
\begin{figure}[!t]
	\centering
	\includegraphics[width=0.33\textwidth]{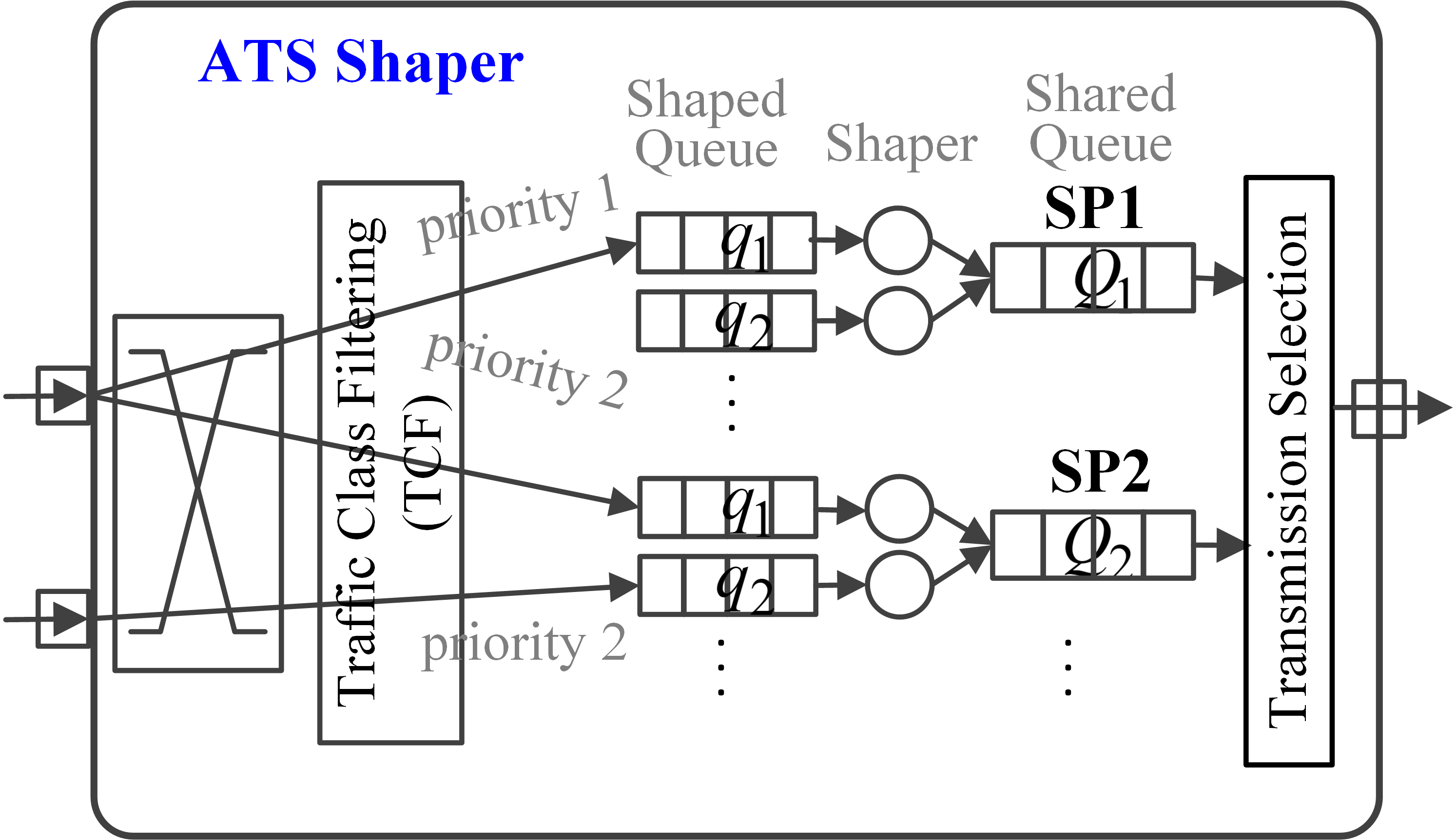}
	\caption{\label{fig:ATSArchitecture} ATS Architecture}
\end{figure}

Asynchronous Traffic Shaping (ATS) is another real-time traffic shaper, standardized by IEEE P802.1Qcr~\cite{802.1Qcr}. It uses asynchronous transmission and, although it does not require a global clock, it uses the local clocks to reshape traffic in each node. The ATS architecture is shown in Fig.~\ref{fig:ATSArchitecture}. Note that although ATS was originally proposed in~\cite{Soheil16} for strict priority queues, the essence of ATS is an interleaved regulator per input port and per traffic class placed after the class-based FIFO system of an upstream node~\cite{LeBoudec18}. It is theoretically possible that ATS can be connected to any class-based FIFO system of upstream nodes, for example in combination used with CBS, which will be discussed in Sect.~\ref{sec:TAS+ATS+CBS}. It contains two levels of queues, (i) shaped queue $q$ and (ii) shared queue $Q$, and the ATS shaping algorithm is located between them. 
Shaped queues are used to pre-store frames, which are waiting to be reshaped into the leaky bucket model by the ATS shaping algorithm.
Shared queues are used for different priority traffic, and there are at most 8 shared queues. The shared queues follow the class-based scheduling mechanism. ATS has been proposed with the goal of avoiding burstiness cascades.
Which shaped queue the frames should enter depends on the queuing schemes as follows~\cite{Soheil16, Zhou19},
\begin{enumerate}[leftmargin=12pt]
	\item[$\bullet$] QAR1: frames from different senders (input ports) should not be assigned to the same shaped queue in the receiver;
	\item[$\bullet$] QAR2: frames from the same sender but with different priority levels are not allowed to be assigned to the same shaped queue;
	\item[$\bullet$] QAR3: frames are not allowed to be stored in the same shaped queue if the frames sent to receivers are in different priority levels.
\end{enumerate}
The number of shaped queues in the receiver is related to the number of senders and the number of priorities assigned to the traffic from the sender to the receiver. Since, in the paper, traffic priority for a flow remains the same at all nodes along its path, the number of shaped queues is only related to the number of senders (used input ports).

The ATS shaping algorithm (Sect.~8.6.11.3 in~\cite{802.1Qcr}) is derived from the Token Bucket Emulation (TBE) algorithm, which implements the committed transmission rate $r_f$ and the committed burst size $b_f$ for each flow by calculating an eligible time for frame transmission. Note that ATS is not implemented with per-flow queuing, but the ATS shaping algorithm needs to record state per flow in order to reshape each flow with the respective constraint~\cite{LeBoudec18}. It also means that the shaping parameters in the ATS shaping algorithm~\cite[\S~8.6.11.3]{802.1Qcr} are for per-flow but not for per-queue.
Frames waiting in each shaped queue are forwarded into the shared queue in FIFO order, following their respective eligible transmission times.

According to the proof in \cite{LeBoudec18} that ATS will not introduce extra overheads to the worst-case delay of the FIFO system, and inspired by~\cite{Mohammadpour18} of the combined ATS and CBS performance analysis based on the NC method, we summarize, for the first time, an NC approach to analyze the performance of ATS used individually as follows in Sect.~\ref{sec:PA_ATS}, which forms the basis of supporting combinations of ATS with other shapers. Moreover, we compare the NC-based method with the closed-form formula proposed by~\cite{Soheil16} in the experiment in Sect.~\ref{sec:CmpNC_nonNC}.

\subsubsection{Performance Analysis -- ATS~\cite{LeBoudec18}}
\label{sec:PA_ATS}
\textbf{\underline{Service Curve $\beta_{Q_i}^{\ATS}\!(t)$ - ATS - Shared Queue}.}
The service for the traffic in the shared queues obeys strict priority scheduling, i.e., flows with low priority can obtain service only when the queues of higher priority traffic are empty. 
Then, by ATS reshaping, the service curve for SP traffic with priority $i$ ($i\in [1,n_{\SP}]$) in the corresponding shared queue $Q_i$ is given by,
\begin{equation}\label{g:aggSer_ShrQ_SP_ATS}
	\beta_{Q_i}^{\ATS}\!(t)=
	C\bigg[ t -
	\frac{\sum_{j=1}^{i-1}\alpha^{\ATS}_{Q_j}\!(t)}{C}-\frac{l_{>i}^{\max}}{C} \bigg]^+,
\end{equation}
where $[x]^+=\max\{0,x\}$, $\alpha^{\ATS}_{Q_j}\!(t)$ (Eq.~(\ref{g:aggArr_SP_ATS+SP})) is the aggregate arrival curve of SP flows after ATS reshaping with the priority $j$ higher than the priority $i$, and $l^{\max}_{>i}=\max_{j>i}\{l^{\max}_{Q_{j}}, l^{\max}_{Q_{\BE}}\}$ that is the maximum frame size of traffic with the priority lower than priority $i$.

\textbf{\underline{Input Arrival Curve $\alpha_{Q_i}^{\ATS}\!(t)$ - ATS - Shared Queue}.} 
The input arrival curve $\alpha_{Q_i}^{\ATS}\!(t)$ of aggregate SP flows with priority $i$ before the shared queue $Q_i$ is related to the total output arrival curves $\alpha^{*}_q(t)$ of individual flows from each previously shaped queue $q$. As mentioned, each flow is reshaped into the leaky bucket model before entering the shared queue. Then, the output arrival curve of an individual flow $f$ from the shaped queue satisfies $r_f\cdot t+b_f$, where $r_f$ and $b_f$ are the committed transmission rate and burst size for the flow $f$ implemented by the ATS shaping algorithm, respectively. Note that, in this paper, if flow $f$ is aperiodic, $b_f$ and $r_f$ are set to the same leaky bucket parameters as before $f$ entered the network, and if $f$ is periodic, we have $b_f=l_f$ and $r_f=l_f/T_f$.
The output arrival curve of aggregate flows from the shaped queue $q$ is the sum of the output arrival curves of individual flows in $q$,
\begin{equation}\label{g:aggOutArrShpQ}
	\alpha^{*}_q(t)=\sum_{f\in q}\left(r_f\cdot t+b_f\right).
\end{equation}
Moreover, according to the queuing schemes, there will be one or more shaped queues connected to the shared queue. Thus, the input arrival curve $\alpha_{Q_i}^{\ATS}\!(t)$ of aggregate flows before the shared queue $Q_i$ is the sum of output arrival curves from all shaped queues $q$ connected to $Q_i$,
\begin{equation}\label{g:aggArr_SP_ATS+SP}
	\alpha_{Q_i}^{\ATS}\!(t)=\sum_{q} \alpha^*_q(t),
\end{equation}
where $\alpha^*_q(t)$ is from Eq.~(\ref{g:aggOutArrShpQ}). Note that frames in all shaped queues $q$ connected to the same shared queue $Q_i$ have the same priority.

By applying $\alpha_{Q_i}^{\ATS}$ and $\beta_{Q_i}^{\ATS}\!(t)$ into Eq.~(\ref{maxDQ}) and Eq.~(\ref{maxBQ}), the upper bound of latency $D^{\ATS}_{Q_i}$ and backlog $B^{\ATS}_{Q_i}$ for SP flows passing through the shared queue $Q_i$ can be given.

\begin{figure}[!t]
	\centering
	\includegraphics[width=0.5\textwidth]{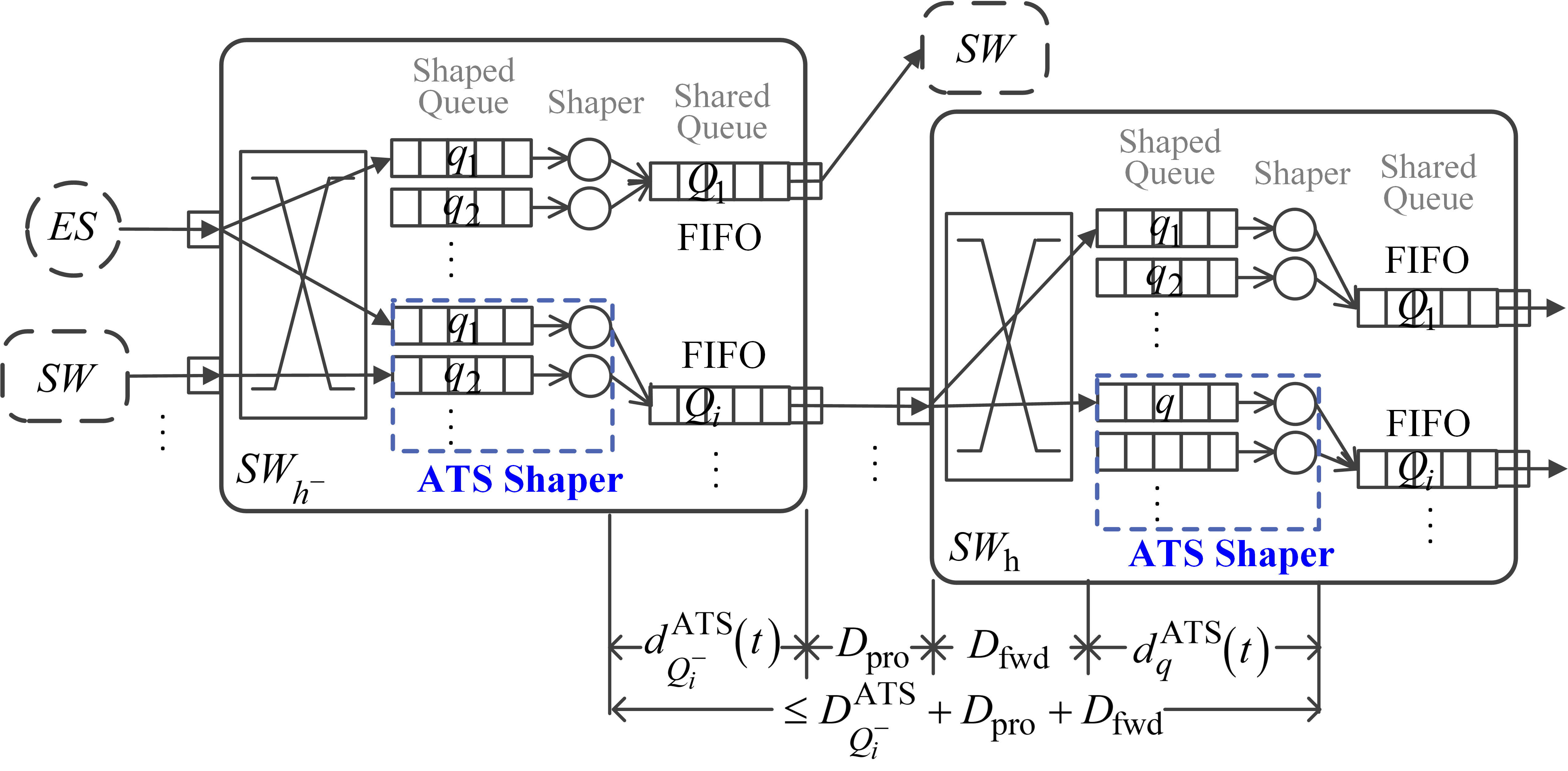}
	\caption{\label{fig:ATSTrans} Latency illustration for ATS Architecture}
\end{figure}

\textbf{\underline{Service Curve $\beta^{\ATS}_q(t)$ - ATS - Shaped Queue}.}
As proved by Theorem 5 in~\cite{LeBoudec18}, the ATS shaper is a kind of minimal interleaved regulator, which has the characteristic that placing it at the back-end of the FIFO system will not introduce extra worst-case delay, i.e., the worst-case delay in the combined system of the front-end FIFO system and the ATS shaper is the same as the worst-case delay of the front-end FIFO system alone. Obviously, the shared queue is served in a FIFO manner. 
Thus, a flow fed to the shaped queue $q$ on the subsequent node will not increase the upper bound of the delay for the flow waiting in the combined element of shared queue $Q_i^-$ on the preceding node and the shaped queue $q$, i.e., $d^{\ATS}_{Q_i^-}(t)+d^{\ATS}_q(t)\leq D^{\ATS}_{Q_i^-}$, where $D^{\ATS}_{Q_i^-}$ is the latency upper bound of SP flows with priority $i$ waiting in the preceding shared queue $Q_i^-$ and can be calculated by applying $\alpha_{Q_i^-}^{\ATS}(t)$ (Eq.~(\ref{g:aggArr_SP_ATS+SP})) and $\beta_{Q_i^-}^{\ATS}\!(t)$ (Eq.~(\ref{g:aggSer_ShrQ_SP_ATS})) to Eq.~(\ref{maxDQ}), as shown for example in Fig.~\ref{fig:ATSTrans} with two switches under ATS architecture in sequence. In the figure, $D_{\text{pro}}$ and $D_{\text{fwd}}$ are respectively propagation and forwarding delays, which are considered as constant in the paper. 
Then, for those flows transmitting from $Q_i^-$ to $q$, as their lower bound of the delay in the shared queue $Q_i^-$ is $l_{q}^{\min}/C$, the maximum latency $D^{\ATS}_{q}$ of SP flows waiting in the shaped $q$ can be given by,
\begin{equation}\label{g:maxDq}
	D^{\ATS}_{q}=D^{\ATS}_{Q_i^-}-l_{q}^{\min}/C.
\end{equation}
Note that not all the flows in the shared queue $Q_i^-$ will enter into the same shaped queue $q$, as they may be forwarded to the other egress port of the subsequent node. Moreover, according to the ATS queuing schemes QAR1 and QAR2, flows queuing in the shaped queue $q$ can only come from the same preceding shared queue $Q_i^-$.

Then, the service curve $\beta^{\ATS}_q(t)$ for aggregate flows in the shaped queue $q$ can be given by means of the pure-delay function~\cite{Mohammadpour18}
\begin{equation}\label{g:aggSer_ShpQ}
	\beta^{\ATS}_q(t)=\delta_{D}^q(t)
\end{equation}
where $\delta_{D}^{q}(t)$ is the pure-delay function~\cite{LeBoudec01} which equals to 0 if $t\leq D$ and $+\infty$ otherwise, while $D=D^{\ATS}_q$ (Eq.~\ref{g:maxDq}) is the delay upper bound of flows in the shaped queue $q$.

\textbf{\underline{Input Arrival Curve $\alpha^{\ATS}_{q}(t)$ - ATS - Shaped Queue}.}
The input arrival curve $\alpha^{\ATS}_{q}(t)$ of aggregate flows before reaching the corresponding shaped queue $q$ is related to the output arrival curve $\alpha^{*}_{Q_i^-}(t)$ when these flows depart the preceding shared queue $Q_i^-$, and can be given by,
\begin{equation}\label{g:aggOutArr_ShpQ_TAS+SP}
	\alpha^*_{Q_i^-}(t)=\sum_{f\in [Q_i^-,q]}\alpha_f^{Q_i^-}\!(t) \oslash\delta_{D}^{Q_i^-}\!(t),
\end{equation}
where $\alpha_f^{Q_i^-}\!(t)=r_f\cdot t+b_f$ is the input arrival curve of the flow $f$ before the shared queue $Q_i^-$, $\delta_{D}^{Q_i^-}\!(t)$ is the pure-delay function of the delay bound $D=D_{Q_i^-}^{\ATS}$ for aggregate SP flows of priority $i$ in the preceding shared queue $Q_i^-$. Note that we use $f\in[Q_i^-,q]$ instead of $f\in Q_i^-$ to emphasize that not all flows queuing in the shared queue $Q_i^-$ will be forwarded to the same shaped queue $q$.

Moreover, different flows sharing the same shaped queue $q$ cannot arrive on the shared queue $Q_i$ at the same time, because flows sharing a common link are serialized. Thus, by taking all flows from $Q_i^-$ to $q$ as a group, their output arrival curve from $Q_i^-$ can be refined by considering the constraint from the physical link with the shaping curve $\sigma^{\link}(t)=C\cdot t$.
According to the ATS queuing schemes QAR1 and QAR2, all flows in the shaped queue $q$ must be from the same preceding shared queue $Q_i^-$. Then, the input arrival curve $\alpha^{\ATS}_{q}(t)$ of aggregate flows before the shaped queue $q$ is given by,
\begin{equation}\label{g:aggArr_ShpQ_TAS+SP}
	\alpha^{\ATS}_{q}(t)=\alpha^{*}_{Q_i^-}(t)\wedge\left(\!\sigma^{\link}(t)\!+\!l_{Q_i^-}^{\max}\!\right),
\end{equation}
where $\alpha^{*}_{Q_i^-}(t)$ is given by Eq.~(\ref{g:aggOutArr_ShpQ_TAS+SP}), $x\wedge y = \min\{x,y\}$, and $l_{Q_i^-}^{\max}$ is the maximum frame size in the queue $Q_i^-$, which needs to be taken into account since the frame is packetized at the switch input.

By applying $\alpha^{\ATS}_{q}(t)$ and $\beta^{\ATS}_q(t)$ into Eq.~\ref{maxBQ}, the upper bound of backlog $B_q^{\ATS}$ in the shaped queue $q$ can be given.

\textbf{\underline{Remark}.}
As discussed above, the ATS shaper only plays a role in reshaping the flows, and will not increase the worst-case delay of the flow in the node transmission. Thus, the end-to-end latency bound for the flow can be obtained by summing up latency bounds $D_{Q_i}^{\ATS}(t)$ only for each shared queue along its path. The backlogs for ATS are the sum of backlogs for the shared queue and the corresponding shaped queue connected to the shared queue,
\begin{equation}\label{g:B_ATS}
	B^{\ATS}_{Q}=B^{\ATS}_{Q_i}+\sum_{q\rightarrow Q_i}B^{\ATS}_{q},
\end{equation}
where $q\rightarrow Q_i$ represents all the shaped queues $q$ connected to the shared queue $Q_i$. This due to that shaped and shared queues are implemented in a single physical queue. It should also note that the maximum backlogs in the shared queue $B^{\ATS}_{Q_i}$ and shaped queues $B^{\ATS}_{q}$ cannot be reached at the same time.

\subsection{Credit-Based Shaper (CBS)}
\label{sec:CBS}
\begin{figure}[!t]
	\centering
	\includegraphics[width=0.5\textwidth]{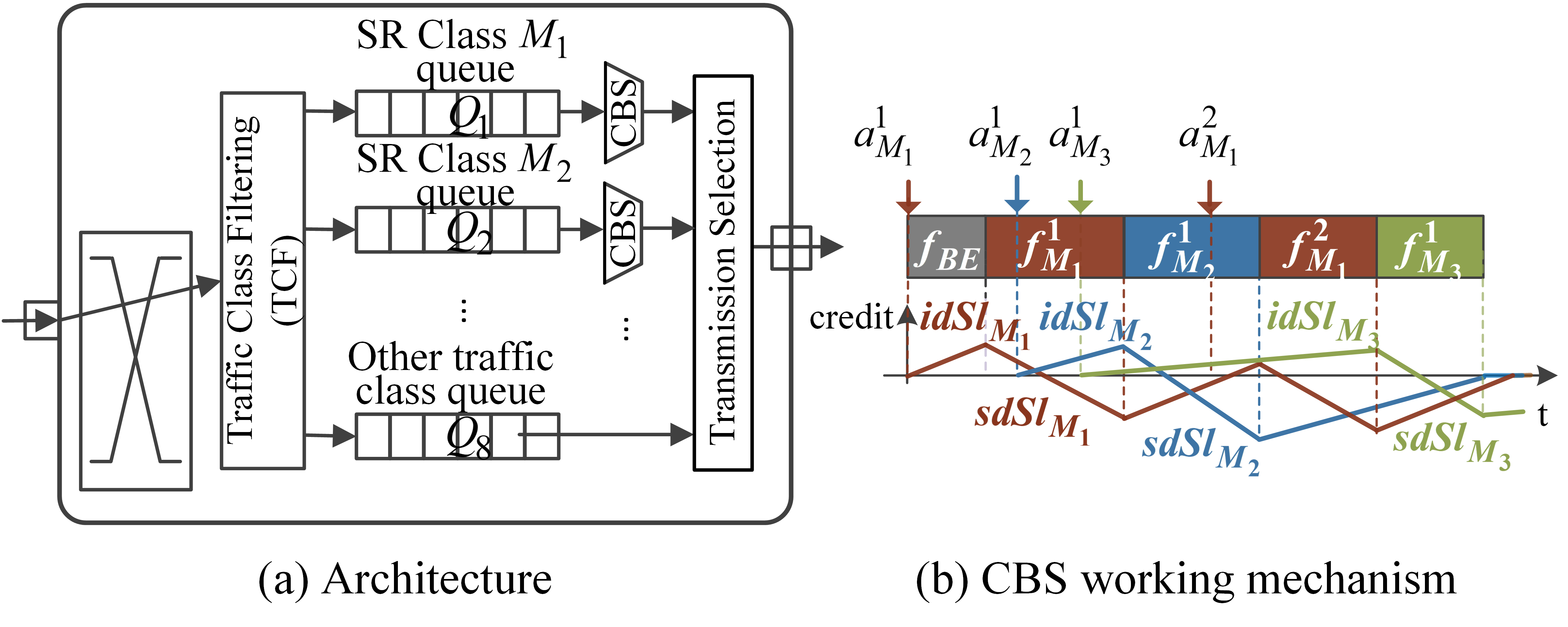}
	\caption{\label{fig:CBSArchitecture} CBS Architecture}
\end{figure}

The Credit-Based Shaper (CBS)~\cite{802.1Qav} is another queuing and forwarding rule proposed for the bandwidth reservation for  Audio-Video Bridging (AVB) traffic. 
Fig.~\ref{fig:CBSArchitecture}(a) depicts a CBS architecture model of an output port of a node. Currently, AVB classes (i.e., Stream Reservation (SR) classes) $M_i$ are expanded from two to more (a maximum of seven, $i\leq 7$) priorities supported by TSN~\cite{802.1Q}. Each AVB class corresponds to a FIFO queue, and has its credit value for the CBS shaper, which is used to control the transmission of AVB frames. For each AVB Class $M_i$, the CBS algorithm has a credit value manipulated by two different parameters called ``idleSlope'' ($idSl_i$) and ``sendSlope'' ($sdSl_i=idSl_{i}-C$). For the AVB traffic of Class $M_i$, $idSl_{i}$ decides its maximum guaranteed bandwidth reservation, of which the minimum value is set according to the actual bandwidth usage of AVB Class $M_i$ traffic.

The frame transmission based on the CBS functionality is shown using an example in Fig.~\ref{fig:CBSArchitecture}(b). The credit is initialized to zero and is increasing with the idleSlope ($idSl_{i}$) when AVB frames are waiting to be transmitted due to other higher priority AVB frames or due to the negative credit and decreasing with the sendSlope ($sdSl_{i}$) during the transmission of an AVB frame. If the credit is positive and no frames are waiting in the corresponding queue, then the credit is set to zero. However, if no frames are waiting in the queue, but the credit of the corresponding queue is negative, it will increase with the idle slope until zero.

Since the standard 802.1Q~\cite{802.1Q} now supports a multiple number of AVB classes, we extend the previous analysis work of supporting two classes~\cite{Azua14} and three classes~\cite{ZhaoLin18} to an arbitrary number of AVB classes, as summarized in Sect.~\ref{sec:WCD_CBS}. The extension proof of credit bounds for an arbitrary number of AVB classes can be found in our previous work~\cite{Zhao20}. Although \cite{Zhao20} is the NC-based performance analysis for combined TAS and CBS, and TT transmission delays AVB traffic, AVB credits will not be affected by TT traffic if credit is frozen during the TT window and the protection interval (``guard band'' (GB)), which is one of the cases discussed in \cite{Zhao20}.

\subsubsection{Performance Analysis -- CBS~\cite{Azua14}}
\label{sec:WCD_CBS}
\textbf{\underline{Service Curve $\beta^{\CBS}_{Q_i}(t)$ - CBS}.}
The service for AVB traffic in the individual CBS architecture depends only on the credit state controlled by CBS. As opposed SP traffic, AVB traffic with low priority can obtain the service even if the queues of AVB traffic of higher priority are not empty. This is because the AVB traffic cannot transmit if the CBS credit of its corresponding class is negative. The guaranteed service for multiple numbers of AVB classes $M_i$ ($i\in[1,n_{\CBS}]$)~\cite{Azua14},
\begin{equation}\label{g:aggSerQ_AVB_CBS}
	\beta^{\CBS}_{Q_i}(t)=idSl_{i}\left[t-\frac{c_{i}^{\max}}{idSl_{i}}\right]^+,
\end{equation}
where $c_{i}^{\max}$ is the credit upper bound for AVB Class $M_i$,
\begin{equation}\label{g:creditMaxF}
	c_{i}^{\max}=idSl_{i}\cdot \frac{\sum_{j=1}^{i-1}c_{j}^{\min}-l^{\max}_{>i}}{\sum_{j=1}^{i-1}idSl_{j}-C},
\end{equation}
where 
$l^{\max}_{>i}=\max_{j>i}\{l^{\max}_{Q_{j}}, l^{\max}_{Q_{\BE}}\}$ is the maximum frame size in the traffic with the priority lower than priority $M_i$, and 
$c_{i}^{\min}$ is the lower bound of the credit of AVB Class $M_i$,
\begin{equation}\label{g:creditMin}
\	c_{i}^{\min}=sdSl_{i}\cdot \frac{l^{\max}_{Q_i}}{C}.
\end{equation}

\textbf{\underline{Input Arrival Curve $\alpha_{Q_i}^{\CBS}(t)$ - CBS}.}
The input arrival curve $\alpha_{Q_i}^{\CBS}(t)$ of aggregate AVB flows of Class $M_i$ before entering the corresponding queue $Q_i$ of the intermediate node is related to the output arrival curve $\alpha^{*}_{Q_i^-}(t)$ of these flows departing the corresponding preceding queues $Q_i^-$ connected to $Q_i$. The output arrival curve of aggregate flows from a preceding queue $Q_i^-$ is,
\begin{equation}\label{g:aggOutArr_AVB_CBS}
	\alpha^*_{Q_i^-}(t)=\sum_{f\in [Q_i^-,Q_i]}\alpha_f^{Q_i^-}\!(t) \oslash\delta_{D}^{Q_i^-}\!(t),
\end{equation}
where $\alpha_f^{Q_i^-}\!(t)$ is the input arrival curve of the AVB flow $f$ before $Q_i^-$, which needs to be iteratively calculated from the node before $Q_i^-$ by Eq.~(\ref{g:outputArr2}) until the source node ES is reached, $\delta_{D}^{Q_i^-}\!(t)$ is the pure-delay function of the delay upper bound $D=D_{Q_i^-}^{\CBS}$ for aggregate AVB flows of Class $M_i$ in the preceding queue $Q_i^-$.
Note that we use $f\!\in\![Q_i^-,Q_i]$ instead of $f\!\in\! Q_i^-$ to emphasize that not all flows queuing in $Q_i^-$ are forwarded to $Q_i$.

Similarly, all AVB flows from the same preceding queue $Q_i^-$ are regarded as a group. On one hand, due to the physical link constraint $\sigma^{\link}(t)$, they cannot arrive on $Q_i$ at the same time. On the other hand, such a group of flows is also constrained by the shaping curve $\sigma^{\CBS}_{Q^-_i}(t)$ of CBS, indicating the effect of CBS on the output of AVB traffic. The CBS shaping curve $\sigma^{\CBS}_{Q_i}\!(t)$ is a non-greedy shaping curve, which is constructed by the upper envelope of output accumulated bits of AVB Class $M_i$ from $Q_i$ in any time interval, 
\begin{equation}\label{g:ShapingCur_CBS}
	\sigma^{\CBS}_{Q_i}\!(t)=idSl_{i}\bigg[t+\frac{c_{i}^{\max}-c_{i}^{\min}}{idSl_{i}}\bigg],
\end{equation}
where $c_i^{\max}$ and $c_i^{\min}$ are upper and lower credit bounds respectively given by Eq.~(\ref{g:creditMaxF}) and Eq.~(\ref{g:creditMin}). The distinction between greedy and non-greedy shaping is clarified in Appendix~\ref{sec:NC}.
Finally, the input arrival curve $\alpha_{Q_i}^{\CBS}(t)$ of aggregate AVB flows of Class $M_i$ before $Q_i$ is given by,
\begin{equation}\label{g:aggArr_AVB_CBS}
	\begin{split}
		\alpha_{Q_i}^{\CBS}(t)\!=\!\sum_{Q_i^-}\!\left[\alpha\!^{*}_{Q_i^-}\!(t)\!\wedge\!\left(\!\sigma^{\link}\!(t)\!+\!l_{Q_i^-}^{\max}\!\right)\!\wedge\! \left(\!\sigma^{\CBS}_{Q_i^-}(t)\!+\!l_{Q_i^-}^{\max}\!\right)\!\right]\!.
	\end{split}
\end{equation}
The use of the term $l_{Q_i^-}^{\max}$ is because the frame is packetized at the switch input.\footnote{The detailed derivations~\cite{Azua14} of CBS service curve in Eq.~(\ref{g:aggSerQ_AVB_CBS}) and CBS shaping curve in Eq.~(\ref{g:ShapingCur_CBS}) are not the contributions of the paper. But they are concluded in the supplementary document with the uniformly symbols: \url{https://zenodo.org/record/6378112\#.YjqQReeZNPY}.}

By applying $\alpha_{Q_i}^{\CBS}(t)$ and $\beta^{\CBS}_{Q_i}(t)$ into Eq.~(\ref{maxDQ}) and Eq.~(\ref{maxBQ}), the upper bound of latency $D^{\CBS}_{Q_i}$ and backlog $B^{\CBS}_{Q_i}$ for AVB flows of Class $M_i$ passing through the queue $Q_i$ can be calculated.

\section{Performance Analysis of Combined Traffic Shapers}
\label{sec:PerformanceCmpComb}
In this section, we discuss the combination of different traffic shapers. We will first show the combination of ATS and CBS used in different queues. Moreover, as we will show in Sect.~\ref{sec:CmpIndi}, from the perspective of latency, jitter and backlog, TAS outperforms ATS, CBS and SP. However, TAS requires the synthesis of optimized GCLs, which does not scale to large networks with many flows. This problem can be mitigated by combining different traffic shapers in the same switch architecture to reduce the number of flows handled by TAS.
Therefore, we believe that the coexistence of time-triggered shapers (TAS) and various event-triggered shapers (ATS, CBS, SP) will be a promising approach in the time-critical and real-time communication networks of the future, to support different performance quality requirements of applications, see also the discussion in~\cite{Gavrilut20}. Three combined traffic shapers investigated in the literature are non-time-triggered-CDT+ATS+CBS~\cite{Mohammadpour18}, TAS+SP inherited from TTEthernet~\cite{Zhao17} and TAS+CBS~\cite{Zhao18,Zhao20}. However, CDT is not a time-triggered traffic type, and the CDT model is not a standard model required by the TSN. Similarly, we add a citation after the subtitles of the performance analysis sections for TAS+SP and TAS+CBS. Inspired by the high performance of TAS, the ATS reshaping function and the existing combined traffic shapers, we are interested in understanding the impact of ATS reshaping on the combined architectures that include TAS. Thus, in the following, we propose additional three architectures of combined traffic shapers, i.e., TAS+SP, TAS+ATS+SP, TAS+ATS+CBS. We extend the NC approach to analyze the worst-case performance of traffic under these architectures for the quantitative performance comparison in Sect.~\ref{sec:CmpComb}.

\begin{figure}[!t]{}
	\centering
	\includegraphics[width=0.48\textwidth]{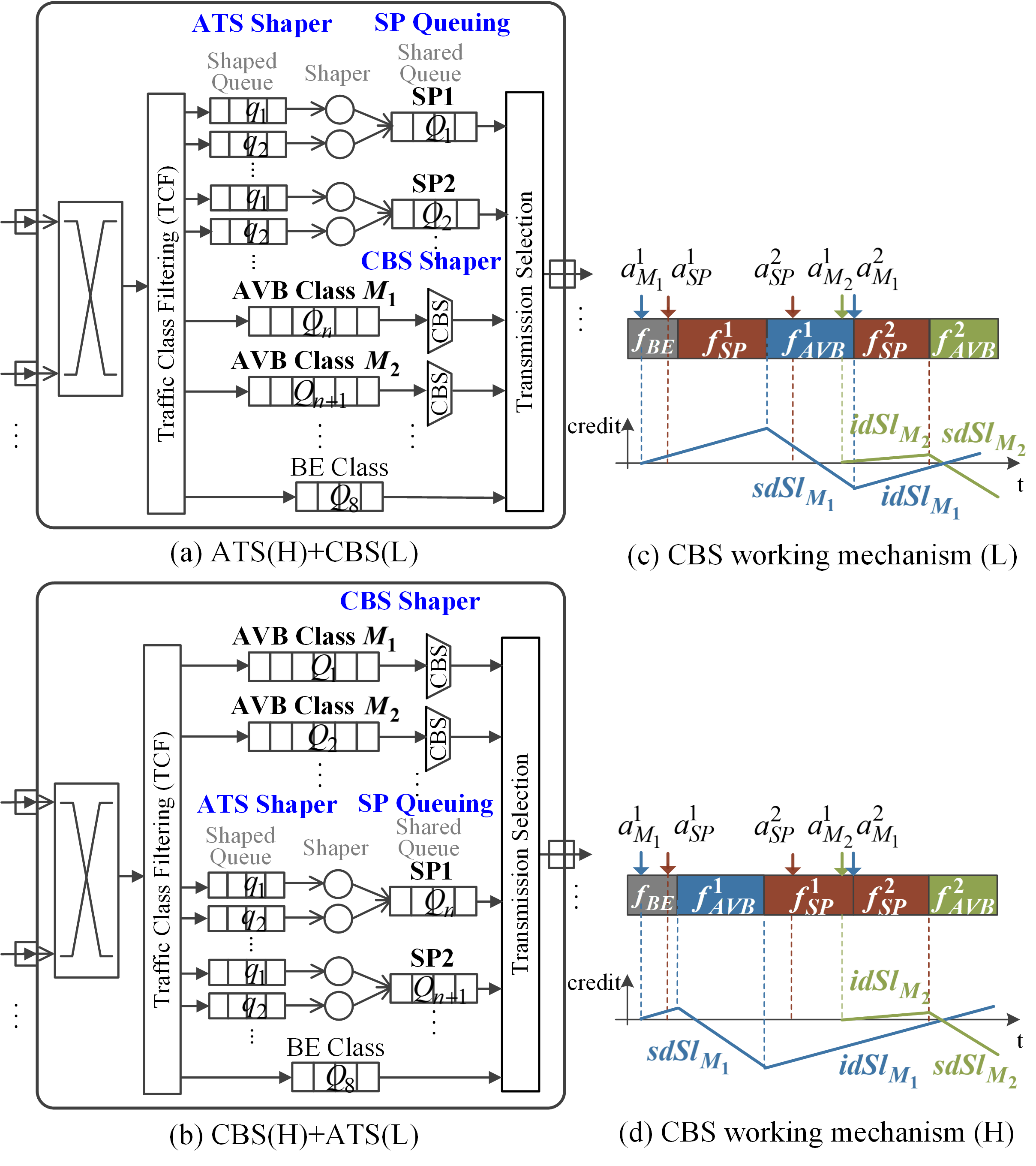}
	\caption{ATS+CBS/CBS+ATS Combined Shaper Architecture (H/L: high/low priority)}
	\label{fig:ATS+CBS^CBS+ATS}
\end{figure}

\subsection{ATS+CBS / CBS+ATS}
\label{sec:ATS+CBS}
In this section, ATS and CBS used for different queues at the same egress port are considered. The working mechanism of ATS and CBS used in combination follows their separate scheduling ways. The only thing that needs to be additionally considered is to set the priority for ATS and CBS queues. In this paper, it is assumed two cases: ATS and CBS respectively, have the higher priority, of which architectures are respectively shown in Fig.~\ref{fig:ATS+CBS^CBS+ATS}(a) and (b). And there are $n_{\SP}$ queues and $n_{\CBS}$ queues. We presume in this paper the non-preemptive integration mode for different traffic types. 
The frame transmission based on the CBS functionality is shown with an example in Fig.~\ref{fig:CBSArchitecture}(c), (d). The credit behavior is very similar to CBS individually used. Note that when CBS is used for low priority traffic, the credit of all AVB classes will be increased when the high priority SP frame is on transmission. 
In the following, the NC approach is expanded to the performance analysis of ATS and AVB traffic in the ATS+CBS and CBS+ATS architectures.

\subsubsection{Performance Analysis -- ATS+CBS}
\label{sec:ATS+CBS_DiffQ}
Since the flows shaped by ATS have the highest priority, and there will be at most one non-preemptive AVB frame that can impact their transmission. Therefore, the performance analysis for ATS is completely the same as ATS individually used, as discussed in Sect.~\ref{sec:PA_ATS}. Note that the only difference is that $l^{\max}_{>i}$ in the service curve $\beta_{Q_i}^{\ATS}$ (Eq.~(\ref{g:aggSer_ShrQ_SP_ATS})) should additionally consider the maximum frame size of AVB traffic with low priority. Then, in this section, we only focus on CBS for low-priority traffic.
%\begin{corollary}\label{Th:analysisATS_ATS+CBS}
%	For high priority traffic shaped by ATS in the architecture ATS+CBS, the NC-based performance analysis expressions, including the service curve $\beta_{Q_i}^{\ATS}\!(t)$ and the arrival curve $\alpha_{Q_i}^{\ATS}\!(t)$ in the share queue and the service curve $\beta^{\ATS}_q(t)$ and the arrival curve $\alpha^{\ATS}_{q}(t)$ in the shaped queue, are the same with ones (Eq.~(\ref{g:aggSer_ShrQ_SP_ATS}), Eq.~(\ref{g:aggArr_SP_ATS+SP}), Eq.~(\ref{g:aggSer_ShpQ}), Eq.~(\ref{g:aggArr_ShpQ_TAS+SP})) discussed in Sect.~\ref{sec:PA_ATS}. Note that $l^{\max}_{>i}$ in the service curve $\beta_{Q_i}^{\ATS}$ should additionally consider the maximum frame size of AVB traffic with low priority.
%\end{corollary}

%\emph{Proof}:
%The corollary can be easily derived as the flows shaped by ATS has the highest priority, and thus at most one non-preemptive AVB frame can impact their transmission.

%\hfill $\blacksquare$

\textbf{\underline{Service Curve $\beta^{\CBS}_{Q_i}(t)$ - CBS}.}
\begin{corollary}\label{Th:aggSerQ_CBS_ATS+CBS}
	With the impact of high priority SP traffic shaped by ATS, the service curve $\beta^{\CBS}_{Q_i}(t)$ for low priority AVB traffic of Class $M_i$ ($i\in[1,n_{\CBS}]$) is same to Eq.~(\ref{g:aggSerQ_AVB_CBS}),
%	\begin{equation}\label{g:aggSerQ_AVB_ATS+CBS}
%		\beta^{\CBS}_{Q_i}(t)=idSl_{i}\left[t-\frac{c_{i}^{\max}}{idSl_{i}}\right]^+,
%	\end{equation}
	where the credit upper bound $c_{i}^{\max}$ for AVB Class $M_i$ is given by,
	\begin{equation}\label{g:creditMaxF_ATS+CBS}
		c_{i}^{\max}=idSl_{i}\cdot\frac{\sum\limits_{j=1}^{i-1}c_{j}^{\min}-\sum\limits_{k=1}^{n_{\SP}}\sum\limits_{q\rightarrow Q_k}\sum\limits_{f\in q}b_f-l_{>i}^{\max}}{\sum\limits_{j=1}^{i-1}idSl_{j}+\sum\limits_{k=1}^{n_{\SP}}\sum\limits_{q\rightarrow Q_k}\sum\limits_{f\in q}r_f-C},
	\end{equation}
	where $c_{j}^{\min}$ is the lower bound of credit of AVB Class $M_j$ given by Eq.~(\ref{g:creditMin}), $q\rightarrow Q_i$ represents all the shaped queues $q$ connected to the shared queue $Q_i$, $b_f$ and $r_f$ are respectively the burst and rate after reshaped by ATS, 
	$l^{\max}_{>i}=\max_{j>i}\{l^{\max}_{Q_{j}}, l^{\max}_{Q_{\BE}}\}$ is the maximum frame size in the traffic with the priority lower than priority $M_i$.
\end{corollary}

\emph{Proof}:
It is assumed that $R_{Q_i}^{\CBS}(t)$ (resp. $R_{Q_i}^{\ATS}(t)$) and $R_{Q_i}^{*\CBS}(t)$ (resp. $R_{Q_i}^{*\ATS}(t)$) are the arrival and departure processes of AVB flows of Class $M_i$ ($i\in[1,n_{\CBS}]$) (resp. SP flows of priority $i$ ($i\in[1,n_{\SP}]$) reshaped by ATS) crossing through an egress port. Let $t\in \mathbb{R}^+$ be a time point when the queue $Q_{i}$ of AVB Class $\M_i$ is backlogged, i.e., $R_{Q_i}^{*\CBS}(t)<R_{Q_i}^{\CBS}(t)$. Then let us define $s=\sup\{u\leq t~\vrule~c_{i}(u)=0, R_{Q_i}^{*\CBS}(u)=R_{Q_i}^{\CBS}(u)\}$, where $c_{i}(t)$ is the credit value of AVB Class $M_i$ at time $t$. 
This implies that $\forall u\in(s,t]$, the queue $Q_i$ is non-empty or $c_{i}(u)< 0$. Otherwise, we can always find another $s<s'\leq t$ that satisfies $c_{i}(s')=0$ and $R_{Q_i}^{*\CBS}(s')=R_{Q_i}^{\CBS}(s')$. Therefore, we define the duration $(s,t]$ the busy period of AVB Class $M_i$.

According to the working mechanism of CBS under ATS+CBS, the interval $\Delta t=t-s$ can be decomposed by
\begin{equation*}\label{g:delta_t}
	\Delta t=\Delta t^{-}+\Delta t^{+},
\end{equation*}
where $\Delta t^{-}$ (resp. $\Delta t^{+}$) is the accumulated length of all periods where the credit is decreasing (resp. increasing). $\Delta t^{-}=\Delta t_{Q_i}^{\CBS}$ represents the transmission duration of AVB Class $M_i$, and $\Delta t^+=\Delta t_{Q_j}^{\CBS}$ is the waiting time duration of AVB Class $M_i$ due to the transmission for higher priority SP flows reshaped by ATS or the transmission for other AVB Classes or due to the credit $c_i(t)<0$.
%$\Delta t^{+}=\Delta t_{M_j}^{j<i}+\Delta t_{M_k}^{k>i}+\Delta t_{GB}+\Delta t_{OH}^{H}+\Delta t_{OH}^{L}$. 
And we have
\begin{equation*}\label{g:delta_t+}
	\Delta t^{+}=\Delta t-\Delta t_{Q_i}^{\CBS}.
\end{equation*}

Due to the definition of $s$, for $\forall u\in(s,t]$, the queue $Q_{i}$ for AVB traffic is not possible to temporarily empty when $c_{i}(u)>0$. Thus, there is no case where the credit of AVB Class $\M_i$ is reduced from some positive value $P$ to 0 due to resets.
Then the variation of credit for AVB Class $M_i$ during the time interval $\Delta t=t-s$ satisfies,
\begin{equation*}\label{g:creditVar}
	\begin{split}
		c_{i}(t)&-c_{i}(s)=c_{i}(t)\\
		&=\Delta t^{+}\cdot idSl_{i}+\Delta t^-\cdot sdSl_{i}\\
		&=(\Delta t-\Delta t_{Q_i}^{\CBS})\cdot idSl_{i}+\Delta t_{Q_i}^{\CBS}\cdot sdSl_{i}\\
		&=\Delta t\cdot idSl_i-\Delta t_{Q_i}^{\CBS}\cdot C,
	\end{split}
\end{equation*}
Thus, it is obtained the expression of service times for AVB Class $\M_i$ in any interval $\Delta t$,
\begin{equation*}\label{g:delta_t_Mi}
	\begin{split}
		\Delta t_{Q_i}^{\CBS}=\frac{\Delta t\cdot idSl_i-c_{i}(t)}{C}.
	\end{split}
\end{equation*}

Since the service could only be supplied for AVB traffic $M_i$ during $\Delta t_{Q_i}^{\CBS}$, i.e., the descent time $\Delta t^-$ of the credit, then over the interval $(s,t]$, we have
\begin{equation*}
	\begin{split}
		R_{Q_i}^{*\CBS}(t)-R_{Q_i}^{*\CBS}(s)=C\cdot\Delta t_{Q_i}^{\CBS}\geq
		\Delta t\cdot idSl_i-c_{i}^{\max},
	\end{split}
\end{equation*}
where $c_{i}(t)\leq c_{i}^{\max}$.
Since we have also $R_{Q_i}^{*\CBS}(t)-R_{Q_i}^{*\CBS}(s)=R_{Q_i}^{*\CBS}(t)-R_{Q_i}^{\CBS}(s)\geq 0$ and $R_{Q_i}^{*\CBS}(t)$ is a wide-sense increasing function, from which we derive
\begin{equation*}
	\begin{split}
		R_{Q_i}^{*\CBS}&(t)\geq R_{Q_i}^{\CBS}(s)+idSl_i\left[t-s-\frac{c_i^{\max}}{idSl_i}\right]^+\\
		&\geq \inf_{0\leq s\leq t}\left\{R_{Q_i}^{\CBS}(s)+idSl_i\left[t-s-\frac{c_i^{\max}}{idSl_i}\right]^+\right\}.
	\end{split}
\end{equation*}
Therefore, the service curve $\beta_{Q_i}^{\CBS}(t)$ for AVB Class $\M_i$ is $\beta_{Q_i}^{\CBS}(t)=
idSl_i\left[t-\frac{c_i^{\max}}{idSl_i}\right]^+$, which is same to the expression in Eq.~(\ref{g:aggSerQ_AVB_CBS}).

In the following, we will derive the credit upper bound $c_i^{\max}$ for each AVB Class $M_i$ ($i\in[1,n_{\CBS}]$). 
Let $t\in \mathbb{R}^+$ be a time point when AVB gates are in the open state, and $c_{i}(t)>0$. Then let us define $s=\sup\{u\leq t~\vrule~\forall Q_{j}\in Q_{\leq i}^{\CBS}, c_{j}(u)\leq 0\}$, where $Q_{\leq i}^{\CBS}$ are the queues with priority no lower than AVB Class $M_i$. 
This implies that $\forall u\in(s,t]$, $\exists Q_{j}\in Q_{\leq i}^{\CBS}$, $c_{j}(u)>0$, \emph{i.e.}, there always exists at least one queue in $Q_{\leq i}^{\CBS}$ with some frame to send. Otherwise, we can always find another $s<s'\leq t$ that satisfies $\forall Q_{j}\in Q_{\leq i}^{\CBS}, c_{j}(s')\leq 0$.

Consider the evolution of the credit value $c_{i}(t)$ of AVB Class $M_i$ between $s$ and $t$. 
The credit $c_{i}(t)$ 1) decreases at speed $sdSl_{i}=idSl_{i}-C$ when the frame in the queue $Q_{i}$ is on transmission (during $\Delta t_{Q_i}^{\CBS}$); 2) increases at speed $idSl_{i}$ when the frame in the queue $Q_{i}$ is waiting when SP flows shaped by ATS is on transmission (during $\Delta t_{\ATS}$) or when AVB flows with higher priority than $M_i$ is on transmission (during $\Delta t_{<i}^{\CBS}$) or a non-preemptive lower priority frame than $M_i$ is on transmission (during $\Delta t_{\LP}$); 3) and may be reduced from some positive value $P$ to 0 due to resets. Thus the variation of $c_{i}(t)$ during $(s,t]$ is,
\begin{equation*}\label{g:deltaMi(i)1}
	\begin{split}
		c_{i}(t)&-c_{i}(s)=\\&\Delta t_{Q_i}^{\CBS}\cdot sdSl_{i}
		+\left(\Delta t_{\ATS}+\Delta t_{<i}^{\CBS}+\Delta t_{\LP}\right)\cdot idSl_{i}-P.
	\end{split}
\end{equation*}
Since $\Delta t_{\ATS}+\Delta t_{<i}^{\CBS}+\Delta t_{\LP}=t-s-\Delta t_{Q_i}^{\CBS}$ and $P\geq0$, The above equation is modified into,
\begin{equation}\label{g:deltaMi(i)}
	\begin{split}
		c_{i}(t)-c_{i}(s)\leq -\Delta t_{Q_i}^{\CBS}\cdot C+(t-s)\cdot idSl_{i}.
	\end{split}
\end{equation}

Let $c_{<i}(t)=\sum_{j=1}^{i-1}c_{j}(t)$ denote the sum of credits of AVB traffic with a priority higher than Class $M_i$. At any instant between $s$ and $t$ the sum of credits $c_{<i}(t)$ 1) increases at most at speed $\sum_{j=1}^{i-1}idSl_{j}$ when a frame of SP traffic shaped by ATS is on transmission (during $\Delta t_{\ATS}$), or a frame of Class $M_i$ uses the link (during $\Delta t_{Q_i}^{\CBS}$), or a low priority frame blocks the link (during $\Delta t_{\LP}$); 
2) decreases at least at speed $\sum_{j=1}^{i-1}idSl_{j}-C$ (all the classes from $Q_{<i}^{\CBS}$ gain credit, except one which loses credit) when a frame from class with higher priority than Class $M_i$ is being sent (during $\Delta t_{<i}^{\CBS}$); 3) be reduced from some positive value $P$ to 0 due to a set of resets. Then the variation of $c_{<i}(t)$ between $s$ and $t$ is,
\begin{equation*}\label{g:c<i1(i)}
	\begin{split}
		c_{<i}(t)-c_{<i}(s)=&\left(\Delta t_{\ATS}+\Delta t_{Q_i}^{\CBS}+\Delta t_{\LP}\right)\cdot \sum_{j=1}^{i-1}idSl_{j}\\
		&+\Delta t_{<i}^{\CBS}\cdot \left(\sum_{j=1}^{i-1}idSl_{j}-C\right)-P\\
	\end{split}
\end{equation*}
Since $\Delta t_{<i}^{\CBS}=t-s-\Delta t_{\ATS}-\Delta t_{Q_i}^{\CBS}-\Delta t_{\LP}$ and $P\geq 0$, the above expression is modified into,
\begin{equation}\label{g:c<i(i)}
	\begin{split}
		c_{<i}(t)-c_{<i}(s)\leq&\left(\Delta t_{\ATS}+\Delta t_{Q_i}^{\CBS}+\Delta t_{\LP}\right)\cdot C\\
		&+(t-s)\cdot \left(\sum_{j=1}^{i-1}idSl_{j}-C\right).
	\end{split}
\end{equation}
For $\Delta t_{\ATS}$, we have,
\begin{equation*}\label{g:arr_ATS_ATS+CBS}
	\begin{split}
		\Delta t_{\ATS}\cdot C=\sum_{k=1}^{n_{\SP}}\left[R_{Q_k}^{\ATS}(t)-R_{Q_k}^{\ATS}(s)\right]\leq\sum_{k=1}^{n_{\SP}}\alpha_{Q_k}^{\ATS}(t-s),
	\end{split}
\end{equation*}
where $\alpha_{Q_k}^{\ATS}(t)$ is the arrival curve of aggregate SP flows of priority $k$ before the shared queue of ATS, given by Eq.~(\ref{g:aggArr_SP_ATS+SP}). Thus, the above equation can be written,
\begin{equation*}\label{g:arr_ATS_ATS+CBS1}
	\begin{split}
		\Delta t_{\ATS}\cdot C\leq\sum_{k=1}^{n_{\SP}}\sum_{q\rightarrow Q_k}\sum_{f\in q}(r_f(t-s)+b_f),
	\end{split}
\end{equation*}
where $q\rightarrow Q_i$ represents all the shaped queues $q$ connected to the shared queue $Q_i$. Moreover, since $\Delta t_{\LP}\cdot C\leq\max_{j\in[i+1,n_{\CBS}]} \{l_{Q_j}^{\max},l_{\BE}^{\max}\}=l_{>i}^{\max}$,Eq.~(\ref{g:c<i(i)}) can be rewritten to,
\begin{equation}\label{g:t-s-deltaTST}
	\begin{split}	
		t-&s \leq \\
		&\frac{c_{<i}(t)-c_{<i}(s)-\Delta t_{Q_i}^{\CBS}\cdot C-\sum\limits_{k=1}^{n_{\SP}}\sum\limits_{q\rightarrow Q_k}\sum\limits_{f\in q}b_f-l_{>i}^{\max}}{\sum\limits_{j=1}^{i-1}idSl_{j}+\sum\limits_{k=1}^{n_{\SP}}\sum\limits_{q\rightarrow Q_k}\sum\limits_{f\in q}r_f-C}.
	\end{split}
\end{equation}
By applying Eq.~(\ref{g:t-s-deltaTST}) into Eq.~(\ref{g:deltaMi(i)}), it is obtained that
\begin{equation*}\label{g:deltaMi(i)2}
	\begin{split}
		c_{i}(t)&-c_{i}(s)\leq\\
		&-\frac{\sum\limits_{j=1}^{i}idSl_{j}+\sum\limits_{k=1}^{n_{\SP}}\sum\limits_{q\rightarrow Q_k}\sum\limits_{f\in q}r_f-C}{\sum_{j=1}^{i-1}idSl_{j}+\sum\limits_{k=1}^{n_{\SP}}\sum\limits_{q\rightarrow Q_k}\sum\limits_{f\in q}r_f-C}\cdot\Delta t_{Q_i}^{\CBS}\cdot C\\
		&+\frac{c_{<i}(t)-c_{<i}(s)-\sum\limits_{k=1}^{n_{\SP}}\sum\limits_{q\rightarrow Q_k}\sum\limits_{f\in q}b_f-l_{>i}^{\max}}{\sum\limits_{j=1}^{i-1} idSl_{j}+\sum\limits_{k=1}^{n_{\SP}}\sum\limits_{q\rightarrow Q_k}\sum\limits_{f\in q}r_f-C}\cdot idSl_{i}\\
		&\leq\frac{c_{<i}(t)-c_{<i}(s)-\sum\limits_{k=1}^{n_{\SP}}\sum\limits_{q\rightarrow Q_k}\sum\limits_{f\in q}b_f-l_{>i}^{\max}}{\sum\limits_{j=1}^{i-1} idSl_{j}+\sum\limits_{k=1}^{n_{\SP}}\sum\limits_{q\rightarrow Q_k}\sum\limits_{f\in q}r_f-C}\cdot idSl_{i}.
	\end{split}
\end{equation*}
By definition of $s$, we have $c_{<i}(s)\leq 0$, $c_{i}(s)\leq 0$, $\sum_{j=1}^{i-1} idSl_{j}+\sum_{k=1}^{n_{\SP}}\sum_{q\rightarrow Q_k}\sum_{f\in q}r_f-C<0$, and $c_{<i}(t)\geq \sum_{j=1}^{i-1}c_{j}^{\min}$, it is concluded that,
\begin{equation*}
	c_{i}(t)\leq idSl_{i}\cdot\frac{\sum\limits_{j=1}^{i-1}c_{j}^{\min}-\sum\limits_{k=1}^{n_{\SP}}\sum\limits_{q\rightarrow Q_k}\sum\limits_{f\in q}b_f-l_{>i}^{\max}}{\sum\limits_{j=1}^{i-1}idSl_{j}+\sum\limits_{k=1}^{n_{\SP}}\sum\limits_{q\rightarrow Q_k}\sum\limits_{f\in q}r_f-C}.
\end{equation*}
where $c_{j}^{\min}$ is the lower bound of credit of AVB Class $M_j$. Since an AVB frame cannot start to be forwarded if the credit is lower than 0, the minimum credit bound is reached when the size of the frame is maximized, which is the same as Eq.~(\ref{g:creditMin}).

\rightline{\rule{6pt}{6pt}}

\textbf{\underline{Input Arrival Curve $\alpha_{Q_i}^{\CBS}(t)$ - CBS}.}
\begin{corollary}\label{Th:aggArrQ_CBS_ATS+CBS}
	With the impact of high priority SP traffic shaped by ATS, the input arrival curve $\alpha_{Q_i}^{\CBS}(t)$ for low priority AVB traffic of Class $M_i$ ($i\in[1,n_{\CBS}]$) is the same to Eq.~(\ref{g:aggArr_AVB_CBS}), where the CBS shaping curve $\sigma^{\CBS}_{Q_i}(t)$ is given from Eq.~(\ref{g:ShapingCur_CBS}) by replacing $c_{i}^{\max}$ with Eq.~(\ref{g:creditMaxF_ATS+CBS}).
\end{corollary}

The proof is the same as the proof of $\sigma^{\CBS}_{Q_i}(t)$ when CBS individually used.

\subsubsection{Performance Analysis -- CBS+ATS}
Similar to the ATS+CBS in conversely used, flows passing through CBS have the highest priority and their performance analyses are the same as the analysis (Sect.~\ref{sec:WCD_CBS}) when CBS individually used. It should also be noted that $l_{>i}^{\max}$ in the credit upper bound $c_i^{\max}$ should take the maximum frame size of flows reshaped by ATS into account. Thus, in this section, we only focus on ATS for low-priority traffic.

\textbf{\underline{Service Curve $\beta_{Q_i}^{\ATS}\!(t)$ - ATS - Shared Queue}.}
\begin{corollary}\label{Th:aggArrSharQ_ATS_CBS+ATS}
With the impact of high priority AVB traffic, the service curve $\beta^{\ATS}_{Q_i}(t)$ for low priority SP traffic reshaped by ATS in the corresponding shared queue $Q_i$ ($i\in[1,n_{\SP}]$) is given by,
\begin{equation}\label{g:aggSer_ShrQ_SP_CBS+ATS}
	\beta_{Q_i}^{\ATS}\!(t)=
	C\left[ t -
	\frac{\sum_{k=1}^{n_{\CBS}}\alpha^{\CBS}_{Q_k}\!(t)+\sum_{j=1}^{i-1}\alpha^{\ATS}_{Q_j}\!(t)}{C}-\frac{l_{>i}^{\max}}{C} \right]^+,
\end{equation}
where $\alpha^{\CBS}_{Q_k}\!(t)$ (Eq.~(\ref{g:aggArr_AVB_CBS})) is the arrival curve of aggregate AVB flows of Class $M_k$ ($k\in[1,n_{\CBS}]$), $\alpha^{\ATS}_{Q_j}\!(t)$ (Eq.~(\ref{g:aggArr_SP_ATS+SP})) is the arrival curve of SP flows after ATS reshaping with the priority $j$ higher than the priority $i$, and $l^{\max}_{>i}=\max_{j\in[i+1,n_{\SP}]}\{l^{\max}_{Q_{j}}, l^{\max}_{Q_{\BE}}\}$ that is the maximum frame size of traffic with the priority lower than priority $i$.
\end{corollary}

\emph{Proof}:
Let $R_{Q_i}^{\ATS}(t)$ (resp. $R_{Q_i}^{\CBS}(t)$) and $R_{Q_i}^{*\ATS}(t)$ (resp. $R_{Q_i}^{*\CBS}(t)$) are the arrival and departure processes of SP flows of priority $i$ ($i\in[1,n_{\SP}]$) reshaped by ATS (resp. AVB flows of Class $M_i$ ($i\in[1,n_{\CBS}]$)). It is assumed that $t\in \mathbb{R}^+$ be a time point when the queue $Q_{i}$ of SP traffic of priority $i$ reshaped by ATS is backlogged, i.e., $R_{Q_i}^{*\ATS}(t)<R_{Q_i}^{\ATS}(t)$. Then, let $s=\sup\{u\leq t~\vrule~\forall k\in[1,n_{\CBS}], R_{Q_k}^{*\CBS}(u)=R_{Q_k}^{\CBS}(u); \forall j\in[1,i], R_{Q_j}^{*\ATS}(u)=R_{Q_j}^{\ATS}(u)\}$.

During the time interval $(s,t]$, flows of priority $i$ in the shared queue $Q_i$ will obtain the service only when the queue for higher priority traffic is empty. Moreover, at most one non-preemptable frame with lower priority is transmitted within $(s,t]$. Thus, we have
\begin{equation}\label{g:cumR_ATS_CBS+ATS}
	\begin{split}
		R_{Q_i}^{*\ATS}(t)-R_{Q_i}^{*\ATS}(s)&=C(t-s)-\sum_{k=1}^{n_{\CBS}} \left(R_{Q_k}^{*\CBS}(t)-R_{Q_k}^{*\CBS}(s)\right)\\&-\sum_{j=1}^{i-1}\left(R_{Q_j}^{*\ATS}(t)-R_{Q_j}^{*\ATS}(s)\right)-\Delta t_{\LP},
	\end{split}
\end{equation}
Since $\Delta t_{\LP}\cdot C\leq\max_{j\in[i+1,n_{\SP}]} \{l_{Q_j}^{\max},l_{\BE}^{\max}\}=l_{>i}^{\max}$, and
\begin{equation*}
	\begin{split}
		R_{Q_k}^{*\CBS}(t)-R_{Q_k}^{*\CBS}(s)&=R_{Q_k}^{*\CBS}(t)-R_{Q_k}^{\CBS}(s)\\
		&\leq R_{Q_k}^{\CBS}(t)-R_{Q_k}^{\CBS}(s)\leq \alpha^{\CBS}_{Q_k}(t-s) 
	\end{split}
\end{equation*}
where $\alpha^{\CBS}_{Q_k}(t)$ given by Eq.~(\ref{g:aggArr_AVB_CBS}) is the arrival curve of aggregate AVB flows of Class $M_k$. Similarly, $R_{Q_j}^{*\ATS}(t)-R_{Q_j}^{*\ATS}(s)\leq \alpha^{\ATS}_{Q_j}(t-s)$, where $\alpha^{\ATS}_{Q_j}(t)$ given by Eq.~(\ref{g:aggArr_SP_ATS+SP}) is the input arrival curve of aggregate flows before the shared queue $Q_j$. Then, Eq.~(\ref{g:cumR_ATS_CBS+ATS}) can be changed into,
\begin{equation*}\label{g:cumR_ATS_CBS+ATS_1}
	\begin{split}
		R_{Q_i}^{*\ATS}(t)-R_{Q_i}^{*\ATS}(s)&\geq C(t-s)-\sum_{k=1}^{n_{\CBS}} \alpha^{\CBS}_{Q_k}(t-s)\\&-\sum_{j=1}^{i-1}\alpha^{\ATS}_{Q_j}(t-s)-l_{>i}^{\max},
	\end{split}
\end{equation*}

Since we have also $R_{Q_i}^{*\ATS}(t)-R_{Q_i}^{*\ATS}(s)=R_{Q_i}^{*\ATS}(t)-R_{Q_i}^{\ATS}(s)\geq 0$ and $R_{Q_i}^{*\ATS}(t)$ is a wide-sense increasing function, from which we derive
\begin{equation*}
	\begin{split}
		R_{Q_i}^{*\ATS}(t)\geq R_{Q_i}^{\ATS}(s)&+\left[C(t-s)-\sum_{k=1}^{n_{\CBS}} \alpha^{\CBS}_{Q_k}(t-s)\right.\\
		&\left.-\sum_{j=1}^{i-1}\alpha^{\ATS}_{Q_j}(t-s)-l_{>i}^{\max}\right]^+\\
		&\geq \inf_{0\leq s\leq t}\left\{R_{Q_i}^{\ATS}(s)+\beta_{Q_i}^{\ATS}(t-s)\right\}.
	\end{split}
\end{equation*}
where the service curve $\beta_{Q_i}^{\ATS}(t-s)=[C(t-s)-\sum_{k=1}^{n_{\CBS}} \alpha^{\CBS}_{Q_k}(t-s)-\sum_{j=1}^{i-1}\alpha^{\ATS}_{Q_j}(t-s)-l_{>i}^{\max}]^+$.

\rightline{\rule{6pt}{6pt}}

Especially, for the	input arrival curve $\alpha_{Q_i}^{\ATS}\!(t)$ in the shared queue, and the service curve $\beta^{\ATS}_q(t)$ and the arrival curve $\alpha^{\ATS}_{q}(t)$ in the shaped queue, are the same with ones (Eq.~(\ref{g:aggArr_SP_ATS+SP}), Eq.~(\ref{g:aggSer_ShpQ}), Eq.~(\ref{g:aggArr_ShpQ_TAS+SP})) discussed in Sect.~\ref{sec:PA_ATS}.

\begin{figure}[!t]{}
	\centering
	\includegraphics[width=0.35\textwidth]{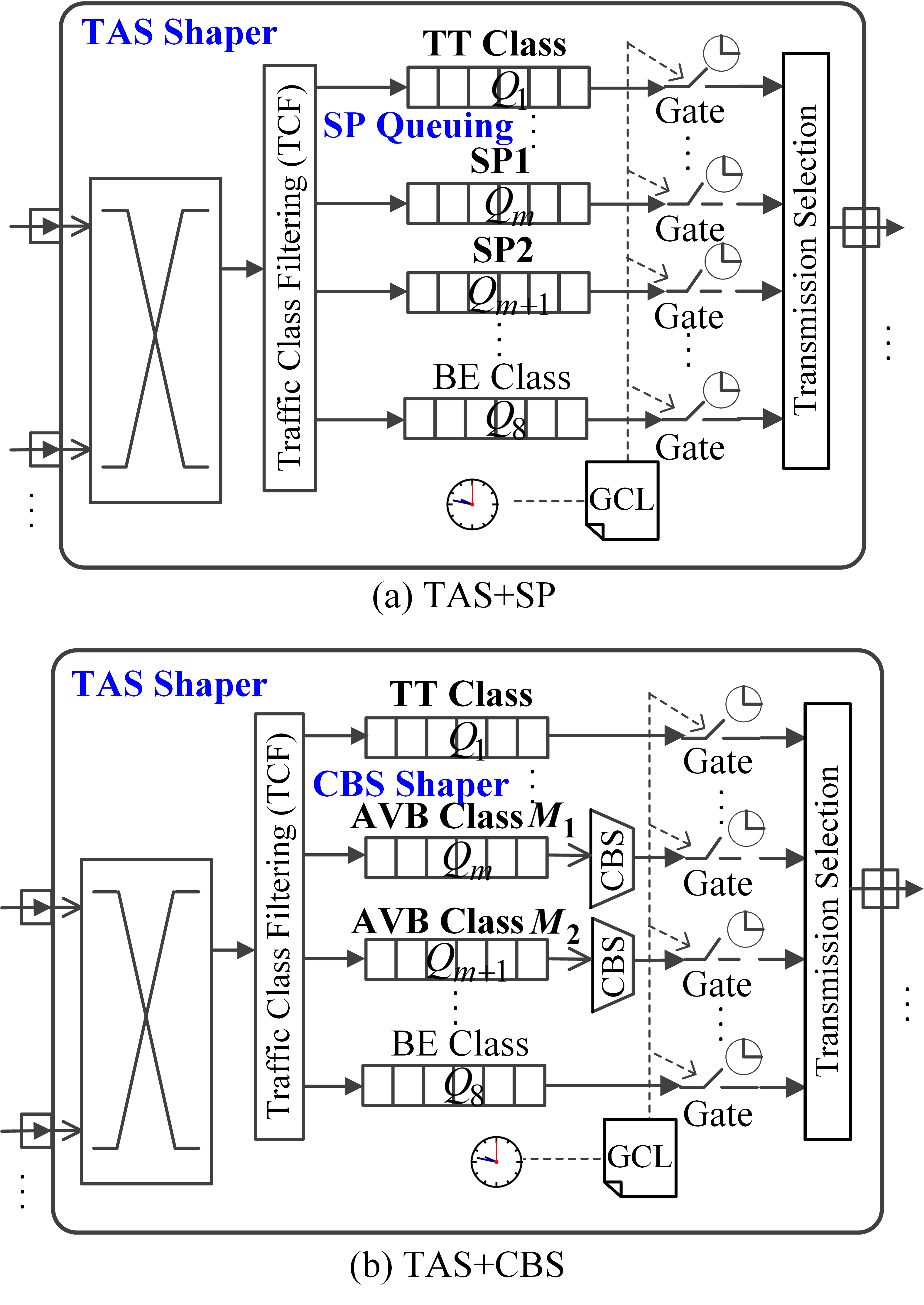}
	\caption{TAS+SP/CBS Combined Shaper Architecture}
	\label{fig:TAS+SP^CBS}
\end{figure}

\subsection{TAS+SP / TAS+CBS}
\label{sec:TAS+SP}
With the combination of the completely deterministic transmission of TAS, we first address the architecture without ATS.
One possible combination is that of the Time-Aware Shaper (TAS) and Strict Priority (SP) queuing (i.e., TAS+SP), shown in Fig.~\ref{fig:TAS+SP^CBS}(a). The combined scheduling mechanisms of TAS+SP are inherited from TTEthernet which supports Time-Triggered (TT) traffic and Rate-Constrained (RC) communication with a strict priority allocation.
The difference in TSN is how the TT frames are transmitted. In TTEthernet, TT communication is implemented by directly controlling the temporal behavior of each individual frame of the TT flows~\cite{Craciunas16:RTSJ}. However, in TSN, TT communication depends on the gate control for corresponding TT queues in egress ports, which requires flow or frame isolation constraints in order to achieve completely deterministic transmission~\cite{Craciunas16:RTNS, Pop16}. 
Another possible combination is that of TAS and the Credit-Based Shaper (CBS) (TAS+CBS)~\cite{Zhao20}. The TAS+CBS architecture is presented in Fig.~\ref{fig:TAS+SP^CBS}(b). 
In any combination, TT traffic implemented by TAS always has the highest priority. Thus, it will have the same high real-time performance as when individually used.
SP/AVB traffic has the secondary priority. Best-Effort (BE) traffic has the lowest priority without timing guarantee requirements.
Different from SP scheduling, which handles the flows based on their priorities, CBS enforces a bandwidth reservation for multiple priorities of AVB traffic. CBS is used to prevent the starvation of lower-priority AVB traffic, and can tolerate a certain degree of degradation in real-time performance for high-priority traffic.

\begin{figure}[!t]{}
	\centering
	\includegraphics[width=0.32\textwidth]{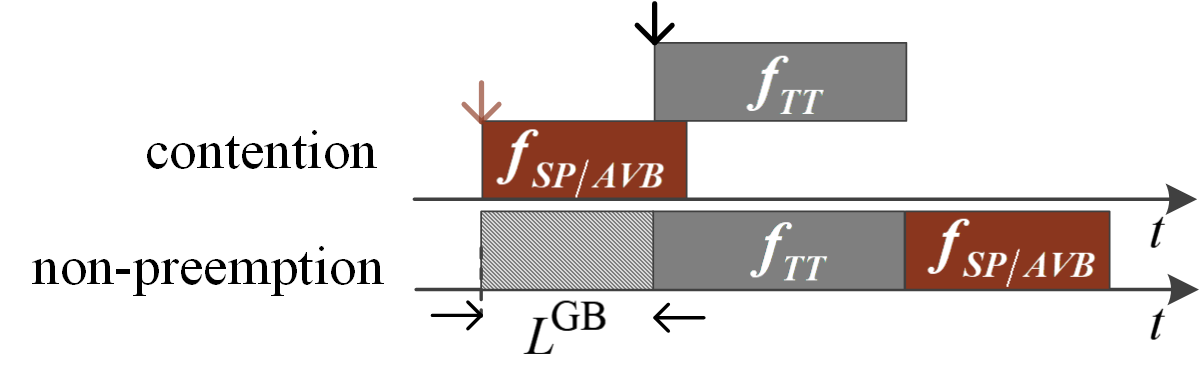}
	\caption{Non-preemption integration modes of TT and SP/AVB Traffic}
	\label{fig:IntegrationModes}
\end{figure}

With TSN, there is a gate for each queue of egress ports. Only when the gate is open, the frames in the corresponding queue can be forwarded. If more than one gate opens at the same time, the frame transmission is based on their priority. In order to keep the completely deterministic transmission for TT traffic, when an associated gate for TT traffic is open, the remaining gates for other traffic types (SP, AVB, etc.) are closed, and vice versa. Thus, lower priority traffic can be prevented from occupying the time slots reserved for TT frames.
In this paper, we consider the non-preemption integration mode~\cite{802.1Qbv} to solve the issue when a SP/AVB frame is already in transmission at the beginning of the time slot reserved for TT traffic, as shown in Fig.~\ref{fig:IntegrationModes}. The non-preemption mode introduces a ``guard band'' (GB) interval before the TT time slot to ensure no additional delay and jitter for TT traffic. The frame is prevented from initiating transmission if there is not sufficient time for the whole frame transmission before the gate is closed. For SP/AVB traffic, the maximum GB ($L^{\GB}$) is related to the transmission time of a maximum SP/AVB frame waiting in the corresponding queue. 

For the combined traffic shapers, there is no need to re-analyze TT traffic shaped by the TAS shaper, as TT traffic is scheduled within pre-allocated time slots and is not interfered with by other traffic types. 
However, for lower priority SP/AVB traffic in the combined traffic shaper TAS+SP/TAS+CBS, the real-time performance is different from the one in the corresponding individual (SP or CBS) traffic shapers. Here, the lower priority traffic (SP/AVB) can only obtain the remaining service after TT frames are forwarded. Therefore, for the TAS+SP/CBS combined traffic shaper, it is necessary to first calculate the arrival curve related to TT traffic, i.e., the maximum accumulated bits within any time interval that could not be served for lower priority traffic (SP/AVB), which was first proposed in~\cite{Zhao17} for TTEthernet and extended to TSN~\cite{Zhao18}.

\textbf{\underline{Arrival Curve $\alpha^{\TAS}_{\text{I}}(t)$ - TAS~\cite{Zhao17,Zhao18}}.} The construction of the arrival curve $\alpha^{\TAS}_{\text{I}}(t)$ for TT traffic needs to be considered under two situations. The subscript $\text{I}=\{\TT, \GB+\TT\}$ is used for distinguishing whether the credit for AVB traffic is frozen or not during GB. If the credit is non-frozen during GB~\cite{Zhao20} inconsistency with the standard~\cite{802.1Q}, the credit will only be frozen during the time slots (windows), occupied by TT frames. Then $\text{I}=\TT$. If the credit is frozen during GB as most literature assumes~\cite{Ashjaei17, Zhao18, Mohammadpour18}, the GB duration and the corresponding TT time slot can be taken as a whole to represent the credit frozen duration. Then $\text{I}=\GB+\TT$. Moreover, whether the GB and the corresponding TT time slot are taken as a whole to construct the arrival curve $\alpha^{\TAS}_{\text{I}}(t)$ also depends on the selection of the scheduling architecture (TAS+SP or TAS+CBS), which will be respectively in Sect.~\ref{sec:PA_TAS+SP} and Sect.~\ref{sec:PA_TAS+CBS}. %More specifically for CBS, it depends on the credit variation (frozen or not) during GB, which will be explained in detail in Sect.~\ref{sec:PA_TAS+CBS}. 
%In the following, we use the subscript $\text{I}=\{\TT, \GB+\TT\}$ for distinguishing two forms of the arrival curve related to TT traffic. 

The time slots for TT traffic are given by GCLs for each egress port, as in the example in Fig.~\ref{fig:GCL-GanttChart}. They appear periodically according to the GCL period $T_{\GCL}$, and the number $N_{\TT}$ of TT time slots within the GCL period is finite. It is assumed that the $n^{\text{th}}$ ($n\in [0, N_{\TT}-1]$) time slot occupied by TT traffic starts at time $o^{\TT}_{n}$ and has the time duration $L_{n}^{\TT}$. Then, the relative offset between the starting times of the $n^{\text{th}}$ and $m^{\text{th}}$ ($m\in [n, (n+N_{\TT}-1)\% N_{\TT}]$) time slots is $o^{\TT}_{m,n}=o_{m}^{\TT}-o_{n}^{\TT}$, if $m\geq n$, and $o^{\TT}_{m,n}=o_{m}^{\TT}+T_{\GCL}-o_{n}^{\TT}$ otherwise. 
Consequently, the arrival curve $\alpha^{\TAS}_{\text{I}}(t)$ related to TT traffic can be given for all $t\in \mathbb{R}^+$~\cite{Zhao20},
\begin{equation}\label{g:alpha_TT}
	\alpha^{\TAS}_{\text{I}}(t)\!=\!\mathop{\max}_{0\leq n\leq N_{\TT}\!-\!1}\!\left\{\sum_{m=n}^{(n\!+\!N_{\TT}\!-\!1)\% N_{\TT}}l_{\text{I},m}^{\TAS}	\left\lceil\frac{t-o_{\text{I},m,n}^{\TAS}}{T_{\GCL}}\!\right\rceil\!\right\},
\end{equation}
where 
\begin{equation*}
	l_{\text{I},m}^{\TAS}\!=\!
	\begin{cases}
		L^{\TT}_m\cdot C, &\text{I}\!=\!\TT \\
		\left(L^{\TT}_m\!+\!L^{\GB}_m\right)\cdot C, &\text{I}\!=\!\GB\!+\!\TT, \\
	\end{cases}
\end{equation*}
and
\begin{equation*}
	o_{\text{I},m,n}^{\TAS}\!=\!
	\begin{cases}
		o^{\TT}_{m,n}, &\text{I}\!=\!\TT \\
		o^{\TT}_{m,n}\!+\!L^{\GB}_n\!-\!L^{\GB}_m, &\text{I}\!=\!\GB\!+\!\TT. \\
	\end{cases}
\end{equation*}
Note that $L^{\GB}_m$ is the minimum value of the maximum frame of lower priority than the flow of interest with priority $i$ and maximum idle time slot between two consecutive TT time slots,
\begin{equation*}
	L_m^{\GB}=\min\left\{l^{\max}_{>i}, o^{\TT}_{m,m-1}+L^{\TT}_{m-1}\right\},
\end{equation*}
where $l^{\max}_{>i}$ is the maximum frame size in traffic with the priority lower than the priority $i$ traffic.

\noindent
\subsubsection{Performance Analysis -- TAS+SP~\cite{Zhao17}}
\label{sec:PA_TAS+SP}
\textbf{\underline{Service Curve $\beta_{Q_i}^{\SP}(t)$ - SP}.}
SP traffic of different priorities competes for the leftover bandwidth after serving TT traffic. Moreover, the service SP traffic obtains also depends on the integration mode selected. Since we consider the non-preemption mode, there will be a GB before each TT window to prevent an SP frame already in transmission from interfering with TT traffic. Then, in the worst case, the time slot that SP traffic cannot occupy will be enlarged to GB + TT. SP traffic with low priority can obtain the service only when the queues of SP traffic of higher priority are empty. Then, the service curve for SP traffic with priority $i$ ($i\in [1,n_{\SP}]$) in the corresponding queue $Q_i$ can be given as follows,
\begin{equation}\label{g:aggSerQ_SP_TAS+SP}
	\beta_{Q_i}^{\SP}(t)=
	C\bigg[t -\frac{\alpha_{\GB+\TT}^{\TAS}(t)}{C} -
	\frac{\sum_{j=1}^{i-1}\alpha^{\SP}_{Q_j}\!(t)}{C}-\frac{l_{>i}^{\max}}{C} \bigg]_\uparrow^+,
\end{equation}
where $[f(t)]_\uparrow^{+}=\max \limits_{0 \leq s \leq t} \{f(s),0\}$, $\alpha_{\GB+\TT}^{\TAS}(t)$ is from Eq.~(\ref{g:alpha_TT}) with $\text{I}\!=\!\GB\!+\!\TT$, $\alpha^{\SP}_{Q_j}(t)$ (Eq.~(\ref{g:aggArr_SP_TAS+SP})) is the arrival curve of aggregate SP flows with the priority $j$ higher than the priority $i$, and $l^{\max}_{>i}$ is the maximum frame size in traffic with the priority lower than the priority $i$.

\textbf{\underline{Input Arrival Curve $\alpha_{Q_i}^{\SP}(t)$ - SP}.} 
The input arrival curve $\alpha_{Q_i}^{\SP}(t)$ of aggregate SP flows with the priority $i$ before entering the corresponding queue $Q_i$ of the intermediate node is related to the total output arrival curve $\alpha^{*}_{Q_i^-}(t)$ of these flows departing the corresponding preceding queues $Q_i^-$ connected to $Q_i$ and the shaping curve $\sigma^{\link}(t)$ of the physical link by taking all SP flows from $Q_i^-$ to $Q_i$ as a group. The calculation of $\alpha^{*}_{Q_i^-}(t)$ can be done considering Eq.~(\ref{g:aggOutArr_AVB_CBS}), by substituting the delay bound in $\delta_{D}^{Q_i^-}\!(t)$ with the delay upper bound $D_{Q_i^-}^{\SP}$ of aggregate SP flows with priority $i$ at the preceding queue $Q_i^-$. 
Then, $\alpha_{Q_i}^{\SP}(t)$ can be given by,
\begin{equation}\label{g:aggArr_SP_TAS+SP}
	\alpha_{Q_i}^{\SP}(t)=\sum_{Q_i^-}\left[\alpha^{*}_{Q_i^-}(t)\wedge\left(\!\sigma^{\link}(t)\!+\!l_{Q_i^-}^{\max}\!\right)\right].
\end{equation}

By applying $\alpha_{Q_i}^{\SP}(t)$ and $\beta_{Q_i}^{\SP}(t)$ in Eq.~(\ref{maxDQ}) and Eq.~(\ref{maxBQ}), the upper bound of latency $D^{\SP}_{Q_i}$ and backlog $B^{\SP}_{Q_i}$ for SP flows of priority $i$ passing through the queue $Q_i$ under the architecture TAS+SP can be determined.

\subsubsection{Performance Analysis -- TAS+CBS~\cite{Zhao20}}
\label{sec:PA_TAS+CBS}
\begin{figure}[!t]{}
	\centering
	\includegraphics[width=0.47\textwidth]{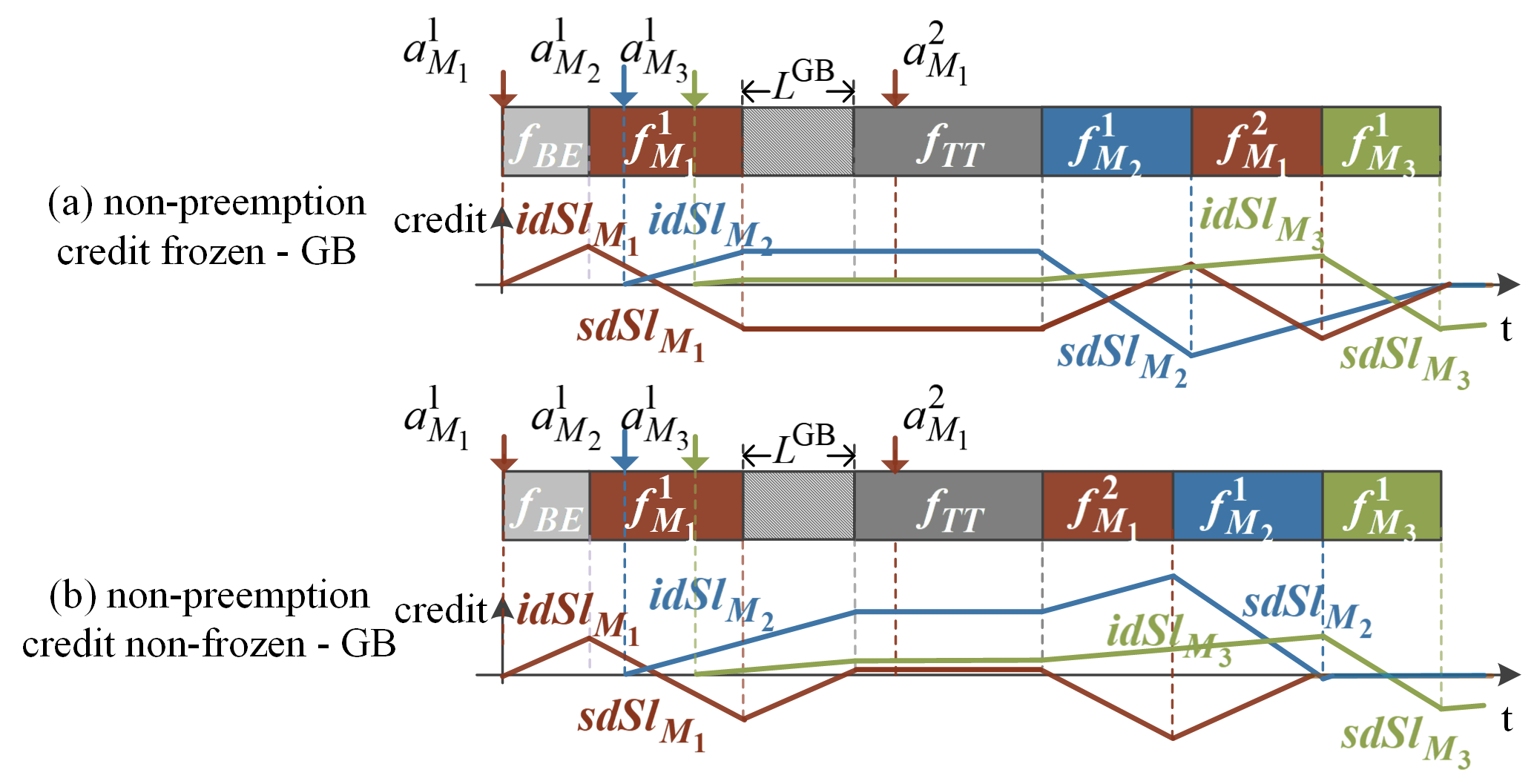}
	\caption{CBS forwarding frames under the impact of TAS}
	\label{fig:creditVar_TAS+X}
\end{figure}

\textbf{\underline{Service Curve $\beta_{Q_i[\M]}^{\CBS}(t)$ - CBS}.}
The service for AVB traffic in the TAS+CBS architecture depends not only on the leftover service after serving TT traffic, but also on the credit state controlled by CBS. AVB traffic with different classes competes for the remaining bandwidth. 
When the gate for the AVB queue is open, the variation of associated credit is the same as in the case CBS is used individually, see Sect.~\ref{sec:CBS}. When the gate for the AVB queue is closed, i.e., during TT transmission, the credit is frozen. In particular, during GB, the gates for all AVB queues are open without any frame transmission, however. Then, the variation of credit during GB has two cases, frozen and non-frozen, which will impact the service for AVB traffic. An example of the CBS working mechanism under the non-preemption integration mode with different assumptions on the variation of credit during GB is shown in Fig.~\ref{fig:creditVar_TAS+X}.
The service curve for AVB Class $M_i$ ($i\in [1,n_{\CBS}]$) in the corresponding queue $Q_i$ is given by~\cite{Zhao20},
\begin{equation}\label{g:aggSerQ_AVB_TAS+CBS}
	\beta_{Q_i[\M]}^{\CBS}(t)=
	idSl_{i}\bigg[ t -\frac{\alpha_{[\M]}^{\TAS}(t)}{C} -
	\frac{c_{i[\M]}^{\max}}{idSl_{i}} \bigg]_\uparrow^+,
\end{equation}
where $\M\in\{\F,\NF\}$ representing the choice of the credit state during GB (F --- frozen credit during GB; NF --- non-frozen credit during GB), $\alpha_{[\M]}^{\TAS}(t)=\{\alpha_{\TT}^{\TAS}(t),\alpha_{\TT+\GB}^{\TAS}(t)\}$ from Eq.~(\ref{g:alpha_TT}), and the credit upper bound $c_{i[\M]}^{\max}=\{c_i^{\max},\overline{c}_i^{\max}\}$ of AVB Class $M_i$. Here $c_i^{\max}$ is the credit upper bound when credit is considered frozen during GB, which equals to the credit upper bound (Eq.~(\ref{g:creditMaxF})) of CBS used individually, and $\overline{c}_i^{\max}$ is the credit upper bound when considering the non-frozen credit during GB,
\begin{equation}\label{g:creditMaxNF}
	c_{i}(t)\leq idSl_{i}\cdot\frac{\sum_{j=1}^{i-1}c_{j}^{\min}-l_{>i}^{\max}-\sigma^{\GB}_{i}}{\rho^{\GB}_{i}+\sum_{j=1}^{i-1}idSl_{j}-C}=\overline{c}_i^{\max}.
\end{equation}
where $c_{j}^{\min}$ is the lower bound of credit of AVB Class $M_j$ (Eq.~(\ref{g:creditMin})), and $\sigma^{\GB}_{i}$ and $\rho^{\GB}_{i}$ are parameters of the linear upper envelope related to GB duration and satisfy $\forall s,t\in\mathbb{R}^+, s\leq t$, $C\cdot\Delta t_{\GB}(s,t)\leq \sigma^{\GB}_{i}+\rho^{\GB}_{i}\cdot(t-s-\Delta t_{\TT}(s,t))$, where $\Delta t_{\TT}(s,t)$ and $\Delta t_{\GB}(s,t)$ respectively represent duration of TT frames emission and of accumulative guard bands during the interval $(s,t]$.

The choice of expressions of $\alpha_{[\M]}^{\TAS}(t)$ and $c_{i[\M]}^{\max}$ depends on the credit state during GB, as follows. %\Red{More details can be found from Theorems 1 and 3 in~\cite{Zhao20}.}
\begin{enumerate}[leftmargin=12pt]
	\item[$\bullet$] M=F: There will be a GB before each TT window to prevent an AVB frame already in transmission interfering with TT traffic. Then, the time slot that AVB traffic cannot occupy will be enlarged to GB+TT. Moreover, 
	since credit is always frozen during GB+TT, the maximum credit value will not be affected by GB+TT time slots and equals the credit upper bound when using CBS individually. Here we take GB and TT slots as a whole and then have $\alpha_{[\M]}^{\TAS}(t)=\alpha_{\TT+\GB}^{\TAS}(t)$ (Eq.~(\ref{g:alpha_TT})), $c_{i[\M]}^{\max}=c_i^{\max}$ (Eq.~(\ref{g:creditMaxF})).
	\item[$\bullet$] M=NF: Although there is a GB before each TT window, and no AVB traffic class can transmit during GB+TT, the credit of the corresponding AVB class will be increased during GB, however. Therefore, when deriving the service curve for AVB traffic, GB and TT slots cannot be taken as a whole as the maximum credit value will be affected by GB duration, which is not equal to the one in individually using CBS anymore.
	Here we have $\alpha_{[\M]}^{\TAS}(t)=\alpha_{\TT}^{\TAS}(t)$ (Eq.~(\ref{g:alpha_TT})), $c_{i[\M]}^{\max}=\overline{c}_i^{\max}$ (Eq.~(\ref{g:creditMaxNF})).
\end{enumerate}

\textbf{\underline{Input Arrival Curve $\alpha_{Q_i[\M]}^{\CBS}(t)$ - CBS}.}
Similar to the CBS used individually, the input arrival curve $\alpha_{Q_i[\M]}^{\CBS}(t)$ of aggregate AVB flows with priority $M_i$ before entering the corresponding queue $Q_i$ of the intermediate node is related to the total output arrival curve $\alpha^{*}_{Q_i^-[\M]}(t)$ of these flows in the preceding queues $Q_i^-$ connected to $Q_i$, to the shaping curve $\sigma^{\link}(t)$ of the physical link by taking all AVB flows from $Q_i^-$ to $Q_i$ as a group, and to the shaping curve $\sigma^{\CBS}_{Q_i[\M]}(t)$ (Eq.~(\ref{g:ShapingCur_TAS+CBS})) of CBS with the consideration of TAS influence,
\begin{equation}\label{g:aggArr_AVB_TAS+CBS}
	\begin{split}
		\alpha_{Q_i[\M]}^{\CBS}\!(t)\!=\!\sum_{Q_i^-}\!&\left[\alpha\!^{*}_{Q_i^-[\M]}\!(t)\!\wedge\!\left(\!\sigma^{\link}\!(t)\!+\!l_{Q_i^-}^{\max}\!\right)\!\wedge\! \left(\!\sigma^{\CBS}_{Q_i^-[\M]}(t)\!+\!l_{Q_i^-}^{\max}\!\right)\!\right].
	\end{split}
\end{equation}
where the calculation of $\alpha^{*}_{Q_i^-[\M]}(t)$ can refer to Eq.~(\ref{g:aggOutArr_AVB_CBS}), and the delay in $\delta_{D}^{Q_i^-}\!(t)$ is the delay upper bound $D_{Q_i^-[\M]}^{\CBS}$ of AVB Class $M_i$ traffic at queue $Q_i^-$.

The CBS shaping curve $\sigma^{\CBS}_{Q_i[\M]}(t)$ is also a non-greedy shaping curve, which is constructed as the upper envelope of output accumulated bits of AVB Class $i$ from $Q_i$ in any time interval. 
%\Red{The detailed derivation of CBS shaping curve considering the influence of TAS can be found in \cite{Zhao20}.} 
Its expression depends on the choice of the credit state during GB and is given by,~\footnote{The detailed derivations~\cite{Zhao20} of CBS service curve in Eq.~(\ref{g:aggSerQ_AVB_TAS+CBS}) under TAS+CBS, credit upper bounds $c_i^{\max}$ and $\overline{c}_i^{\max}$ for arbitrary number of AVB classes, expressions for $\sigma^{\GB}_{i}$ and $\rho^{\GB}_{i}$, and CBS shaping curve in Eq.~(\ref{g:ShapingCur_TAS+CBS}) are not the contributions of the paper. But they are concluded in the supplementary document with the uniform symbols: \url{https://zenodo.org/record/6378112\#.YjqQReeZNPY}.}
\begin{equation}\label{g:ShapingCur_TAS+CBS}
	\sigma^{\CBS}_{Q_i[\M]}(t)=idSl_{i}\bigg[t-\frac{\beta^{\TAS}_{\TT}(t)}{C}+\frac{c_{i[\M]}^{\max}-c_{i}^{\min}}{idSl_{i}}\bigg]_\uparrow^+,
\end{equation}
where $\M\in\{\F,\NF\}$ represents the choice of the credit state during GB, $c_{i[\M]}^{\max}=\{c_i^{\max},\overline{c}_i^{\max}\}$ with $c_i^{\max}$ from Eq.~(\ref{g:creditMaxF}) and $\overline{c}_i^{\max}$ from Eq.~(\ref{g:creditMaxNF}), of which the expression selection depends on the credit state during GB, and $\beta_{\TT}^{\TAS}(t)$ represents the minimum amount of service obtained by TT traffic in any interval, and is given as follows, 
\begin{equation}\label{g:StrSerTT}
	\beta^{\TAS}_{\TT}\!(t)\!=\!
	\mathop{\min}_{0\!\leq n\leq\! N_{\TT}\!-\!1}\!\left\{\sum_{m=n}^{n\!+\!N_{\TT}\!-\!1}\beta_{\TDMA}(t\!+\!t_0,L^{\TT}_{m})\!\right\},
\end{equation}
where
\begin{equation*}
	\begin{split}
		\beta_{\text{TDMA}}&(t, L)\!=\!C\!\cdot\! \max\bigg\{\bigg\lfloor\frac{t}{T_{\GCL}}\bigg\rfloor L, t\!-\!\bigg\lceil\frac{t}{T_{\GCL}}\bigg\rceil(T_{\GCL}-L)\bigg\},
	\end{split}
\end{equation*}
and
\begin{equation*}
	t_0\!=\!T_{\GCL}\!-\!L^{\TT}_{m}\!-\!o^{\TT}_{m}\!+\!o^{\TT}_{n\!-\!1}\!+\!L^{\TT}_{n\!-\!1}.
\end{equation*}

By applying $\alpha_{Q_i[\M]}^{\CBS}(t)$ and $\beta_{Q_i[\M]}^{\CBS}(t)$ into Eq.~(\ref{maxDQ}) and Eq.~(\ref{maxBQ}), the upper bound of latency $D_{Q_i[\M]}^{\CBS}$ and backlog $B_{Q_i[\M]}^{\CBS}$ for AVB flows of Class $M_i$ passing through the queue $Q_i$ under the architecture TAS+CBS can be calculated for two cases of credit during GB, respectively.

\subsection{TAS+ATS+SP / TAS+ATS+CBS}
\label{sec:TAS+ATS+CBS}
An ATS shaper is a type of minimal interleaved regulator~\cite{LeBoudec18}, used to reshape traffic before entering into the queue for each egress port of the middle node in the network. In this section, the hybrid architectures TAS+ATS+SP/CBS are presented, aiming to evaluate the reshaping influence of ATS on the real-time performance of other event-triggered shapers under the effect of the time-triggered shaper (TAS). Note that the combination of ATS+CBS on the same queue is not supported by the standards, since the TSN standards allow a queue to have only one transmission selection algorithm, either CBS or ATS. However, the combination of ATS and CBS used for the same queue is also worthwhile to be investigated, since it could be relevant for industrial use cases. 
%It could also be beneficial for low priority class of AVB traffic under the architecture of ATS+CBS without TAS, and a similar architecture and conclusion can be found from the ATS used with SP from Sect.~\ref{sec:ATS} and Sect.~\ref{sec:CmpTrafficLoadATSSP}. However, the ATS+CBS architecture suffers from the same issue of incompatibility with the current TSN standards. 
In Sect.~\ref{sec:CmpIndi}, we will find that ATS used alone is not always better than SP and CBS. But the advantage of the reshaping effect of ATS for lower priority traffic under the hybrid architecture is greater than that of using ATS alone, as will be shown in Sect.~\ref{sec:Experiment}.

\begin{figure}[!t]{}
	\centering
	\includegraphics[width=0.41\textwidth]{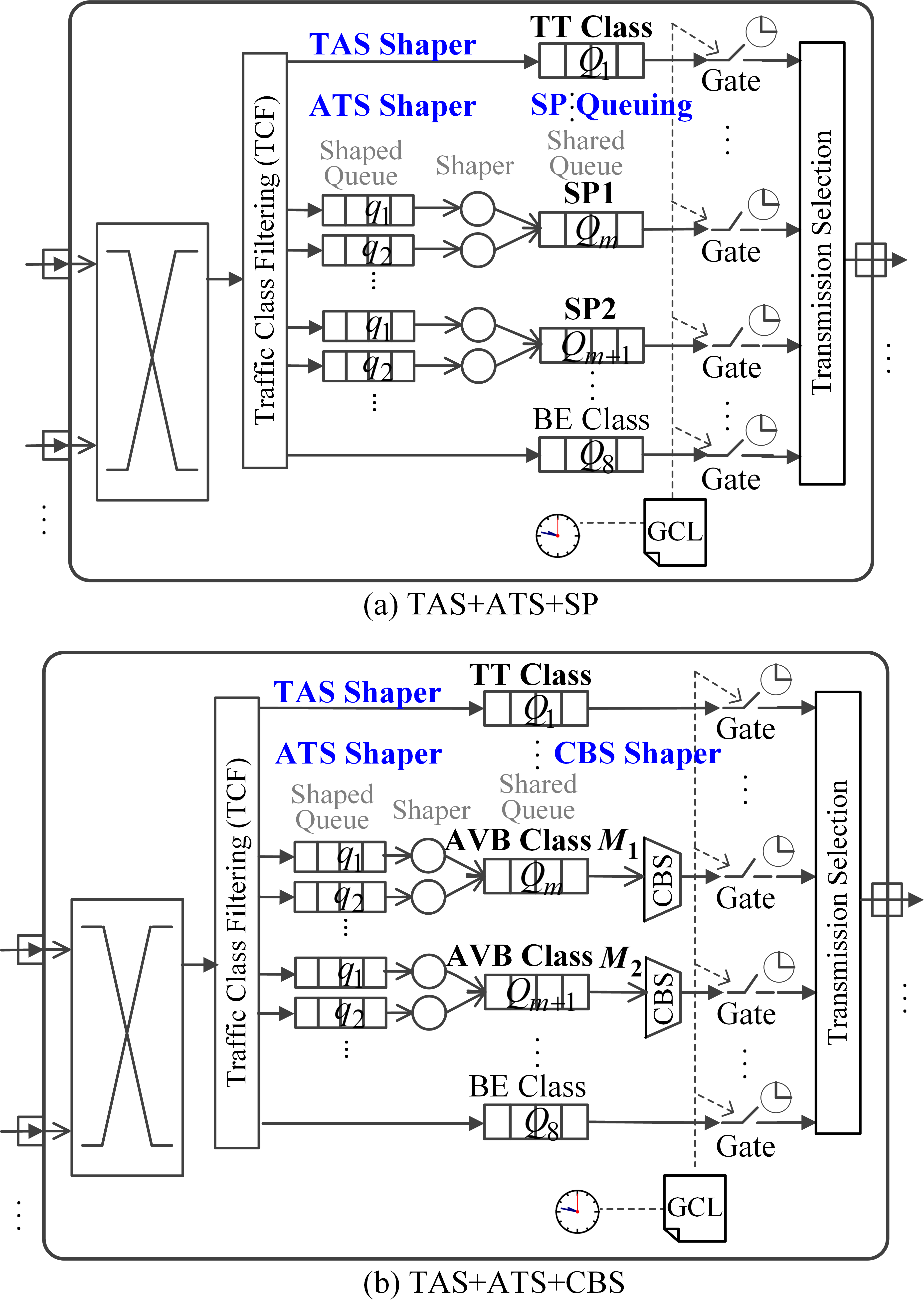}
	\caption{TAS+ATS+SP/CBS Combined Shaper Architecture}
	\label{fig:TAS+ATS+SP_CBS}
\end{figure}

Different from the architecture of CDT+ATS+CBS proposed by~\cite{Mohammadpour18}, which assumes that CDT with the highest priority satisfies the leaky bucket model, here we consider the more general model for TAS, i.e., satisfying arbitrary time-triggered slots. For this mode, the arrival curve of TT traffic satisfies the non-linear staircase function from Eq.~(\ref{g:alpha_TT}).
The architectures of TAS+ATS+SP and TAS+ATS+CBS are shown in Fig.~\ref{fig:TAS+ATS+SP_CBS}(a) and (b), respectively. Compared with the TAS+SP and TAS+CBS architectures in Fig.~\ref{fig:TAS+SP^CBS}, there are additional shaped queues and the ATS shaping algorithm is used to reshape SP/AVB flows before admitting them into their corresponding priority queues (shared queues). The queuing schemes for frames entering the shaped queues and the ATS shaping algorithm are the same as the ATS used individually. Moreover, the gate operation is the same as in the architecture of TAS+SP/CBS without ATS, i.e., TT traffic has the exclusive gate opening. SP/AVB frames are allowed to be transmitted only when the TT gate is closed and their corresponding gates are open. As earlier, we use the non-preemption integration mode with the GB duration.
Since the TT traffic controlled by the TAS has the highest priority, the traffic reshaping for SP/AVB flows by ATS will not affect the transmission of TT traffic.

In the following, we extend the NC approach to TAS+ATS+SP/CBS architectures for the quantitative performance comparison in Sect.~\ref{sec:CmpComb}.

\subsubsection{Performance Analysis -- TAS+ATS+SP}
\textbf{\underline{Input Arrival Curve $\alpha_{Q_i}^{\ATS+\SP}\!(t)$ - SP - Shared Queue}.}
As the output of each SP flow $f$ departing the shaped queue $q$ is constrained by the committed transmission rate $r_f$ and the committed burst size $b_f$, the output arrival curve of $f$ from the shaped queue $q$ satisfies $r_f\cdot t+b_f$. It is also the input arrival curve of $f$ before entering into the shared queue $Q_i$. Therefore, the input arrival curve $\alpha_{Q_i}^{\ATS+\SP}(t)$ of aggregate SP flows with priority $i$ before the shared queue $Q_i$ is the sum of output arrival curves from all the previous shaped queues $q$ connected to $Q_i$, which is the same as the situation of ATS used individually (Eq.~{\ref{g:aggArr_SP_ATS+SP}}),
\begin{equation}\label{g:aggArr_SP_TAS+ATS+SP}
	\alpha_{Q_i}^{\ATS+\SP}\!(t)=\sum_{q\rightarrow Q_i} \sum_{f\in q}\left(r_f\cdot t+b_f\right),
\end{equation}
where $q\rightarrow Q_i$ represents all the shaped queues connected to the shared queue $Q_i$.

\textbf{\underline{Service Curve $\beta_{Q_i}^{\ATS+\SP}\!(t)$ - SP - Shared Queue}.}
\begin{corollary}\label{Th:aggSerQ_SP_TAS+ATS+SP}
	The service curve $\beta_{Q_i}^{\ATS+\SP}\!(t)$ for SP traffic with priority $i$ ($i\in[1,n_{\SP}]$) in the shared queue $Q_i$ under the TAS+ATS+SP architecture is
	\begin{equation}\label{g:aggSerQ_SP_TAS+ATS+SP}
		\beta_{Q_i}^{\ATS+\SP}(t)=
		C\bigg[t -\frac{\alpha_{\GB+\TT}^{\TAS}(t)}{C} -	\frac{\sum_{j=1}^{i-1}\alpha^{\ATS+\SP}_{Q_j}\!(t)}{C}-\frac{l_{>i}^{\max}}{C} \bigg]_\uparrow^+,
	\end{equation}
	where $\alpha^{\ATS+\SP}_{Q_j}(t)$ is the arrival curve of aggregate SP flows from Eq.~(\ref{g:aggArr_SP_TAS+ATS+SP}), $\alpha^{\TAS}_{\GB+\TT}(t)$ is the arrival curve related to TT traffic give by Eq.~(\ref{g:alpha_TT}), and $l^{\max}_{>i}$ is the maximum frame size with priority lower than $i$.
\end{corollary}

\emph{Proof}:
Let $R_{Q_i}(t)$ and $R^*_{Q_i}(t)$ be respectively the input and output cumulative function for SP traffic with priority $i$ ($i\in [1,n_{\SP}]$) before and after the shared queue $Q_i$. Moreover, let $R_{\TT}(t)$ (resp. $R_{\GB}(t)$) and $R^*_{\TT}(t)$ (resp. $R^*_{\GB}(t)$) be the arrival and departure processes of TT traffic (resp. occupation of guard bands) due to GCL implementation by TAS.

It is assumed that $t\in \mathbb{R}^+$ is a time point when the queue $Q_i$ is non-empty, i.e., $R^*_{Q_i}(t)<R_{Q_i}(t)$. Let $s$ be the start of the last busy period for the traffic with a priority higher than or equal to the priority $i$ of SP traffic, i.e., $s=\sup\{u\leq t~\vrule~R^*_{\TT}(u)=R_{\TT}(u),R^*_{\GB}(u)=R_{\GB}(u),R^*_{Q_j}(u)=R_{Q_j}(u),\forall j\in[1,i]\}$. Since SP traffic in $Q_i$ cannot be served during the TT/GB time slots, and when queues for other SP traffic with higher priority are non-empty or a non-preemptive lower priority frame $l_L$ as well, then over the interval $\Delta t=t-s$, we have
\begin{equation}\label{g:derSPSer}
	\begin{split}
		R^*_{Q_i}&(t)-R^*_{Q_i}(s)=C(t-s)-(R^*_{\TT}(t)-R^*_{\TT}(s))-\\
		&(R^*_{\GB}(t)-R^*_{\GB}(s))-\sum_{j=1}^{i-1}\left(R^*_{Q_j}(t)-R^*_{Q_j}(s)\right)-l_L.
	\end{split}
\end{equation}

With the TAS+ATS+SP architecture, ATS is only used to reshape SP traffic but without the influence on TT traffic type. Moreover, the integration modes between SP and TT traffic are not influenced by ATS. In this paper, in order to compare with the existing work, we also assume the non-preemption integration mode. Therefore, the upper bounds of $\left(R^*_{\TT}(t)-R^*_{\TT}(s)\right)$ and $\left(R^*_{\GB}(t)-R^*_{\GB}(s)\right)$ are the same as the case under the TAS+SP architecture, and can be taken as a whole, i.e.,
\begin{equation}\label{g:derTT}
	\begin{split}
		(R^*_{\TT}(t)&-R^*_{\TT}(s))+(R^*_{\GB}(t)-R^*_{\GB}(s))\\
		&\leq (R_{\TT}(t)-R_{\TT}(s))+(R_{\GB}(t)-R_{\GB}(s))\\
		&\leq\alpha^{\TAS}_{\GB+\TT}(t),
	\end{split}
\end{equation}
where $\alpha^{\TAS}_{\GB+\TT}(t)$ is the arrival curve related to TT traffic given by Eq.~(\ref{g:alpha_TT}). The non-preemption lower priority frame is maximized by 
\begin{equation}\label{g:derSPL}
	l_L\leq\max_{\substack{i+1\leq j\leq n_{\SP},\\Q_j\in Q_{\BE}}}\left\{l_j^{\max}\right\}=l_{>i}^{\max},
\end{equation}
and is not affected by ATS either.

However, since each SP flow before entering the shared queue $Q_j$ is reshaped to the committed transmission rate $r_f$ and the committed burst size $b_f$, the upper bound of $\sum_{j=1}^{i-1}\left(R^*_{Q_j}(t)-R^*_{Q_j}(s)\right)$ satisfies,
\begin{equation}\label{g:derSPH}
	\begin{split}
		\sum_{j=1}^{i-1}&\left(R^*_{Q_j}(t)-R^*_{Q_j}(s)\right)\leq \sum_{j=1}^{i-1}\left(R_{Q_j}(t)-R_{Q_j}(s)\right)\\
		&\leq \sum_{j=1}^{i-1}\sum_{q\rightarrow Q_j}\sum_{f\in q}(r_f\cdot t+b_f)=\sum_{j=1}^{i-1}\alpha_{Q_j}^{\ATS+\SP}(t).
	\end{split}
\end{equation}
This is different from the case under the TAS+SP architecture. Then, by inducing Eq.~(\ref{g:derTT}), Eq.~(\ref{g:derSPL}) and Eq.~(\ref{g:derSPH}) into Eq.~(\ref{g:derSPSer}), we can obtain the service curve (Eq.~(\ref{g:aggSerQ_SP_TAS+ATS+SP})) for SP traffic in the shared queue under the TAS+ATS+SP architecture.

\hfill $\blacksquare$

Therefore, even under the architecture of TAS+ATS+SP, the service curve for SP traffic in the shared queue is similar to the one (Eq.~(\ref{g:aggSerQ_SP_TAS+SP})) under the TAS+SP in Sect.~\ref{sec:PA_TAS+SP}. The only difference is that, due to the ATS shaper, the rate and burst of SP traffic are reshaped and restricted before entering the shared queue.

By applying $\alpha_{Q_i}^{\ATS+\SP}(t)$ and $\beta_{Q_i}^{\ATS+\SP}(t)$ into Eq.~(\ref{g:maxDh}) and Eq.~(\ref{g:maxBv}), we can determine the upper bound of latency
\begin{equation}\label{g:DelaySP}
	D^{\ATS+\SP}_{Q_i}=h(\alpha_{Q_i}^{\ATS+\SP}(t),\beta_{Q_i}^{\ATS+\SP}(t)),
\end{equation}
and backlog 
\begin{equation}\label{g:BacklogSP}
	B^{\ATS+\SP}_{Q_i}=v(\alpha_{Q_i}^{\ATS+\SP}(t),\beta_{Q_i}^{\ATS+\SP}(t))
\end{equation}
for SP flows of priority $i$ passing through the shared queue $Q_i$ under the architecture TAS+ATS+SP.

\textbf{\underline{Input Arrival Curve $\alpha^{\ATS+\SP}_{q}(t)$ - SP - Shaped Queue}.}
\begin{corollary}\label{Cor:aggArr_ShpQ_TAS+ATS+SP}
	The input arrival curve $\alpha^{\ATS+\SP}_{q}(t)$ of aggregate SP flows before entering the shaped queue $q$ is,
	\begin{equation}\label{g:aggArr_ShpQ_TAS+ATS+SP}
		\alpha^{\ATS+\SP}_{q}(t)=\alpha^{*}_{Q_i^-}(t)\wedge\left(\sigma^{\link}(t)\!+\!l_{Q_i^-}^{\max}\right).
	\end{equation}
	where the output arrival curve $\alpha^{*}_{Q_i^-}(t)$ is from Eq.~(\ref{g:aggOutArr_ShpQ_TAS+SP}), by substituting delay upper bound $D$ in $\delta_{D}^{Q_i^-}(t)$ with $D=D_{Q_i^-}^{\ATS+\SP}=h(\alpha_{Q_i^-}^{\ATS+\SP}(t),\beta_{Q_i^-}^{\ATS+\SP}(t))$ of SP traffic in the preceding shared queue $Q_i^-$, in which $\alpha_{Q_i^-}^{\ATS+\SP}\!(t)$ is determined by Eq.~(\ref{g:aggArr_SP_TAS+ATS+SP}) and $\beta_{Q_i^-}^{\ATS+\SP}\!(t)$ is from Eq.~(\ref{g:aggSerQ_SP_TAS+ATS+SP}).
\end{corollary}

\emph{Proof}:
In the TAS+ATS+SP architecture, compared with the ATS individually used, SP flows have additional waiting time for higher priority traffic, which causes the maximum latency in the preceding shared queue $Q_i^-$ changed to $D_{Q_i^-}^{\ATS+\SP}=h(\alpha_{Q_i^-}^{\ATS+\SP}(t),\beta_{Q_i^-}^{\ATS+\SP}(t))$, where $\alpha_{Q_i^-}^{\ATS+\SP}(t)$ and $\beta_{Q_i^-}^{\ATS+\SP}(t)$ are respectively the input arrival (Eq.~(\ref{g:aggArr_SP_TAS+ATS+SP})) and service curves (Eq.~\ref{g:aggSerQ_SP_TAS+ATS+SP}) for SP flows in the shaped queue $Q_i^-$. 
Therefore, according to the basic NC theory, the output arrival curve $\alpha_{f}^{*Q_i^-}\!(t)$ of a SP flow $f$ departure from the preceding shared queue $Q_i^-$ is $\alpha_f^{Q_i^-}\!(t) \oslash\delta_{D}^{Q_i^-}\!(t)$, where $D=D_{Q_i^-}^{\ATS+\SP}$. Then the aggregate arrival curve of all SP flows departure from $Q_i^-$ to $q$ can be represented by $\alpha_{Q_i^-}^*(t)=\sum_{\substack{f\in Q_i^-,\\Q_i^-\rightarrow q}}\alpha_{f}^{*Q_i^-}\!(t)$, where $D=D_{Q_i^-}^{\ATS+\SP}$.

Moreover, before a shaped queue $q$, the phenomenon of serialization of flows from the same shared queues $Q_i^-$ still exists. And according to the ATS queuing schemes QAR1 and QAR2, flows in the shaped queue $q$ are all from the same preceding shared queue $Q_i^-$.
These characteristics will not be impacted by the integration of TAS. Then we can conclude the corollary.

\hfill $\blacksquare$

\textbf{\underline{Service Curve $\beta^{\ATS+\SP}_q(t)$ - SP - Shaped Queue}.}
\begin{corollary}\label{Cor:aggSer_ShpQ_TAS+ATS+SP}
	The service curve $\beta^{\ATS+\SP}_q(t)$ for aggregate SP flows in the shaped queue $q$ is,
	\begin{equation}\label{g:aggSerq_SP_TAS+ATS+SP}
		\beta^{\ATS+\SP}_q(t)=\delta_{D}^q(t),
	\end{equation}
	where $\delta_{D}^q(t)$ is the pure-delay function with $D=D^{\ATS+\SP}_q=D^{\ATS+\SP}_{Q_i^-}-l_{q}^{\min}/C$, and $D^{\ATS+\SP}_{Q_i^-}$ is from Eq.~(\ref{g:DelaySP}).	
\end{corollary}

\emph{Proof}:
Since the shared queue for each SP priority in the TAS+ATS+SP architecture is served in a FIFO manner, the ATS shaper will not introduce extra overheads to the worst-case delay of such a FIFO system~\cite{LeBoudec18}.

Thus, an SP flow fed to the shaped queue $q$ on the subsequent node will not increase the upper bound of the delay for the flow waiting in the combined element of the shared queue $Q_i^-$ on the preceding node and the shaped queue $q$, i.e., $d^{\ATS+\SP}_{Q_i^-}(t)+d^{\ATS+\SP}_q(t)\leq D^{\ATS+\SP}_{Q_i^-}$. Here, $D^{\ATS+\SP}_{Q_i^-}$ is the latency bound of SP flows with priority $i$ waiting in the preceding shared queue $Q_i^-$ and can be calculated from Eq.~(\ref{g:DelaySP})). Since the lower bound of the delay in the shared queue $Q_i^-$ for all flows traversing through $Q_i^-\rightarrow q$ is $l_{q}^{\min}/C$, the maximum latency $D^{\ATS}_{q}$ of SP flows waiting in the shaped $q$ is given by $D^{\ATS+\SP}_{q}=D^{\ATS+\SP}_{Q_i^-}-l_{q}^{\min}/C$. 
Then, we can conclude the corollary.

\hfill $\blacksquare$
\subsubsection{Performance Analysis -- TAS+ATS+CBS}
\textbf{\underline{Input Arrival Curve $\alpha_{Q_i}^{\ATS+\CBS}\!(t)$ - CBS - Shared Queue}.}
Correspondingly, by ATS reshaping, the input arrival curve $\alpha_{Q_i}^{\ATS+\CBS}(t)$ of aggregate AVB flows with priority $i$ before the shared queue $Q_i$ is the sum of output arrival curves satisfying $r_f\cdot t+b_f$ from all the previous shaped queues $q$,
\begin{equation}\label{g:aggArr_CBS_TAS+ATS+CBS}
	\alpha_{Q_i}^{\ATS+\CBS}\!(t)=\sum_{q\rightarrow Q_i} \sum_{f\in q}\left(r_f\cdot t+b_f\right).
\end{equation}

\textbf{\underline{Service Curve $\beta_{Q_i[\M]}^{\ATS+\CBS}(t)$ - CBS - Shared Queue}.}
\begin{corollary}\label{Th:aggSerQ_CBS_TAS+ATS+CBS}
	The service curve for AVB traffic of Class $M_i$ ($i\in[1,n_{\CBS}]$) in the shared queue $Q_i$ under the TAS+ATS+CBS architecture is same to the service curve for AVB traffic under the TAS+CBS architecture, i.e.,
	\begin{equation}\label{g:aggSerQ_AVB_TAS+ATS+CBS}
		\beta_{Q_i[\M]}^{\ATS+\CBS}(t)=
		idSl_{i}\bigg[ t -\frac{\alpha_{[\M]}^{\TAS}(t)}{C} -
		\frac{c_{i[\M]}^{\max}}{idSl_{i}} \bigg]_\uparrow^+,
	\end{equation}
	where $\M\in\{\F,\NF\}$ representing the choice of the credit state during GB (F --- frozen credit during GB; NF --- non-frozen credit during GB), $\alpha_{[\M]}^{\TAS}(t)=\{\alpha_{\TT}^{\TAS}(t),\alpha_{\TT+\GB}^{\TAS}(t)\}$ is the arrival curve related to TT traffic, given by Eq.~(\ref{g:alpha_TT}); $c_{i[\M]}^{\max}=c_i^{\max}$ (Eq.~(\ref{g:creditMaxF})) is the credit upper bound of AVB Class $M_i$ if credit is considered frozen during GB, and $c_{i[\M]}^{\max}=\overline{c}_i^{\max}$ (Eq.~(\ref{g:creditMaxNF})) is the credit upper bound if credit is considered non-frozen during GB.
\end{corollary}

\emph{Proof}:
Let $R_{Q_i}(t)$ and $R^*_{Q_i}(t)$ be respectively the input and output cumulative function for AVB traffic of Class $M_i$ ($i\in [1,n_{\CBS}]$) before and after the shared queue $Q_i$. Moreover, let $R_{\TT}(t)$ (resp. $R_{\GB}(t)$) and $R^*_{\TT}(t)$ (resp. $R^*_{\GB}(t)$) be the arrival and departure processes of TT traffic (resp. occupation of guard bands) due to GCL implementation by TAS.

It is assumed that $t\in \mathbb{R}^+$ is a time point when the queue $Q_i$ is non-empty, i.e., $R^*_{Q_i}(t)<R_{Q_i}(t)$. Then let $s=\sup\{u\leq t~\vrule~R^*_{\TT}(u)=R_{\TT}(u),R^*_{\GB}(u)=R_{\GB}(u),R^*_{Q_i}(u)=R_{Q_i}(u),c_i(u)=0\}$. Since AVB traffic in $Q_i$ obtained service during $\Delta t^-$ with the associated credit decreased, and be blocked during $\Delta t^+$ and $\Delta t^0$ with the associated credit increased and frozen respectively. Then over the interval $\Delta t=t-s$, we have
\begin{equation*}
	\begin{split}
		c_i(t)-c_i(s)=(\Delta t-\Delta t^--\Delta t^0)idSl_i+\Delta t^- sdSl_i.
	\end{split}
\end{equation*}
Then, since $idSl_i-sdSl_i=C$, and $c_i(s)=0$,
\begin{equation}\label{g:derSerCBS}
	\begin{split}
		R^*_{Q_i}(t)-R^*_{Q_i}(s)=(t-\Delta t^0)idSl_i-c_i(t).
	\end{split}
\end{equation}
Since
\begin{equation}\label{g:derDeltat0CBS}
	\Delta t^0\leq \alpha_{[\M]}^{\TAS}(t)/C, 
\end{equation}
in which $\alpha_{[\M]}^{\TAS}(t)$ (Eq.~(\ref{g:alpha_TT})) is only related to the TT traffic and the credit behavior during GB, $\Delta t^0$ is not affected by ATS reshaping. 

Moreover, $c_i(t)$ in Eq.~(\ref{g:derSerCBS}) can be lower and upper bounded by,
\begin{equation}\label{g:derCboundCBS}
	c_i^{\min}\leq c_i(t)\leq c_{i[\M]}^{\max}, 
\end{equation}
where $c_i^{\min}$ is given by Eq.~(\ref{g:creditMin}), which is only related to the maximum frame size in $Q_i$. 
$c_{i[\M]}^{\max}=c_i^{\max}$ is given by Eq.~(\ref{g:creditMaxF}) for the case of credit frozen during GB, which is associated with the credit lower bound $c_j^{\min}$ of AVB traffic with the priority $M_j$ higher than $M_i$, and the maximum frame size $l_{>i}^{\max}$ of AVB traffic with the priority lower than $M_i$. 
$c_{i[\M]}^{\max}=\overline{c}_i^{\max}$ is given by Eq.~(\ref{g:creditMaxNF}) for the case of credit non-frozen during GB, which besides the above-mentioned parameters in $c_i^{\max}$, is also related to the linear upper envelope for GB duration. 
As can be found, $c_i(t)\leq c_{i[\M]}^{\max}$ is not related to flows arrival pattern of AVB flows before the share queue $Q_i$. 

By the analysis of Eq.~(\ref{g:derDeltat0CBS}) and Eq.~(\ref{g:derCboundCBS}), it is known that ATS reshaping on AVB traffic does not change the ability to serve AVB traffic of different priorities in the shared queue, compared with the service capability for AVB traffic under the TAS+CBS architecture. Then, by inducing Eq.~(\ref{g:derDeltat0CBS}) and Eq.~(\ref{g:derCboundCBS}) into Eq.~(\ref{g:derSerCBS}), we can obtain the service curve (Eq.~(\ref{g:aggSerQ_AVB_TAS+ATS+CBS})) for AVB traffic in the shared queue under the TAS+ATS+CBS architecture, the same as the one (Eq.~(\ref{g:aggSerQ_AVB_TAS+CBS}) in Appendix~\ref{sec:PA_TAS+CBS}) under the TAS+AVB architecture.

\hfill $\blacksquare$

By applying $\alpha_{Q_i}^{\ATS+\CBS}(t)$ and $\beta_{Q_i[\M]}^{\ATS+\CBS}(t)$ into Eq.~(\ref{g:maxDh}) and Eq.~(\ref{g:maxBv}), we can determine the upper bound of latency
\begin{equation}\label{g:DelayCBS}
	D^{\ATS+\CBS}_{Q_i[\M]}=h(\alpha_{Q_i}^{\ATS+\CBS}(t),\beta_{Q_i[\M]}^{\ATS+\CBS}(t)),
\end{equation}	
and backlog 
\begin{equation}\label{g:BacklogCBS}
	B^{\ATS+\CBS}_{Q_i[\M]}=v(\alpha_{Q_i}^{\ATS+\CBS}(t),\beta_{Q_i[\M]}^{\ATS+\CBS}(t))
\end{equation}
for SP flows of priority $M_i$ passing through the shared queue $Q_i$ under TAS+ATS+CBS.

\textbf{\underline{Input Arrival Curve $\alpha^{\ATS+\CBS}_{q[\M]}(t)$ - CBS - Shaped Queue}.}
\begin{corollary}\label{Cor:aggArr_ShpQ_TAS+ATS+CBS}
	The input arrival curve $\alpha^{\ATS+\CBS}_{q[\M]}(t)$ of aggregate AVB flows before the shaped queue $q$ is,
	\begin{equation}\label{g:aggArr_ShpQ_TAS+ATS+CBS}
		\begin{split}
			\alpha^{\ATS+\CBS}_{q[\M]}(t)\!=&\!\alpha\!^{*}_{Q_i^-}\!(t)\!\wedge\!\left(\!\sigma^{\link}(t)\!+\!l_{Q_i^-}^{\max}\!\right)\!\wedge\! 	\left(\!\sigma\!^{\ATS+\CBS}_{Q_i^-[\M]}(t)\!+\!l_{Q_i^-}^{\max}\!\right)\!.
		\end{split}
	\end{equation}
	where $\sigma^{\ATS+\CBS}_{Q_i^-[\M]}(t)$ is same to Eq.~(\ref{g:ShapingCur_TAS+CBS}) of the case without ATS, and the output arrival curve $\alpha^{*}_{Q_i^-}(t)$ is from Eq.~(\ref{g:aggOutArr_ShpQ_TAS+SP}), by replacing the delay bound $D$ in $\delta_{D}^{Q_i^-}(t)$ with $D=D_{Q_i^-[\M]}^{\ATS+\CBS}=h(\alpha_{Q_i^-}^{\ATS+\CBS}(t),\beta_{Q_i^-[\M]}^{\ATS+\CBS}(t))$ of AVB traffic in the preceding shared queue $Q_i^-$, in which $\alpha_{Q_i^-}^{\ATS+\CBS}(t)$ and $\beta_{Q_i^-[\M]}^{\ATS+\CBS}(t)$ are respectively from Eq.~(\ref{g:aggArr_CBS_TAS+ATS+CBS}) and Eq.~(\ref{g:aggSerQ_AVB_TAS+ATS+CBS}).
\end{corollary}

\emph{Proof}:
In the TAS+ATS+CBS architecture, in addition to the shaping curve $\sigma^{\link}(t)$ of the physical link due to serialization of all AVB flows from $Q_i^-$ to $q$, the input arrival curve $\alpha^{\ATS+\CBS}_{q[\M]}(t)$ of aggregate AVB flows before entering the shaped queue $q$ is also related to the CBS shaping curve $\sigma^{\ATS+\CBS}_{Q_i^-[\M]}(t)$. 

As can be seen from the CBS shaping curve under the TAS+CBS architecture in Eq.~(\ref{g:ShapingCur_TAS+CBS}), it is related to, on the one hand, the service curve $\beta^{\TAS}_{\TT}(t)$ (Eq.~\ref{g:StrSerTT}) supplied to TT traffic, which is only related to the time slots reserved for TT frames. On the other hand, the credit lower bound $c_{i}^{\min}$ (Eq.~\ref{g:creditMin}) and upper bound $c_{i[\M]}^{\max}=\{c_i^{\max},\overline{c}_i^{\max}\}$ for AVB flows in $Q_i^-$, which have been discussed in Corollary~\ref{Th:aggSerQ_CBS_TAS+ATS+CBS}, are not related to the arrival pattern of AVB flows before the shared queue $Q_i^-$. Since ATS only has an impact on the flows' arrival pattern before shared queues $Q_i^-$, and shaped queue $q$, the CBS shaping curve $\sigma^{\ATS+\CBS}_{Q_i^-[\M]}(t)$ in the TAS+ATS+CBS architecture is the same as the CBS shaping curve $\sigma^{\CBS}_{Q_i^-[\M]}(t)$ in the TAS+CBS architecture.

The discussion for the aggregate arrival curve $\alpha^{*}_{Q_i^-}(t)$ of AVB flows departure from the shared queue $Q_i^-$ to the shaped queue $q$ is similar to the one in Corollary~\ref{Cor:aggArr_ShpQ_TAS+ATS+SP}.

\hfill $\blacksquare$

\textbf{\underline{Service Curve $\beta^{\ATS+\CBS}_{q[\M]}(t)$ - CBS - Shaped Queue}.}
\begin{corollary}\label{Cor:aggSer_ShpQ_TAS+ATS+CBS}
	The service curve $\beta^{\ATS+\CBS}_{q[\M]}(t)$ for aggregate AVB flows in the shaped queue $q$ is
	\begin{equation}\label{g:aggSerq_AVB_TAS+ATS+CBS}
		\beta^{\ATS+\CBS}_{q[\M]}(t)=\delta_{D}^q(t),
	\end{equation}
	where $\delta_{D}^q(t)$ is the pure-delay function with $D=D^{\ATS+\CBS}_{q[\M]}=D^{\ATS+\CBS}_{Q_i^-[\M]}-l_{q}^{\min}/C$, and $D^{\ATS+\CBS}_{Q_i^-[\M]}$ is from Eq.~(\ref{g:DelayCBS}).
\end{corollary}

\emph{Proof}:
The discussion is similar to Corollary~\ref{Cor:aggSer_ShpQ_TAS+ATS+SP}, by considering the latency upper bound $D^{\ATS+\CBS}_{Q_i^-[\M]}$ in Eq.~(\ref{g:DelayCBS}) instead of $D^{\ATS+\SP}_{Q_i^-}$ in Eq.~(\ref{g:DelaySP}).

\hfill $\blacksquare$

\section{Performance Comparison Evaluation}
\label{sec:Experiment}
In this section, in order to compare the performance evaluation of individual traffic shapers and their combinations, we use a large set of synthetic test cases~\footnote{Details of flows, routes and GCLs for all the test cases can be downloaded from \url{https://zenodo.org/record/6378112\#.YjqQReeZNPY}} with different topologies and a realistic test case, i.e., the Orion Crew Exploration Vehicle (CEV) from NASA~\cite{Tamas-Selicean14}.
\subsection{Individual Traffic Shapers}
\label{sec:CmpIndi}
\subsubsection{Comparison of NC and non-NC approaches for ATS evaluation}
\label{sec:CmpNC_nonNC}
Before discussing the various traffic shapers, we first compare the two different methods used for ATS evaluation, i.e., the Network Calculus (NC) used in this article and a non-NC approach proposed in~\cite{Soheil16}.
By comparing the upper bound of the delay obtained by the two methods, the latency bound for a flow $f$ in an egress port calculated by NC is $\Delta D_f^Q$ more pessimistic than the result calculated by the non-NC approach~\cite{Soheil16},
\begin{equation}\label{g:delta}
	\Delta D_f^Q=\max_{\forall f'\in Q(f)}\left(\frac{l_{f'}}{C-\sum_{f''\in Q_H}r_{f''}}-\frac{l_{f'}}{C}\right),
\end{equation}
where $l_f'$ is the frame size of a flow with the same priority level as the flow $f$ of interest, $l_f''$ is the frame size of flows with a higher priority than $f$, and $r_{f''}$ is the committed burst size of the flow $f''$ supported by ATS.

\begin{figure}[!t]{}
	\centering
	\includegraphics[width=0.49\textwidth]{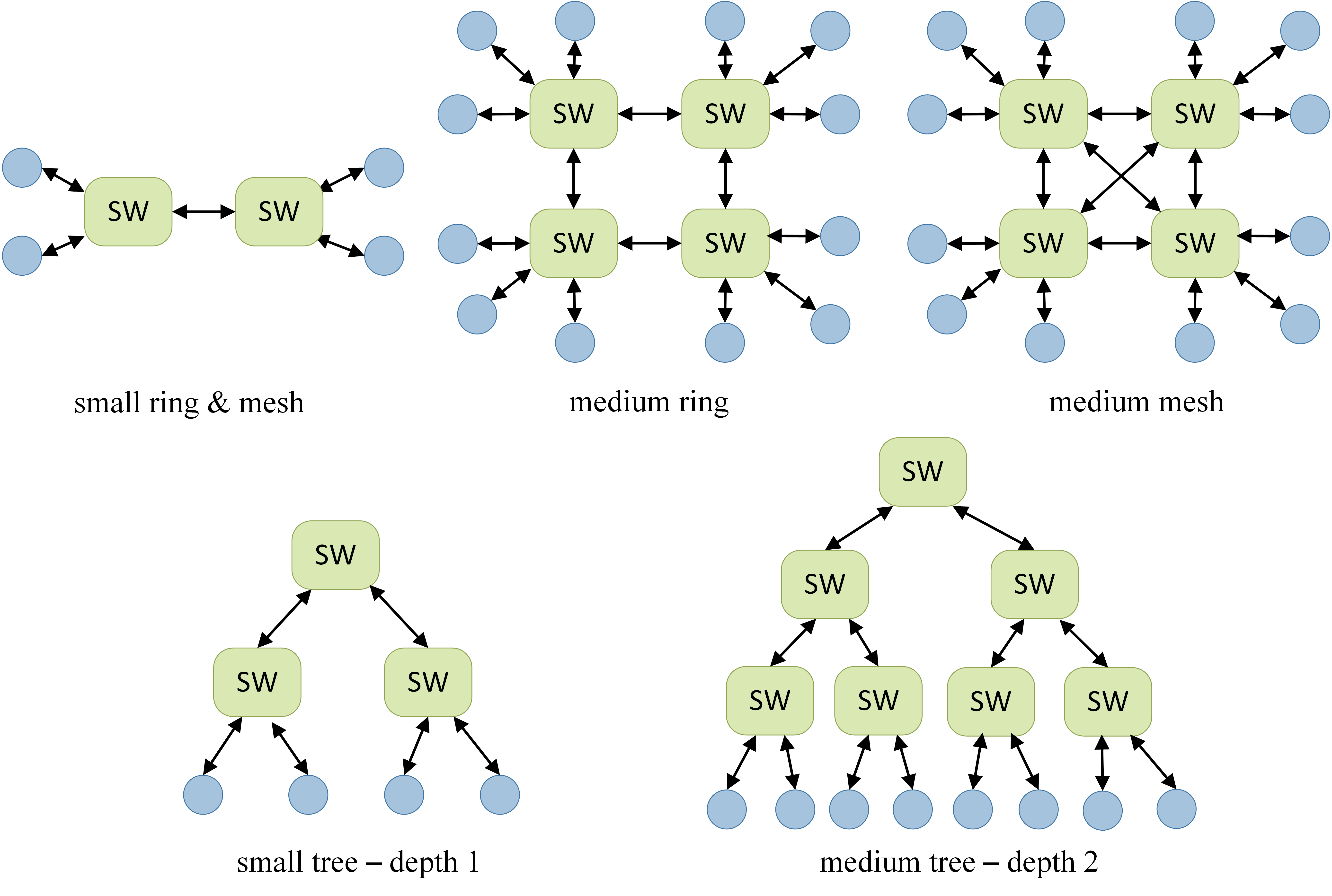}
	\caption{Network topologies of synthetic test cases}
	\label{fig:SynTC_topo}
\end{figure}

\begin{figure}[!t]{}
	\centering
	\includegraphics[width=0.47\textwidth]{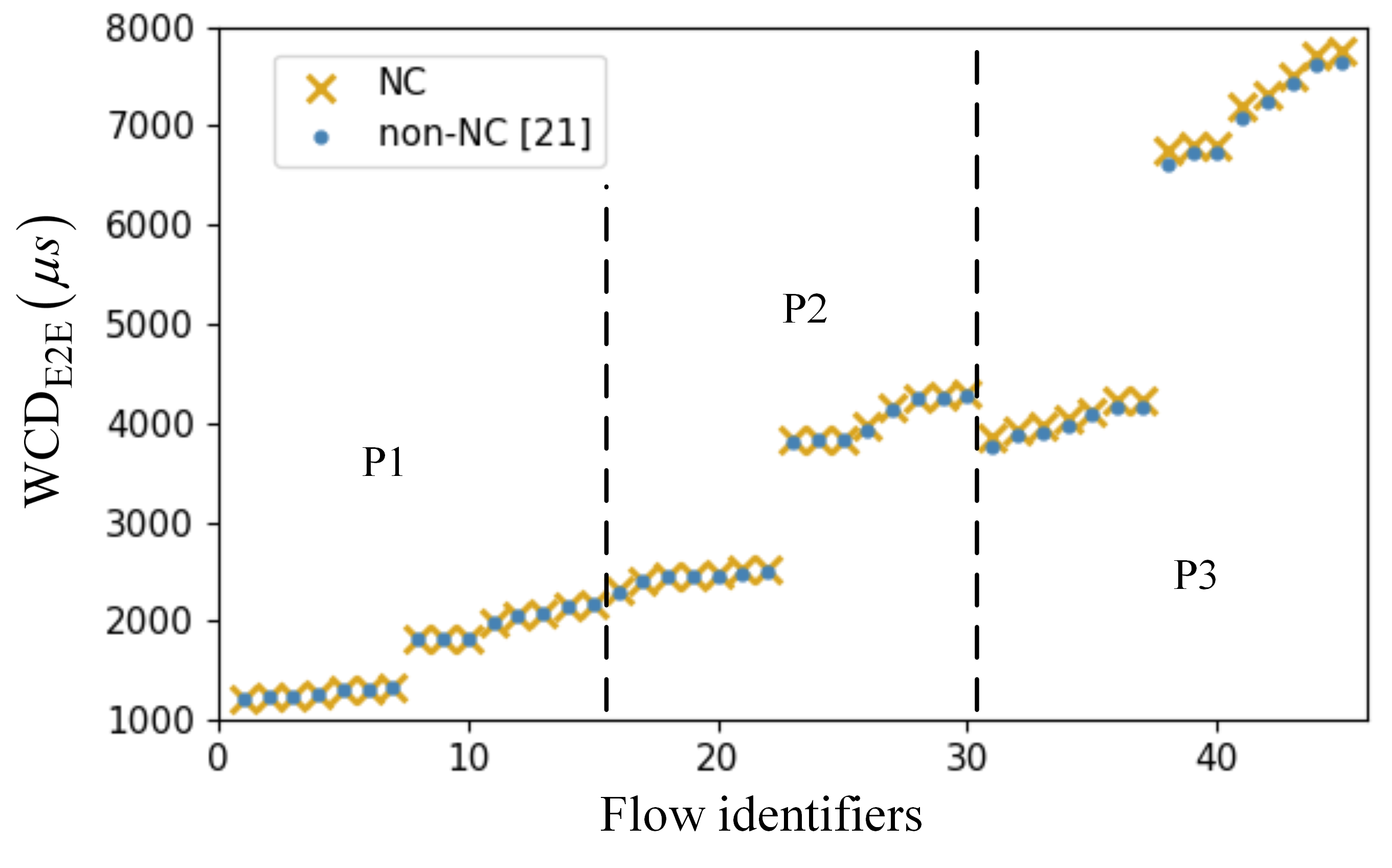}
	\caption{Comparison of NC and non-NC methods for ATS evaluation}
	\label{fig:cmpNC-nonNC}
\end{figure} 

For the evaluation of the two approaches, we use a synthetic test case where the topology is a medium mesh (Fig.~\ref{fig:SynTC_topo}), including 45 flows with 3 priorities. ATS is applied to all the priorities. The average traffic load is around 70\%, and the physical link rate is set to 100 Mb/s. We show a comparison of the two methods in Fig.~\ref{fig:cmpNC-nonNC}, where the value on the x-axis represents the identifiers of each flow, and the y-axis shows the upper bound of end-to-end latency in microseconds. The obtained results are grouped by priority, denoted by vertical dotted lines, and sorted in increasing order by results within each priority.
As we can see from Fig.~\ref{fig:cmpNC-nonNC}, the performance evaluation by the NC analysis of ATS is very close to the analysis from~\cite{Soheil16}, which is as expected according to Eq.~(\ref{g:delta}). We note that with a decrease in priority, the gap between the two will increase slightly. This is because the denominator of the first term of Eq.~(\ref{g:delta}) is related to the sum of the rates of all high-priority flows. The lower the priority of the flow of interest, the greater the rate accumulated by the high-priority flow. Thus, the first term of Eq.~(\ref{g:delta}) is becoming larger. For example, for the highest priority flows, the results from the two approaches are the same. For the lower priority 1 and 2, the evaluation results calculated by NC are 0.6\% and 1.4\% slightly more pessimistic on average than the results by the non-NC approach, respectively. 

However, the non-NC approach proposed by~\cite{Soheil16} is focused on ATS in isolation, and thus it is not applicable and cannot be extended to combinations of traffic shapers. Hence, in this paper, we consider the Network Calculus approach for evaluating various traffic shapers and their combinations. All the following evaluation results are based on the Network Calculus approach introduced in this article.

\subsubsection{Performance comparison among TAS, ATS, SP and CBS}
\label{sec:CmpIndiDiffTopo}
In the first set of experiments, we are interested in comparing the performance from the perspective of the upper bounds of end-to-end latency, jitter and backlog without the frame loss for each individual traffic shapers (including TAS, ATS, CBS and SP) under different network topologies. The network topologies are respectively small ring \& mesh (SRM), medium ring (MR), medium mesh (MM), small tree-depth 1 (ST) and medium tree-depth 2 (MT) which are inspired by industrial application requirements~\cite{Craciunas16:RTSJ}, as shown in Fig.~\ref{fig:SynTC_topo}. There are 100 test cases (TCs) randomly generated. 
For each test case, there are 15 flows. The frame size $l_f$ of each flow is randomly chosen between the minimum (64 bytes) and the maximum ($1,522$ bytes) Ethernet frame size, and flows can be periodic or sporadic\footnote{Flows served by the TAS are periodic, and ATS and AVB support both periodic and sporadic flows.}. For the periodic flow, the periods are uniformly selected from the set $T_f=\{1\,000, 2\,000, 5\,000, 10\,000\}$ $\mu s$. For the sporadic flows, it is assumed that each flow satisfies the leaky bucket model with the burst $b_f=l_f$ and rate $r_f=l_f/T_f$, where $T_f$ is the minimum interval between two consecutive frames. Since TT flows manipulated by the TAS have no priority division, it is assumed that all flows are assigned to the same priority level for each use case in this experiment. The GCLs for TAS are generated according to~\cite{Pop16}. All the test cases are applied to the above five topologies, respectively. The routes of flows are generated according to the routing optimization strategy proposed for TT traffic~\cite{Pop16}. The idle slope for AVB traffic is set to the default value of 75\%. For each test case, we considered the average hops of flow and the average traffic load under each topology. Table~\ref{tab:SynTCPara} gives the statistics over the 100 test cases under each topology. 
The physical link rate is set to $C$=100~Mb/s.

\begin{table}[!t]
	\caption{Statistical hops and traffic load for 100 test cases}
	\label{tab:SynTCPara}
	\centering	
	\begin{tabular}{c|c|c|c|c|c|}
		\cline{1-6}
		& SRM & MR & MM & ST & MT  \\ \hline
		Average Hops & 2.7 & 4.2 & 3.8 & 3.5 & 5.5  \\
		Average Traffic Load & 28.9\% & 20.5\% & 17.4\% & 29.0\% & 19.7\%  \\
		Max Traffic Load & 47\% & 40\% & 38\% & 47\% & 30\%  \\
		Min Traffic Load & 13\% & 8\% & 6\% & 13\% & 10\%   \\
		\hline
	\end{tabular}
\end{table}

\begin{figure}[t!]
	\centering
	\subfigure[End-to-end latency bounds]{
		\includegraphics[width=0.47\textwidth]{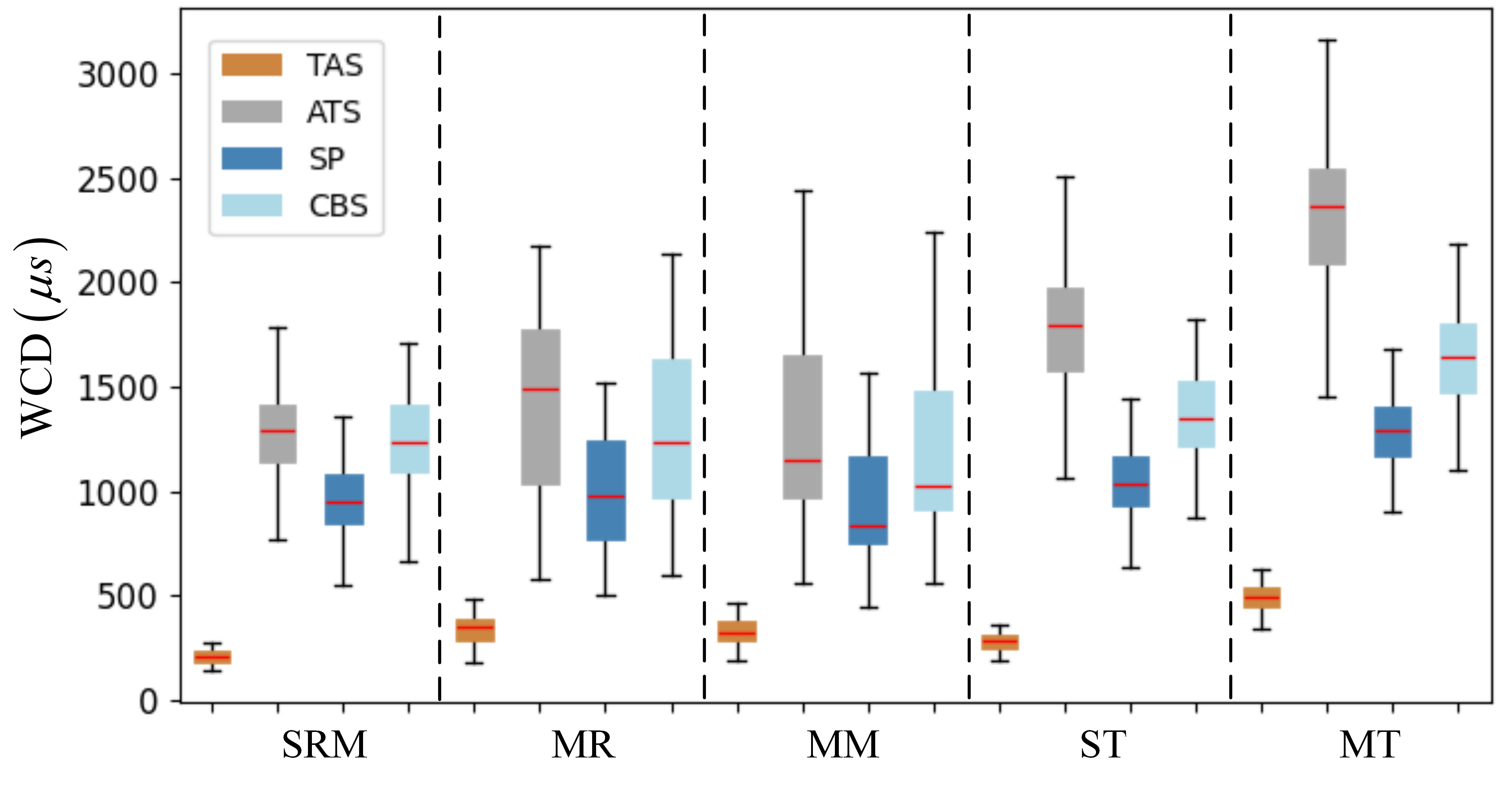}
	}
	\subfigure[Backlog bounds]{
		\includegraphics[width=0.47\textwidth]{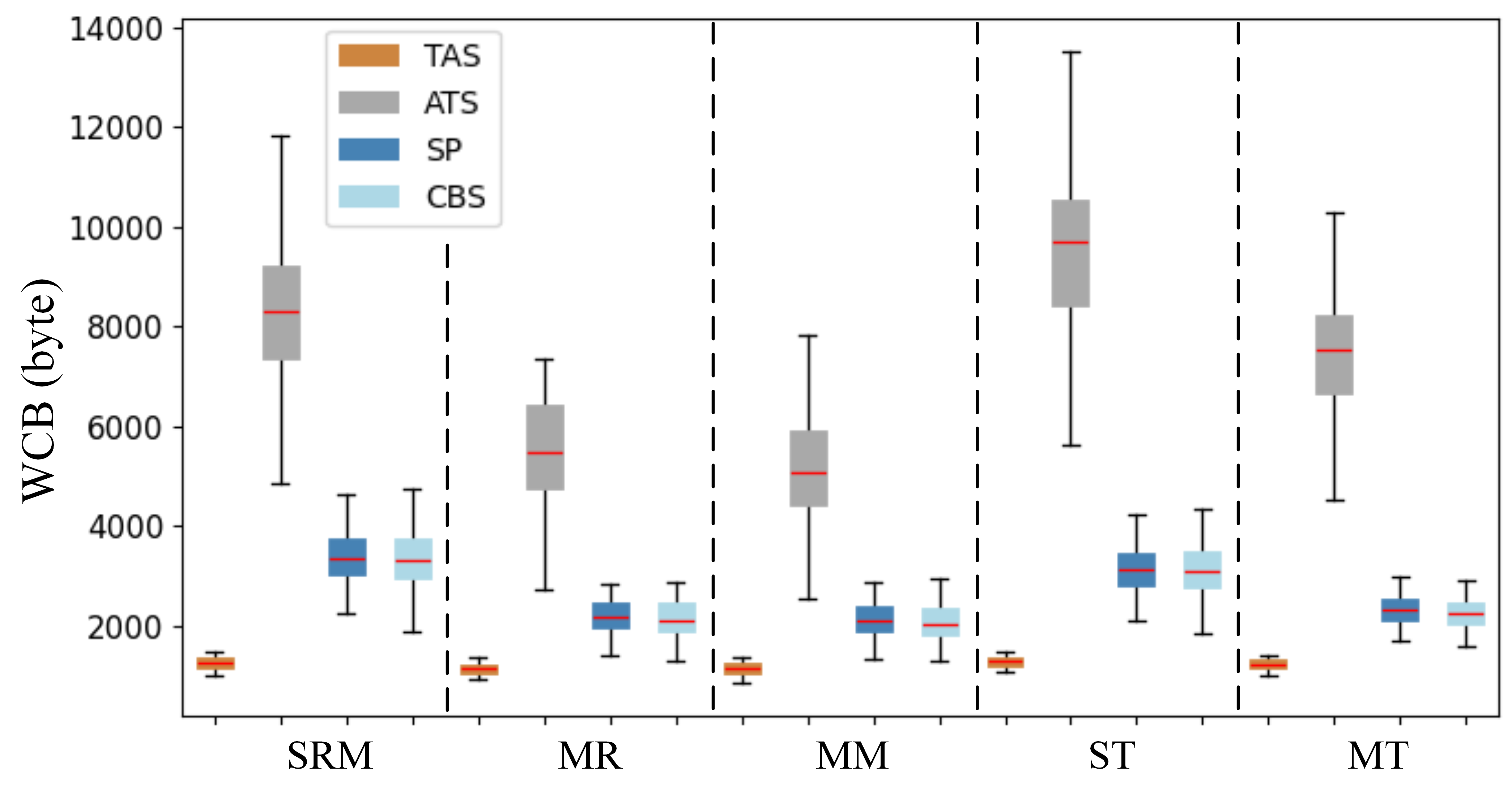}
	}
	\subfigure[Jitter bounds]{
		\includegraphics[width=0.47\textwidth]{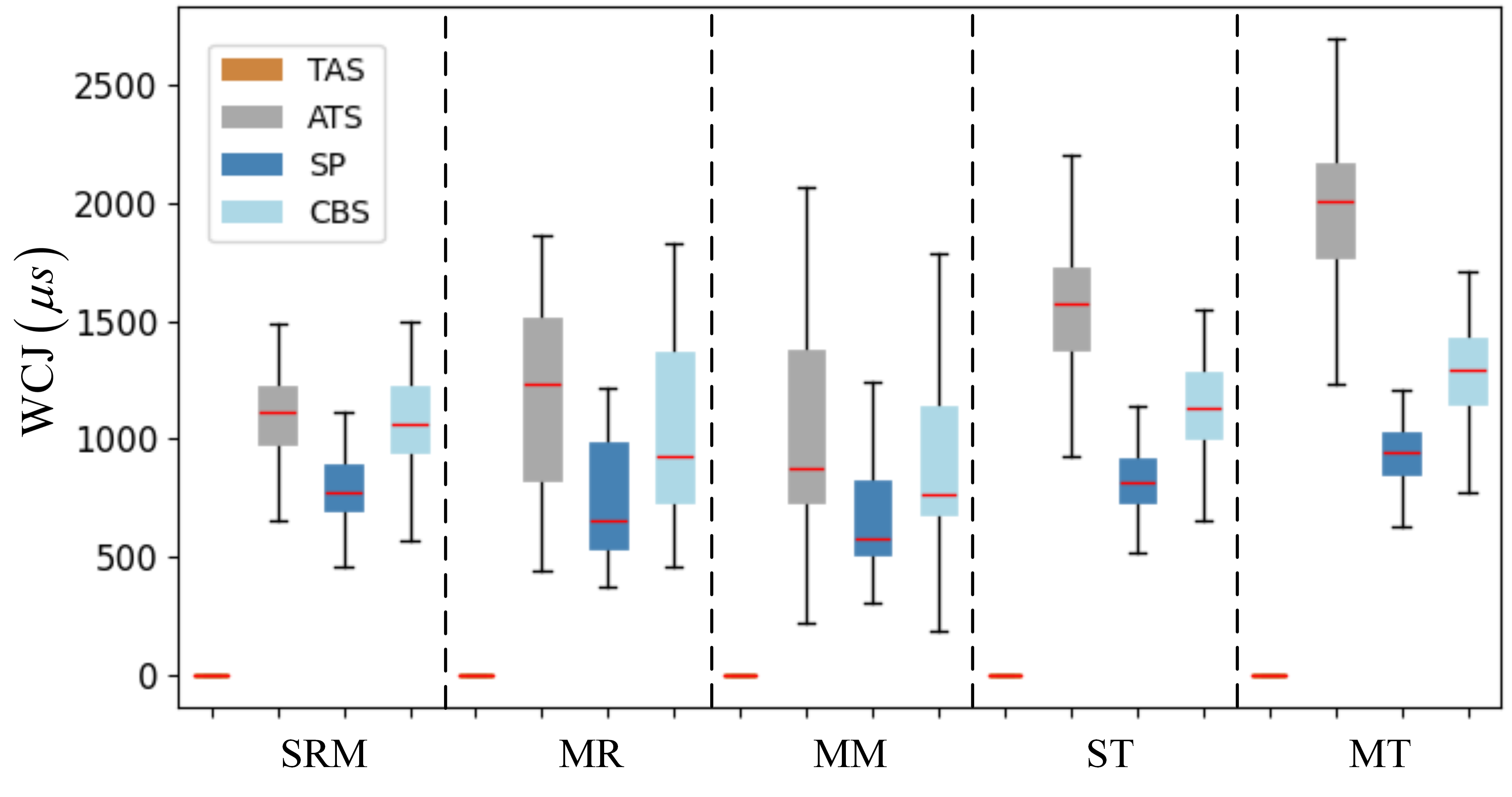}
	}
	\caption{\label{fig:CmpIndi} Comparison of different individual traffic shapers under different topologies}
\end{figure}

For each test case under a given topology, we evaluate the quality of service for different individual traffic shapers, i.e., TAS, ATS, CBS and SP, respectively. 
By applying the NC approach, we can get the two evaluation metrics of each flow under different individual traffic shapers, i.e., the upper bound of end-to-end delay and jitter; and one evaluation metric for each egress port, i.e., the upper bound of the backlog without the frame loss. 
Fig.~\ref{fig:CmpIndi} presents the performance evaluation of the network from the perspective of the upper bounds of end-to-end delay, backlog and jitter of different individual traffic shapers under different topologies. 
For each test case, we use the average value of the corresponding evaluation metric of all flows to represent the metric value under the current test case. Therefore, for each individual traffic shaper under each topology, we obtain 100 values of the corresponding metric, and we use box plots to present these results.
Fig.~\ref{fig:CmpIndi}(a) shows the evaluation of the end-to-end latency bounds. The x-axis represents different topologies, and the y-axis shows the upper bound of the end-to-end latency (WCD) in microseconds. Fig.~\ref{fig:CmpIndi}(b) and Fig.~\ref{fig:CmpIndi}(c) show the evaluation results on the upper bounds of worst-case backlog (WCB) in bytes and jitter (WCJ) in microseconds, respectively.

As can be seen in the figure,  
TAS performs best with the lowest latency and backlog, and provides zero-jitter. Such performance is in line with expectations. Since TAS realizes a completely deterministic time-triggered transmission through flow-based scheduling, it avoids the collision of frames of its own traffic type and avoids the collision with frames of other traffic types as well. Thus, flows shaped by TAS can achieve ultra-low latency, backlog and jitter.
For the other three traffic shapers, i.e., ATS, SP and CBS, the overall trend of their performance comparison can be inferred from the figure. However, it is difficult to see their comparison on an individual test case from the figure.
Thus, we additionally use the difference ratio
\begin{equation}\label{g:DiffRatio}
	X_i=(X_i^{Y_1}-X_i^{Y_2})/X_i^{Y_2}
\end{equation}
to capture the performance comparison of traffic shapers $Y_1$ and $Y_2$ ($Y\in\{\ATS,\CBS,\SP\}$), where $X_i$ can represent the end-to-end latency bound or jitter bound of the flow $f_i$ or the upper bound of the backlog for the queue $Q_i$ at the egress port. Then, we take the average value $X=average(X_i)$ as the comparison result of two traffic shapers under the current use case. The comparison results are shown in Table~\ref{tab:SynTCIndiCmp}.

\begin{figure*}[t!]
	\centering
	\subfigure[End-to-end latency bounds (no interference)]{
		\includegraphics[width=0.44\textwidth]{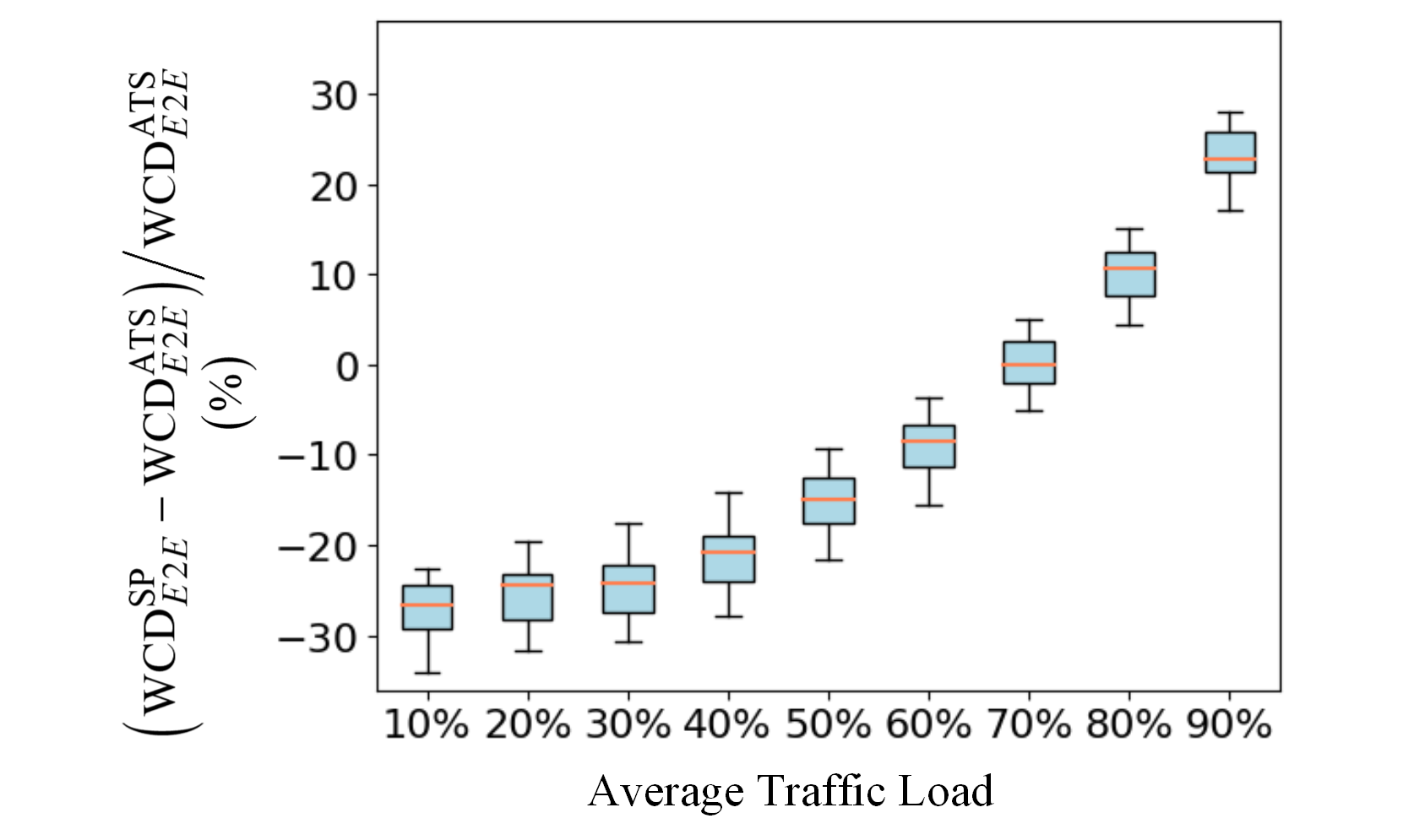}
	}
	\subfigure[Backlog bounds (no interference)]{
		\includegraphics[width=0.44\textwidth]{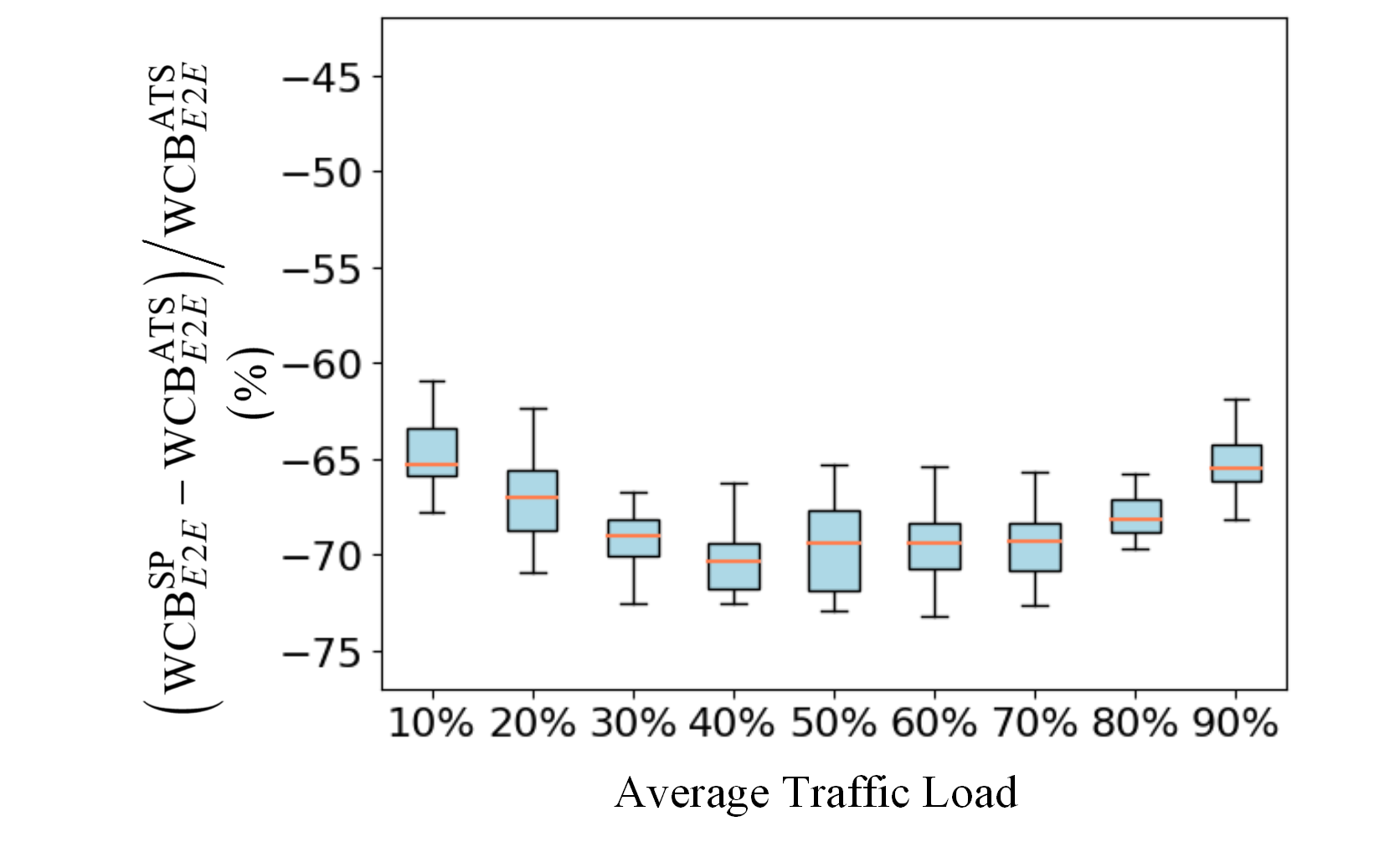}
	}
	\subfigure[End-to-end latency bounds ($l_{\text{BE}}=1522$ B]{
		\includegraphics[width=0.44\textwidth]{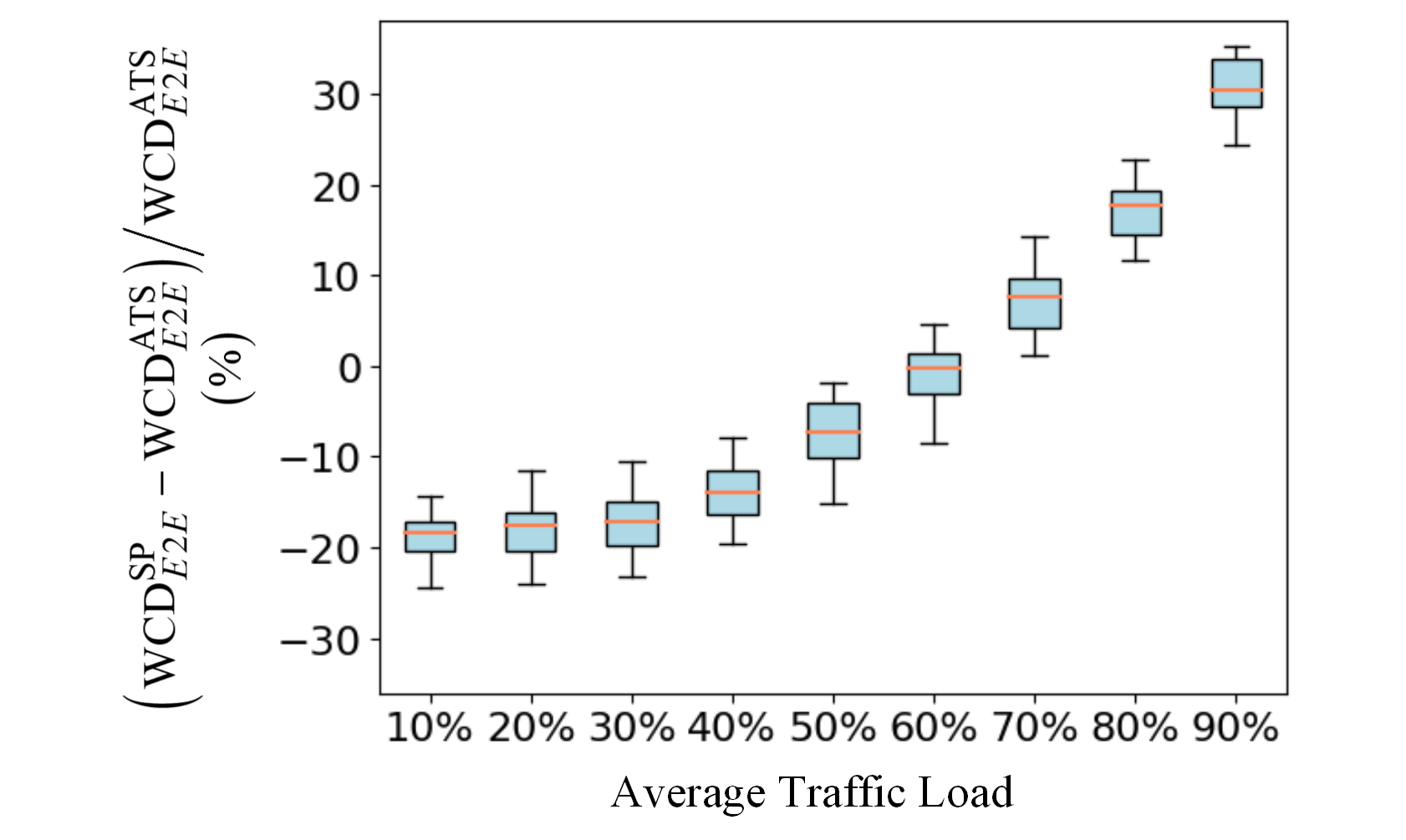}
	}
	\subfigure[Backlog bounds ($l_{\text{BE}}=1522$ B)]{
		\includegraphics[width=0.44\textwidth]{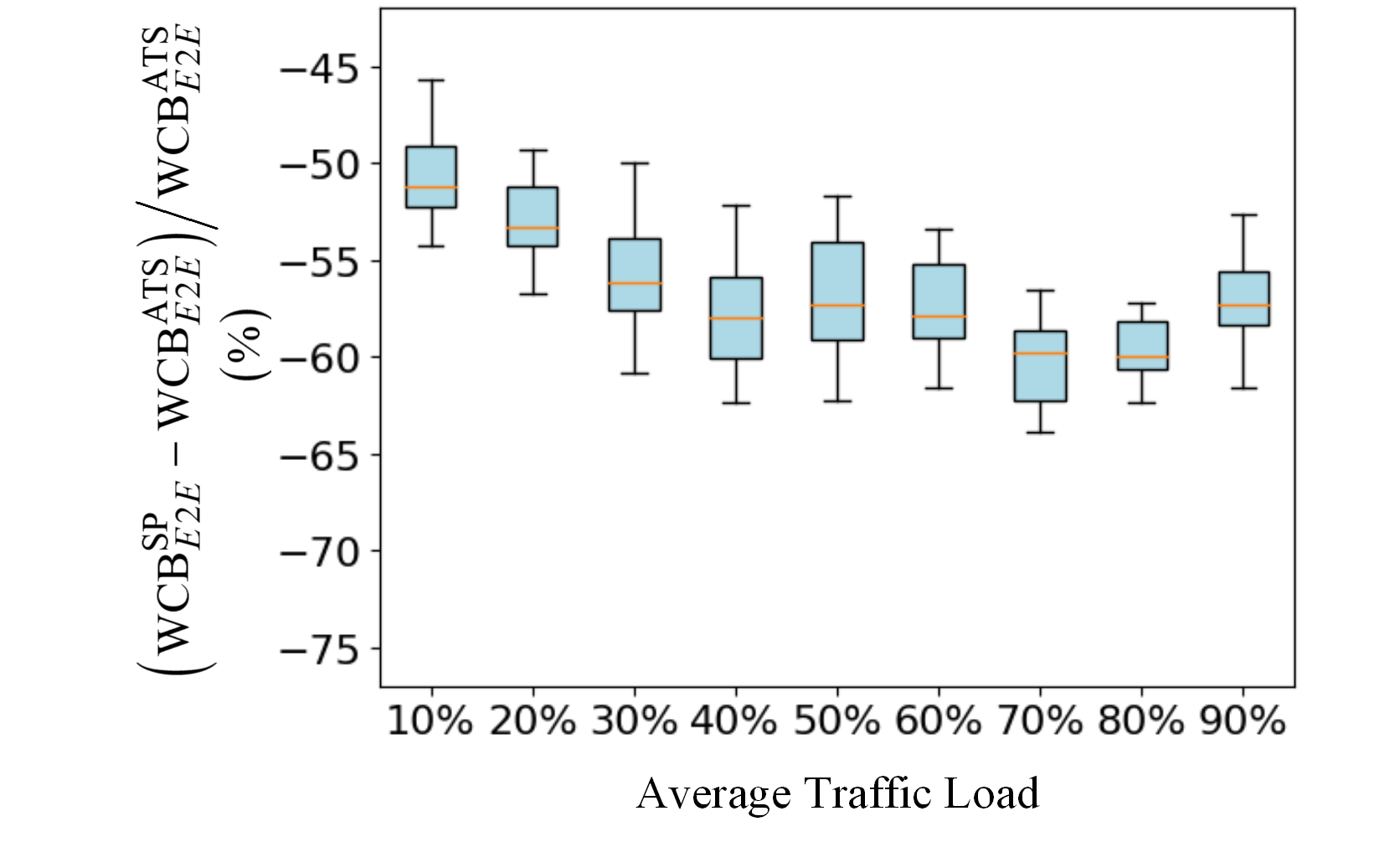}
	}
	\subfigure[End-to-end latency bounds (two priorities)]{
		\includegraphics[width=0.44\textwidth]{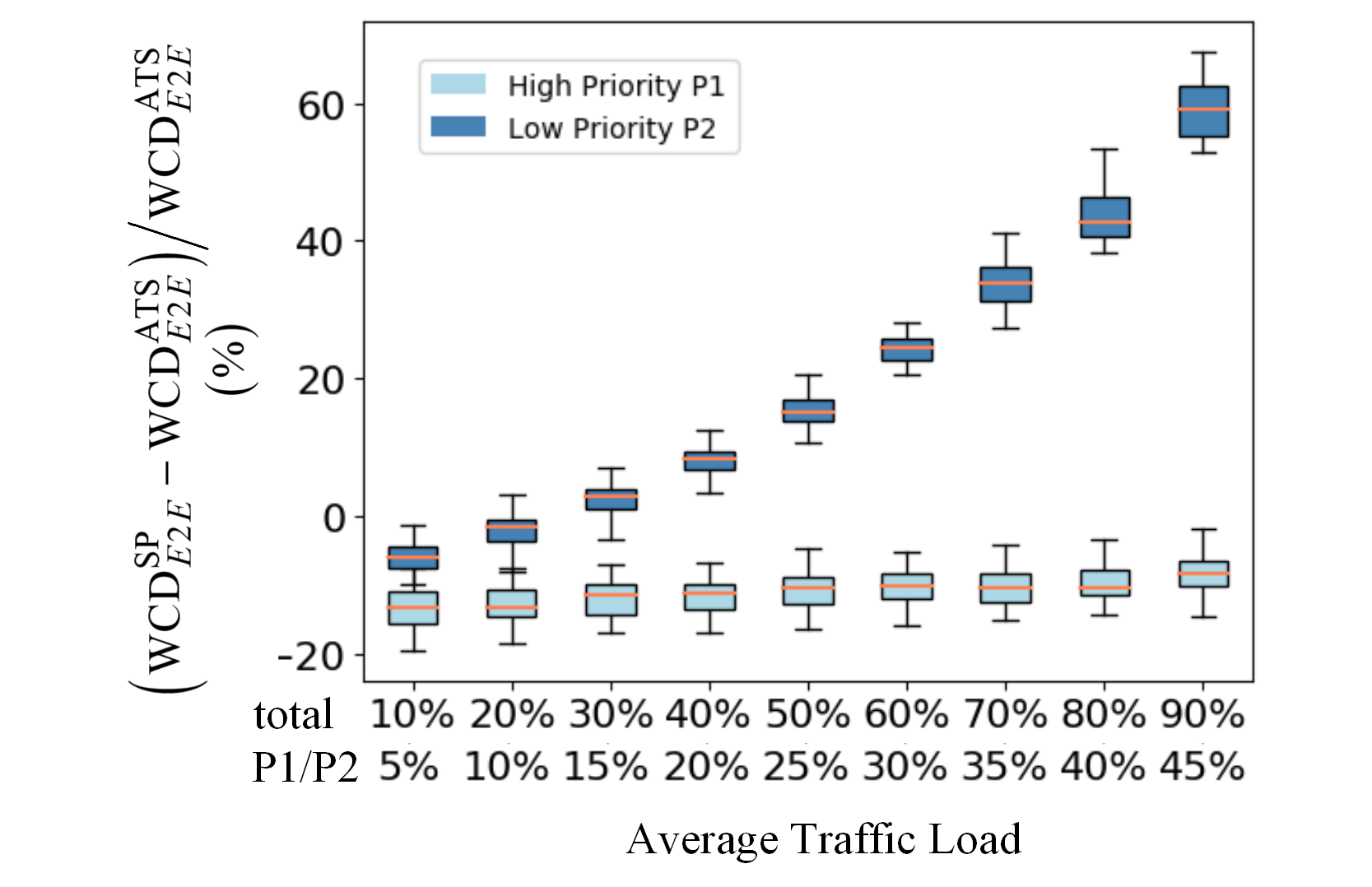}
	}
	\subfigure[Backlog bounds (two priorities)]{
		\includegraphics[width=0.44\textwidth]{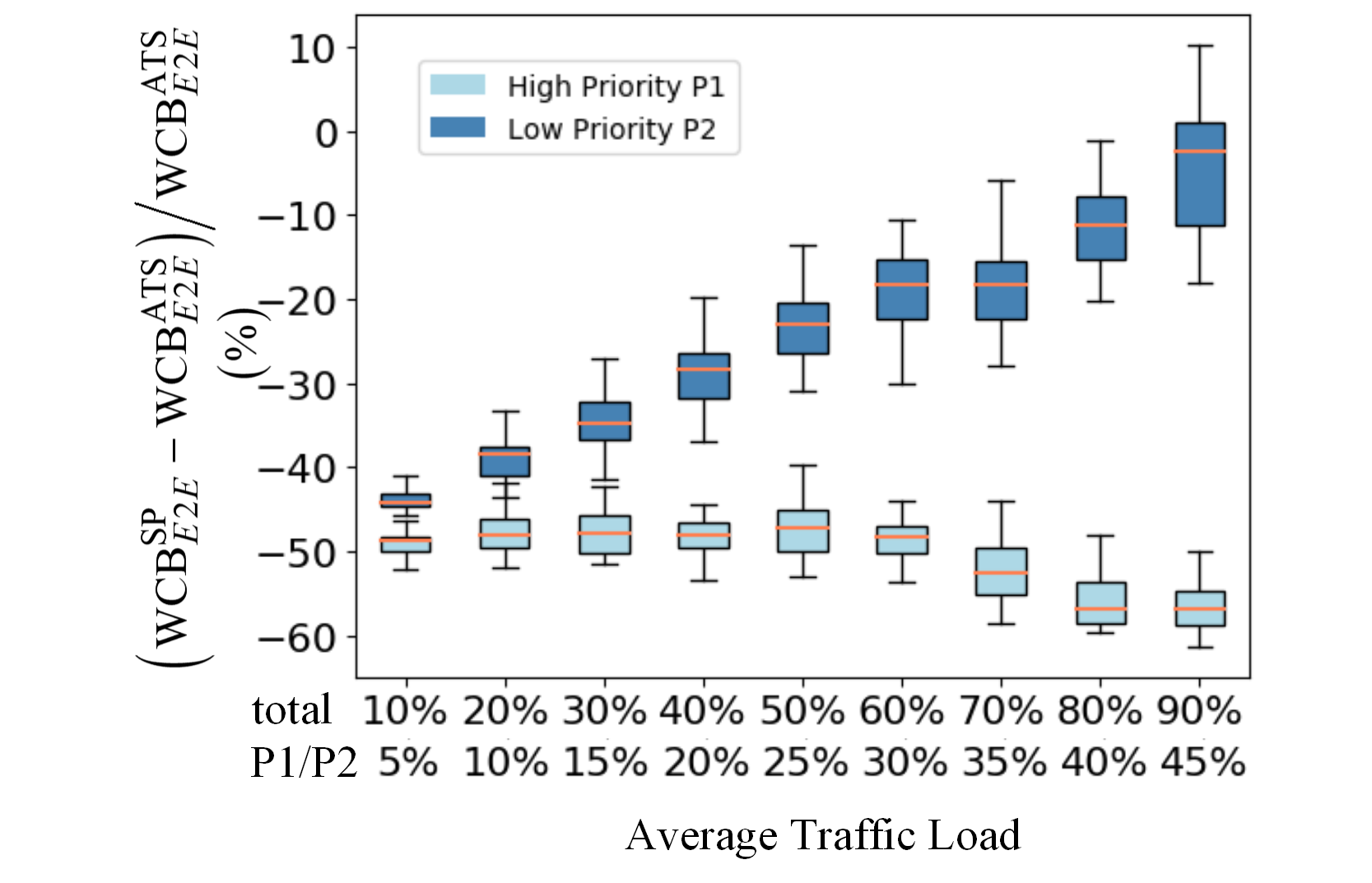}
	}
	\caption{\label{fig:CmpATSSP} Comparison of ATS and SP under different traffic load}
\end{figure*}

Regarding the delay and jitter upper bound, SP performs better than CBS, as shown in Fig.~\ref{fig:CmpIndi}(a) and (c), and the ``Average WCD'' and ``Average WCJ'' of the third column in Table~\ref{tab:SynTCIndiCmp}.
This is because CBS has a bandwidth reservation mechanism for traffic. In our case, only 75\% of the bandwidth is available for AVB traffic. Therefore, compared to SP, the service bandwidth obtained for flows shaped by CBS is lower. 
Concerning the backlog bounds, CBS may perform better than SP, as presented in Fig.~\ref{fig:CmpIndi}(b) and the ``Average WCB'' of the third column in Table~\ref{tab:SynTCIndiCmp}. 
This is because although the waiting time of the flow in the corresponding priority queue has been prolonged by CBS through controlling the credit, it nevertheless reduces the long-term rate of arrival of flows. Thus, the backlog upper bounds of AVB traffic are probably lower than for SP traffic.
\begin{table}[!t]
	\caption{Difference ratio on metrics of two individual traffic shapers}
	\label{tab:SynTCIndiCmp}
	\centering	
	\begin{tabular}{|c|c|c|c|c|}
		\cline{1-5}
		&  & \tabincell{c}{$(\CBS-\SP)$ \\ $/\SP$} & \tabincell{c}{$(\ATS-\SP)$ \\ $/\SP$} & \tabincell{c}{$(\ATS-\CBS)$ \\ $/\CBS$}   \\ \hline\hline
		\multirow{5}*{\tabincell{c}{Average \\ WCD}} & SRM & 30.7\% & 34.1\% & 2.8\%   \\
		& MR & 28.2\% & 43.9\% & 12.5\%   \\
		& MM & 26.2\% & 35.7\% & 7.7\%   \\
		& ST & 31.2\% & 72.3\% & 31.4\%   \\
		& MT & 27.3\% & 79.1\% & 40.8\%   \\ \hline
		\multirow{5}*{\tabincell{c}{Average \\ WCB}} & SRM & -1.3\% & 145.4\% & 148.8\%   \\
		& MR & -2.5\% & 151.3\% & 157.7\%   \\
		& MM & -3.2\% & 142.7\% & 150.6\%   \\
		& ST & -0.8\% & 206.1\% & 208.7\%   \\
		& MT & -3.3\% & 221.0\% & 231.9\%   \\ \hline
		\multirow{5}*{\tabincell{c}{Average \\ WCJ}} & SRM & 37.1\% & 41.4\% & 3.4\%   \\
		& MR & 38.0\% & 60.7\% & 16.6\%   \\
		& MM & 30.3\% & 43.0\% & 10.4\%   \\
		& ST & 39.5\% & 91.9\% & 37.7\%   \\
		& MT & 37.3\% & 108.3\% & 51.9\%   \\
		\hline
	\end{tabular}
\end{table}

What is more interesting is that, for all the current use cases, the performance of ATS is not better than that of SP and CBS, a finding that has not been reported in the related work. Due to the reshaping function of ATS, which avoids burstiness cascades, we initially hypothesized that ATS would improve the performance of flows. However, as it can be observed in Fig.~\ref{fig:CmpIndi}(a), (b), (c) and the fourth and fifth columns in Table~\ref{tab:SynTCIndiCmp}, in terms of upper bounds of delay, backlog or jitter, ATS did not perform superiorly. 

In the following, we will only focus on the comparison of ATS and SP, as ATS compared with CBS can be inferred from the comparison of SP and CBS discussed above. 
From the current results, we can draw a conclusion and a hypothesis. The conclusion is that the advantages of ATS are getting worse with the increase in the concentration of flows transmission and the number of hops (under the same concentration of flows transmission). This is because the more concentrated the flows are in the network, the more obvious is the serialization of flows in transmission, which increases the determinism of traffic transmission at subsequent nodes. Moreover, in the case of the constant concentration of flows transmission, as the number of hops increases, serialization leads to the same determinism of traffic transmission, but the time cost of ATS reshaping increases accordingly. The hypothesis is that as the load increases, the advantages of ATS will also increase, which will be discussed in the next section. The reasoning behind this hypothesis is that the time cost of ATS reshaping traffic cannot offset the queuing time of traffic with the burst cascade without ATS. Thus, ATS has a negative effect when the traffic load has not reached a certain level. Therefore, it is reasonable to infer that when the traffic load increases, the impact of ATS will gradually turn positive.

Next, we analyze the results on column 4 in Table~\ref{tab:SynTCIndiCmp}. Since each test case is applied under five topologies, for some use cases, the five topologies have the same flows, but just with different routes. Moreover, the traffic load in Table~\ref{tab:SynTCPara} reflects the degree of dispersion of flows transmission in the network. For example, the traffic load under the MM topology is the lowest. This is because the MM has the largest number of selectable paths, and thus the flows are more dispersed than the traffic in other topologies. Hence, although the traffic load in MM is lower than that in SRM, the performance comparison between ATS and SP is not much different from that in SRM. Compared with MR, the traffic load in MM is close to that of MR, but the transmission of traffic in MR is more concentrated than in MM. Therefore, the advantage of ATS under the MM topology is higher than under the MR topology. For the ST topology, although its traffic load is close to the traffic load in SRM, its number of hops is higher, which leads to more times of ATS reshaping along the flow's route. Thus, compared with SRM, the performance of ATS under the ST topology is far worse than SP. A similar explanation can be extended to the results for the MT topology.

\subsubsection{Comparison between ATS and SP under changing traffic load}
\label{sec:CmpTrafficLoadATSSP}
In the previous section, the traffic load of all test cases is in the range from 6\% to 47\%, and in all these test cases, ATS does not show its advantages in real-time performance compared with TAS, CBS and SP. In order to rule out the influence of traffic dispersion degree and number of hops under different topologies, for the second set of experiments, we chose the same topology (MM) to study the comparative performance between ATS and SP mechanisms when the average traffic load is increasing from 10\% to 90\%.
We used 20 random synthetic test cases under each traffic load. The frame sizes and intervals are selected in the same way as for the use cases in the previous section. As the comparison trend of individual shapers on the upper bounds of latency and jitter is very similar, we will only show the comparison of the upper bounds of delay and backlog in the following.

We still use the difference ratio from Eq.~(\ref{g:DiffRatio}) to represent the performance comparison result of ATS and SP under the current use case, where $X_i$ can represent the upper bound of the end-to-end latency of flow $f_i$ or the upper bound of the backlog for the queue $Q_i$ at the egress port.

In order to test the performance of ATS in isolation, we first assume that all flows have the same priority, and that there is no interference from other traffic types. This situation can also be similar to the network adopting the preemption integration mode. The comparison results of ATS versus SP are shown in Fig.~\ref{fig:CmpATSSP}(a) and (b). As it can be seen, for the upper bound of end-to-end delay (Fig.~\ref{fig:CmpATSSP}(a)), with the increase of average traffic load, the performance relationship between ATS and SP changes. When the average traffic load is lower than 70\%, SP performs better than ATS. ATS shows its superiority only when the average traffic load increases by more than 70\%.
For the upper bound of the backlog (Fig.~\ref{fig:CmpATSSP}(b)), the difference ratio between ATS and SP does not change significantly, and when considering ATS used individually, the backlog performance is always worse than that of SP. This is due to the fact that ATS is a kind of non-work conserving scheduler that implements reshaping through frame waiting. Moreover, as noted, shaped queues and the shared queue are implemented in a single physical queue, and the backlog for ATS are the sum of backlogs in shaped and shared queues.
It is furthermore assumed that all flows still have the same priority, but there exists the interference of BE traffic with the maximum Ethernet frame size of $1,522$ bytes. If the preemption integration mode is considered, the compared results will be similar to the discussion above (Fig.~\ref{fig:CmpATSSP}(a), (b)). If the non-preemption integration mode is taken into account (as considered in this paper), there will be at most one BE frame interference when the flow of interest obtains the service. The results with and without ATS reshaping are shown in Fig.~\ref{fig:CmpATSSP}(c) and (d). As can be observed from the figure, low-priority non-preemptible frames have no significant impact on the performance with or without ATS from the perspective of upper bounds of end-to-end delays. Latency performance comparison results still mainly depend on the traffic load. However, the performance with ATS and without ATS from the perspective of backlog upper bounds is significantly changed, even though ATS still performs worse than without ATS. This implies that the non-preemptible frames from low priority traffic have a larger impact on the backlog performance of egress ports without ATS reshaping compared to ports with ATS reshaping. From here, we can also see that the backlog bounds comparison between individual ATS and SP (one priority) is more related to the frame length.

Moreover, we assign high and low priorities to the traffic, and the high and low priority traffic each account for 50\% of the overall traffic load. The comparison results of ATS vs SP in terms of upper bound of the delays and backlogs for the high and low priority traffic, with and without reshaping by ATS, are shown in Fig.~\ref{fig:CmpATSSP} (e) and (f), respectively, using light blue and dark blue box plots. It can be seen from the figure that for the high priority traffic, the results are similar to the top 40\% compared results in Fig.~\ref{fig:CmpATSSP} (c) and (d), and similarly ATS does not show its superiority for high-priority traffic. This is because the transmission of the high priority traffic will only be interfered by at most one low priority frame whose frame length ranges from the minimum (64 bytes) to the maximum ($1,522$ bytes) Ethernet frame size, which is same as the case if all flows have the same priority and are interfered by a BE frame. Moreover, the maximum average load of the high-priority traffic is only 45\%. Nevertheless, for low priority, ATS shows its performance advantage from the perspective of latency bounds when the overall average traffic load reaches 30\% while low priority average traffic load reaches 15\%. From the perspective of backlog, no matter for the high-priority or the low-priority traffic, the backlogs of ATS are still worse than that without ATS when the overall average traffic load is increasing from 10\% to 80\% (the average traffic load for low priority is from 5\% to 40\% accordingly). The ATS backlogs of a few cases shows its superiority when the average load is as high as 90\%.
%ATS reshaping for low priority traffic always shows its superiority when the overall average traffic load is increasing from 10\% to 90\% (while the average traffic load for low priority is from 5\% to 45\% accordingly). The reason is that the low priority traffic can only be transmitted when the high priority queue is empty so that the burst cascade of low priority flows is more obvious than that of high priority flows. It also means that the time cost of reshaping low priority traffic by ATS is lower than the waiting time caused by the burst cascade of low-priority traffic.

Therefore, when the ATS is used individually, its positive effect on latency upper bounds can be highlighted only when the average traffic load in the network reaches a certain high level, around 70\% to 80\%, as in Fig.~\ref{fig:CmpATSSP}(a) and (c). Moreover, it always shows a negative impact on the upper limit of the backlog, as in Fig.~\ref{fig:CmpATSSP}(b) and (d). When ATS is used for traffic of multiple priorities, ATS shows a positive impact on latency upper bounds for low-priority traffic even if the average low-priority traffic load is not high. However, its impact on the high priority traffic is still negative, see Fig.~\ref{fig:CmpATSSP}(e). However, ATS still shows a negative impact on the backlog bound on both high and low priority, unless the average traffic load for high priority is reached at least 90\%, as shown in Fig.~\ref{fig:CmpATSSP}(f). Based on the above findings, this is why we believe that the combined use of ATS with TAS will make ATS play a more active role. At the same time, TAS will perfectly maintain its advantages of ultra-low latency and jitter. %In the next section, we will see that ATS has a positive role when combined with other traffic shapers.

\begin{figure}[t!]
	\centering
	\subfigure[Comparison of latency upper bounds of ATS with high priority and CBS with low priority]{
		\includegraphics[width=0.42\textwidth]{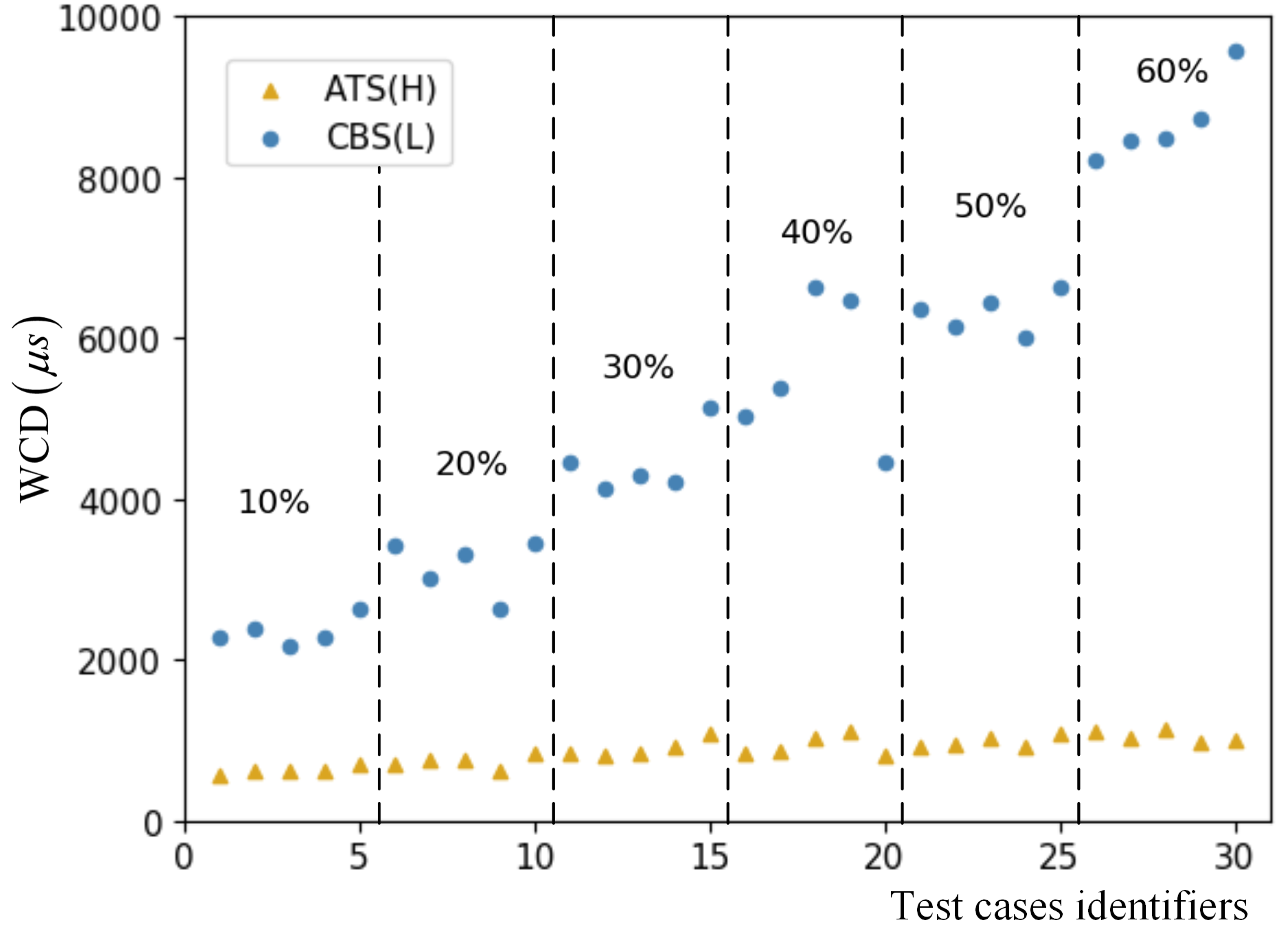}
	}
	\subfigure[Comparison of latency upper bounds of ATS with low priority and CBS with high priority]{
		\includegraphics[width=0.42\textwidth]{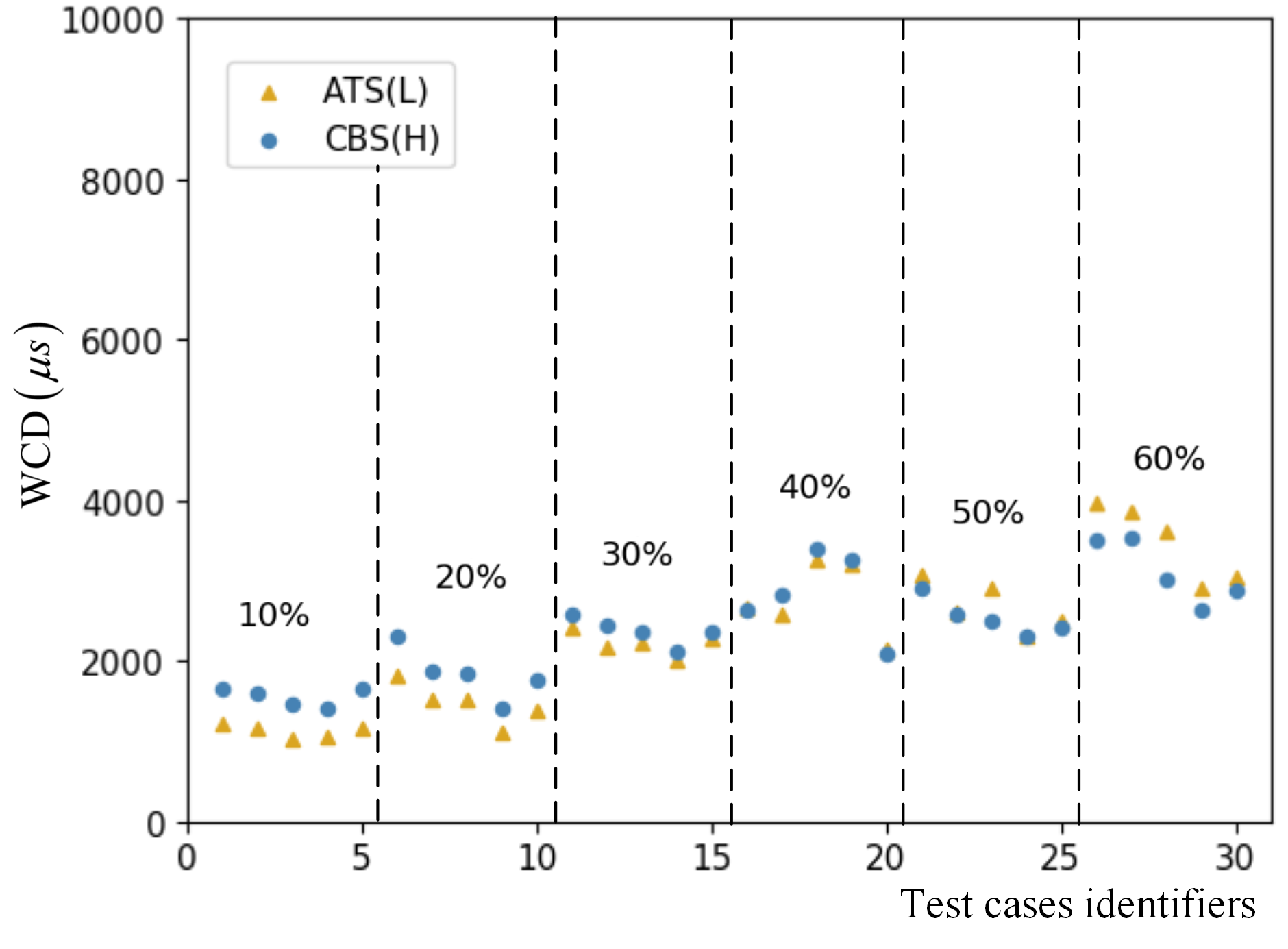}
	}
	\subfigure[Comparison of latency upper bounds of overall flows under ATS(H)+CBS(L) and CBS(H)+ATS(L)]{
		\includegraphics[width=0.42\textwidth]{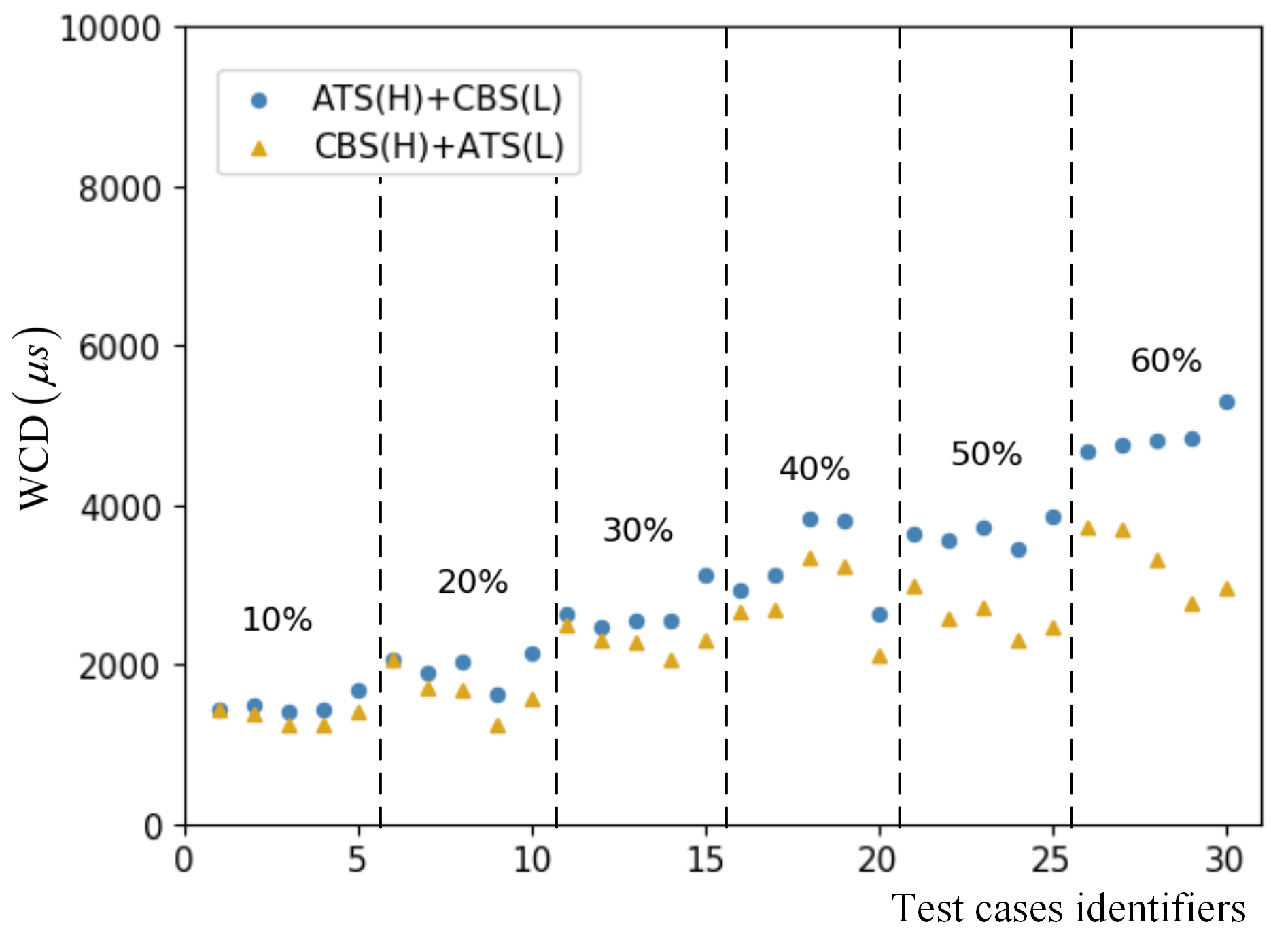}
	}
	\caption{\label{fig:CmpATSCBS} Comparison of ATS(H)+CBS(L) and CBS(H)+ATS(L) under different traffic load}
\end{figure}

\subsection{Combined Traffic Shapers}
\label{sec:CmpComb}
\subsubsection{Evaluation on Synthetic Test Cases}
\label{sec:EvaSynTC}
In this section, we are interested in the influence of ATS on the real-time performance of combined traffic shapers.

We first show the evaluation performance of the combination use of ATS and CBS in different queues, as per the architecture discussed in Sect.~\ref{sec:ATS+CBS_DiffQ}. %The first case is ATS+CBS with ATS for two-level of high priorities and CBS for two-level of low priorities, and the second case CBS+ATS is conversely. 
We will use the synthetic test cases adapted from Sect.~\ref{sec:CmpTrafficLoadATSSP} under MM topology. The average traffic load is increasing from 10\% to 60\%, including five test cases per load case. For each test case, flows are divided into four sets, two of which are configured as high- and low-priority SP traffic served by ATS, and two of which are high- and low-priority AVB traffic served by CBS. The idle slopes for each AVB Class are respectively 45\% and 30\%. 
%It is first assumed the case when flows reshaped by ATS have higher priority than AVB flows under the architecture ATS+CBS. Then, the second case is conversely under the architecture CBS+ATS, with the same flows but with assumption of AVB flows have higher priority than flows reshaped by ATS. 
We discuss the mutual influence of ATS and CBS under the two architectures ATS+CBS (ATS has higher priority than CBS, named ``ATS(H)+CBS(L)'') and CBS+ATS (CBS has higher priority than ATS, named ``CBS(H)+ATS(L)''). It is difficult to say which architecture is better, but we compare and explain some phenomena using the following Fig.~\ref{fig:CmpATSCBS}.
%The results are sorted in ascending order by result gaps. 

\begin{figure*}[t!]
	\centering
	\subfigure[End-to-end latency bounds (TT - 20\%)]{
		\includegraphics[width=0.44\textwidth]{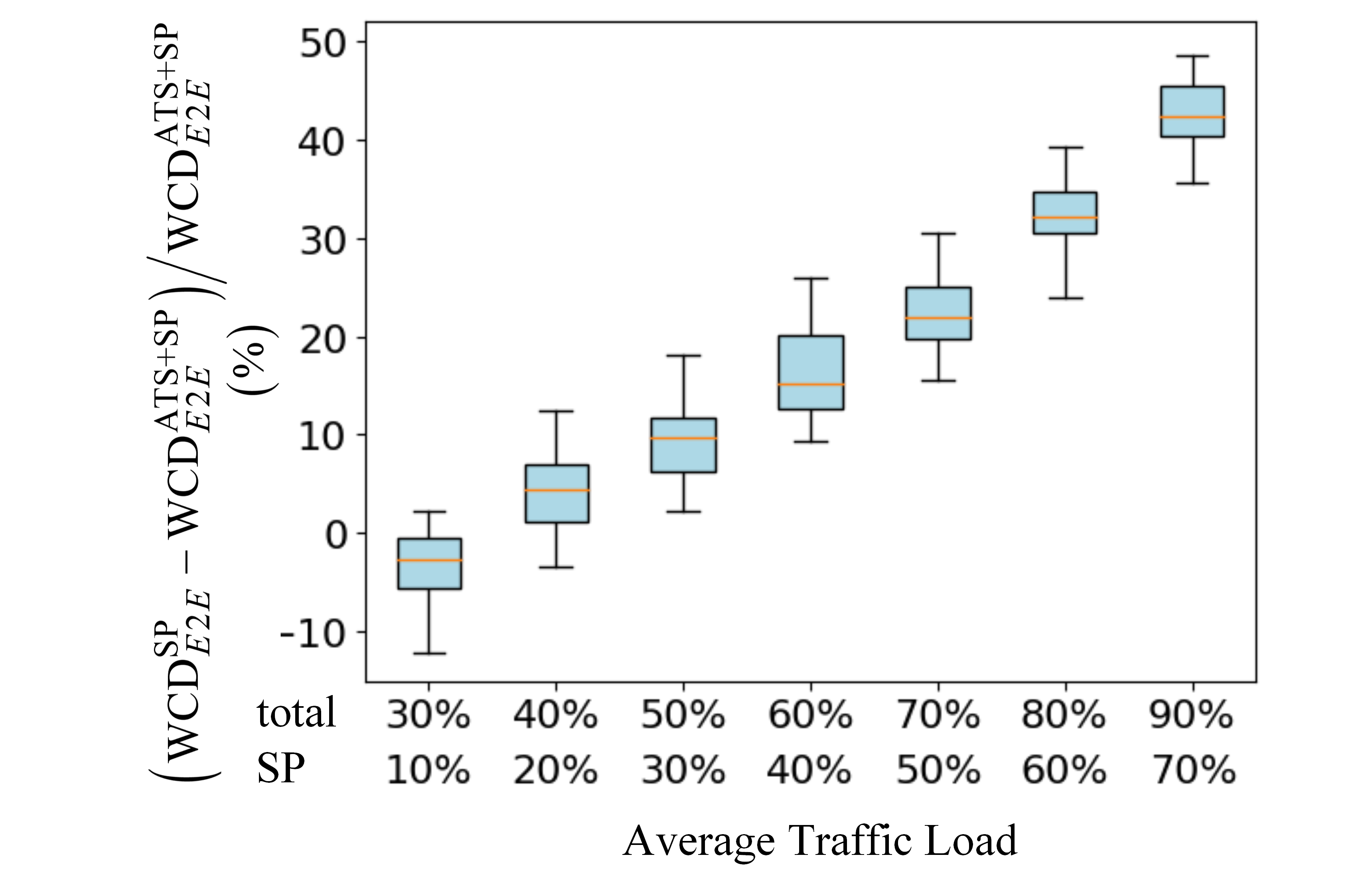}
	}
	\subfigure[Backlog bounds (TT - 20\%)]{
		\includegraphics[width=0.44\textwidth]{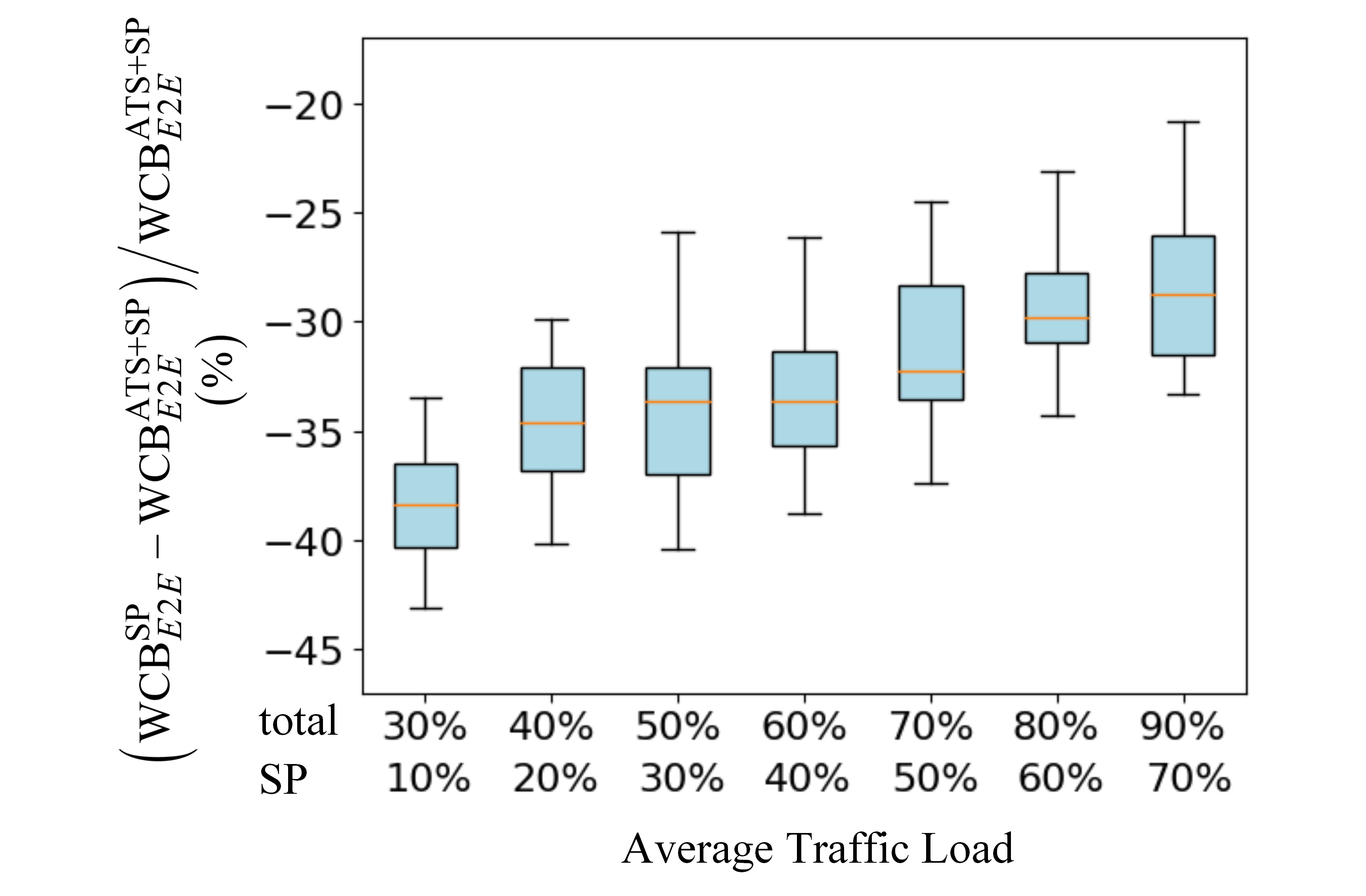}
	}
	\subfigure[End-to-end latency bounds (TT - 30\%)]{
		\includegraphics[width=0.44\textwidth]{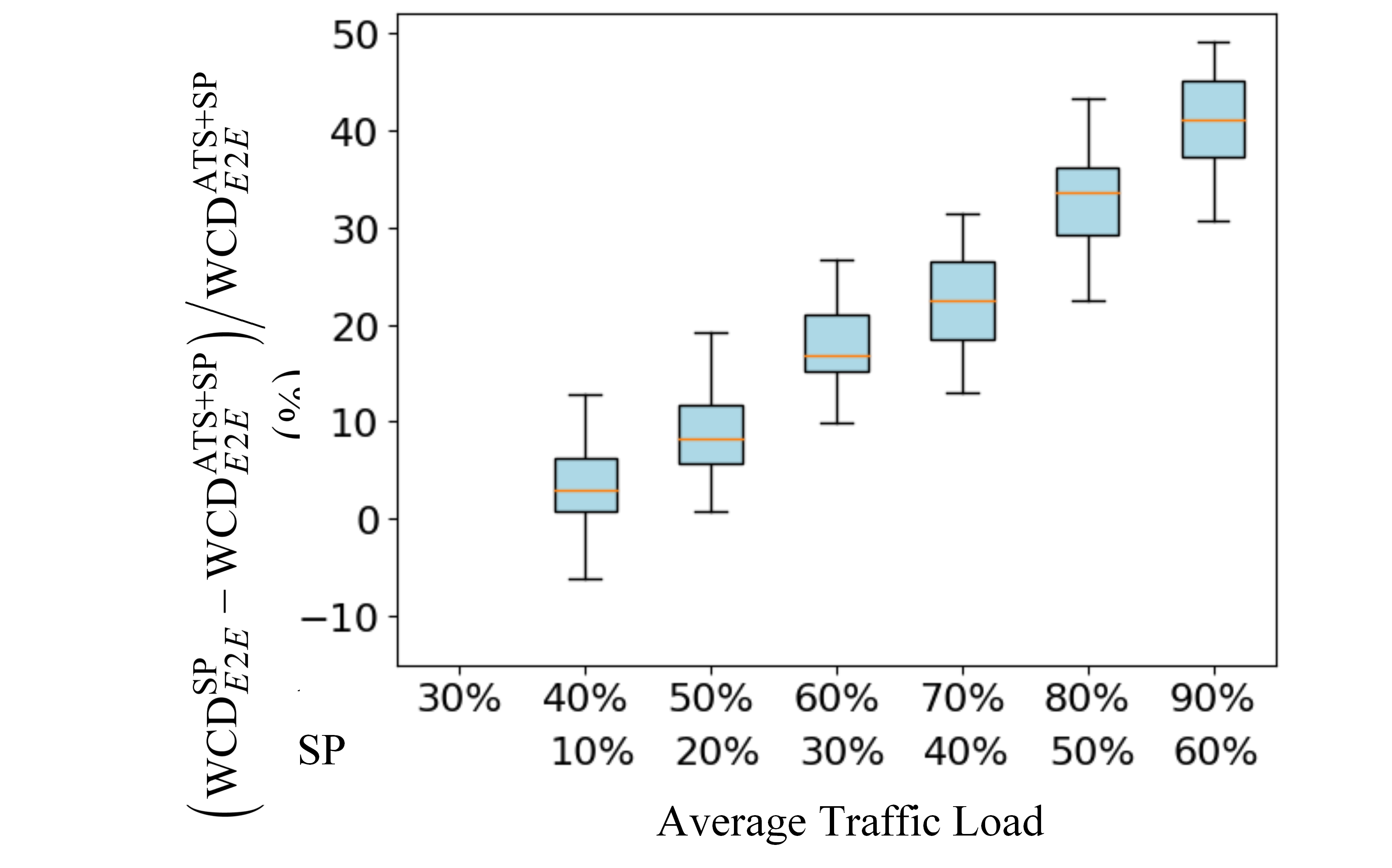}
	}
	\subfigure[Backlog bounds (TT - 30\%)]{
		\includegraphics[width=0.44\textwidth]{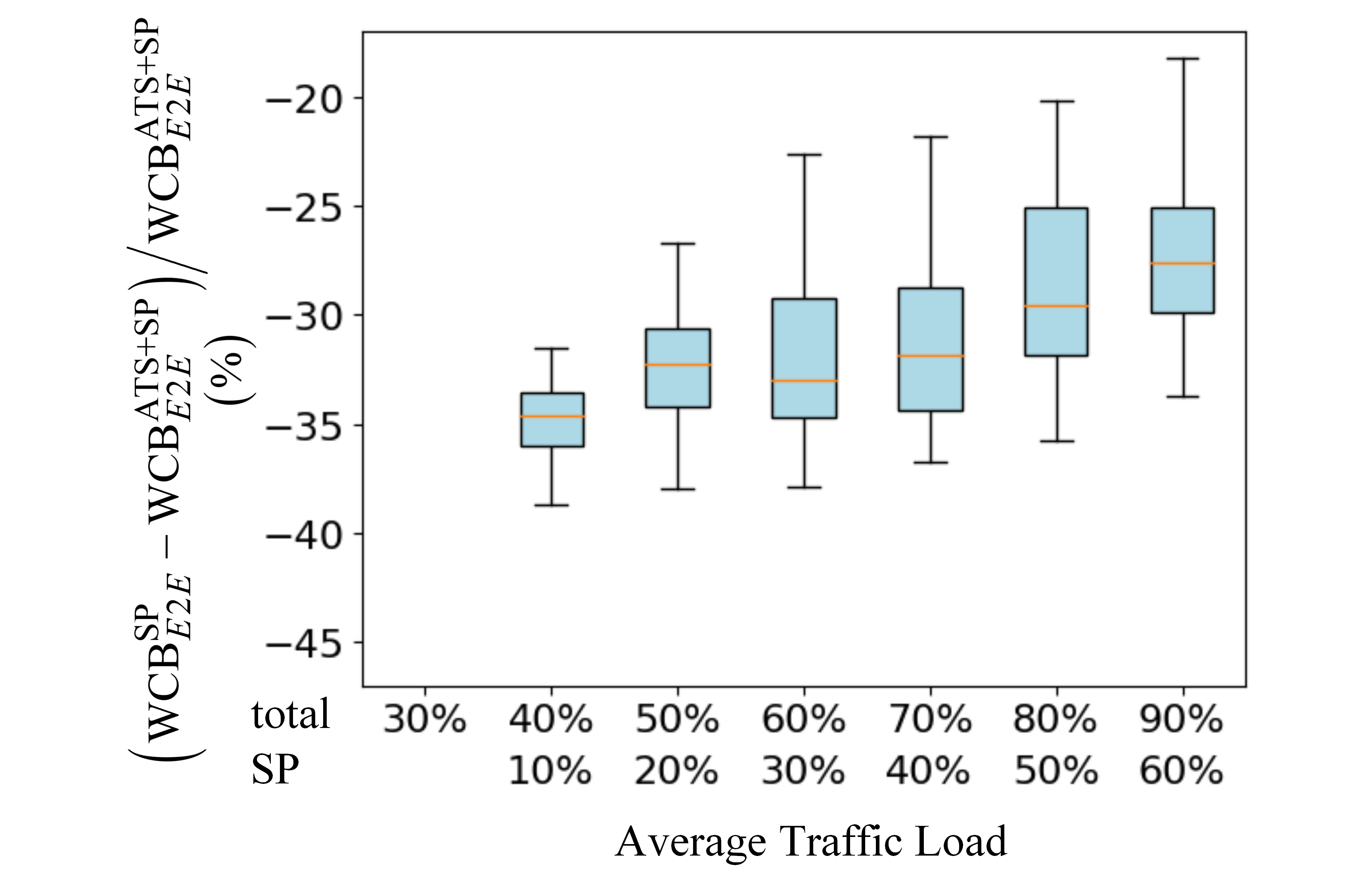}
	}
	\caption{\label{fig:CmpTASATSSP} Comparison of TAS+ATS+SP and TAS+SP under different traffic load}
\end{figure*}

\begin{figure*}[t!]
	\centering
	\subfigure[End-to-end latency bounds (TT - 20\%)]{
		\includegraphics[width=0.44\textwidth]{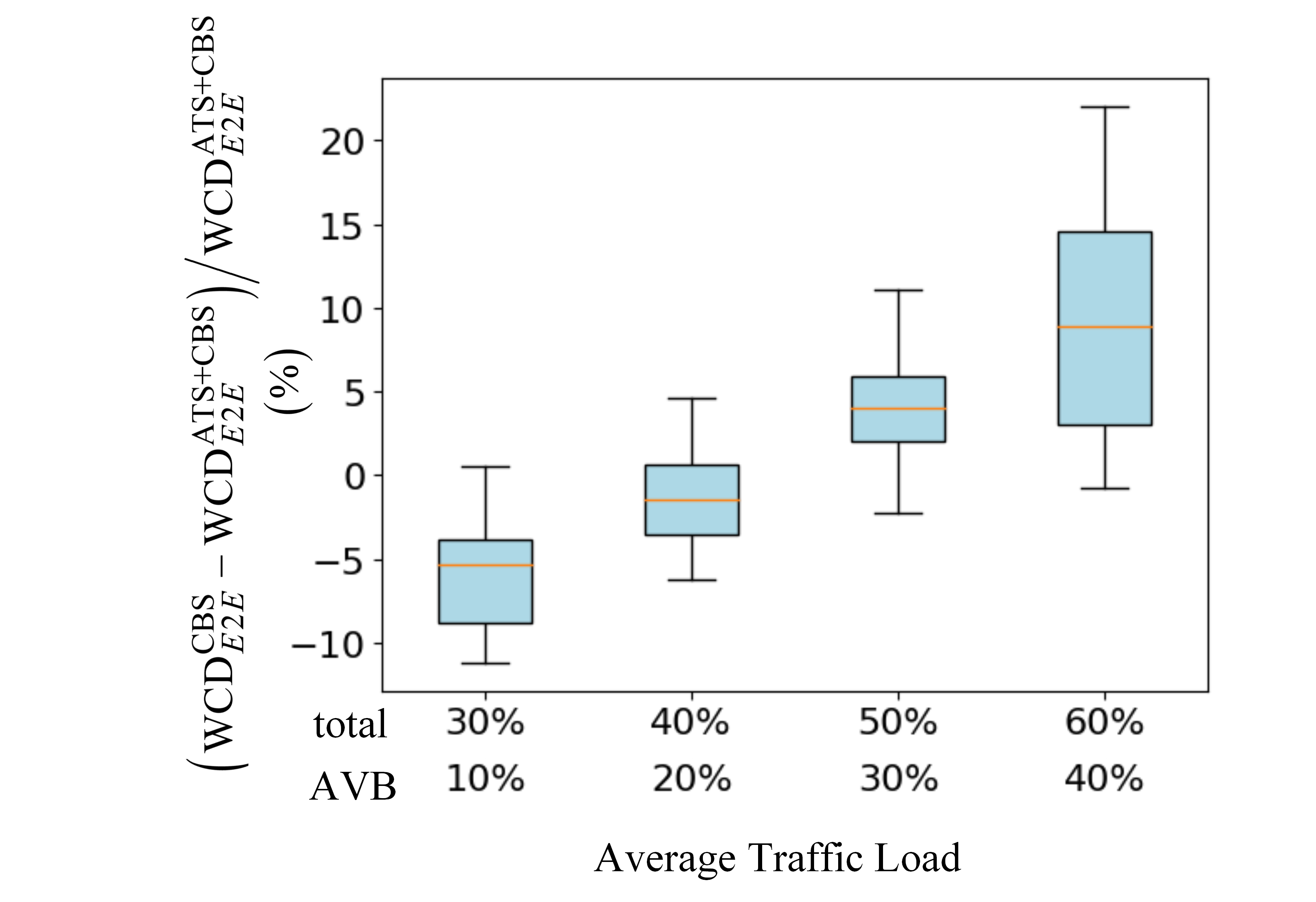}
	}
	\subfigure[Backlog bounds (TT - 20\%)]{
		\includegraphics[width=0.44\textwidth]{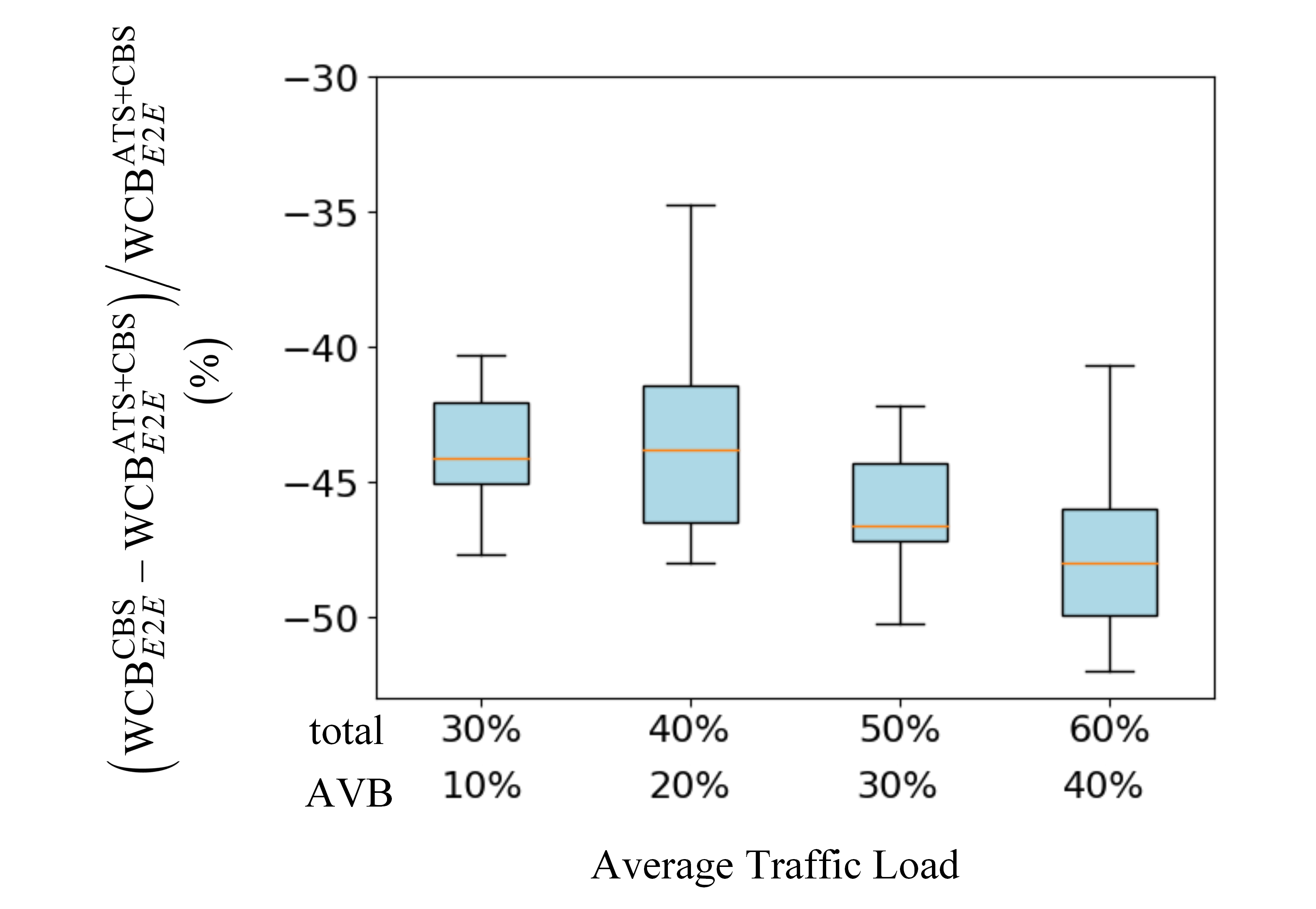}
	}
	\caption{\label{fig:CmpTASATSCBS} Comparison of TAS+ATS+CBS and TAS+CBS under different traffic load}
\end{figure*}

Fig.~\ref{fig:CmpATSCBS}(a) and (b) respectively present the average worst-case end-to-end delays of flows scheduled via ATS and CBS, when ATS and CBS are respectively treated as high priority. The results for ATS are depicted with yellow triangles and for CBS are with blue circles, and identifiers on the x-axis are divided by traffic load.
As can be seen from the figure, for these sets of test cases, no matter when ATS is used for high- or low-priority, its delay is always better than the delay of traffic scheduled by CBS when CBS is at the same priority. This is different from the results of ATS and CBS individually used under MM topology in Fig.~\ref{fig:CmpIndi}. It is due that all flows in Fig.~\ref{fig:CmpIndi} are assumed to have the same priority. So the bandwidth utilization (75\%) of CBS is all used for AVB flows of one priority, which means that the speed of credit recovery is fast. Moreover, the average traffic load in the Fig.~\ref{fig:CmpIndi} is relatively low. Thus, the cost of ATS shaping approximately offsets the delay drag caused by the non-full bandwidth usage of CBS. However, in this experiment, the bandwidth utilization (75\%) is assigned to two AVB classes, which makes the service time interval that can be assigned to each AVB class lengthened. This is because the speed of credit recovery slows down.
Another interesting finding is that the average worst-case end-to-end delays of all the flows (including both AVB flows and SP flows shaped by ATS) under the architecture CBS(H)+ATS(L) may be better than under the architecture ATS(H)+CBS(L), especially when the average traffic load is high. The compared results are shown in Fig.~\ref{fig:CmpATSCBS}(c), in which for ATS(H)+CBS(L) are depicted with blue circles and for CBS(H)+ATS(L) are with yellow triangles. The reason is that the variation in worst-case end-to-end latencies of ATS from high priority to low priority is not as large as that of CBS. From the previous experimental results, we conclude that, on the one hand, when the average traffic load is low, the positive effect of ATS is not prominent, but only when the traffic load reaches a certain level, its advantages will become more and more obvious (Fig.~\ref{fig:CmpATSSP}(a), (c)). On the other hand, ATS has a greater positive effect on low-priority traffic (Fig.~\ref{fig:CmpATSSP}(e).
 
In the leftover section, we will see the impact of ATS on the combined traffic shapers (TAS+SP or TAS+CBS) under different levels of traffic load, as discussed for individual traffic shapers. The architectures of TAS+ATS+SP and TAS+SP without ATS are first compared. We still use the same topology (MM) as in the Sect.~\ref{sec:CmpTrafficLoadATSSP}. In the first set of experiments, it is assumed that there is 20\% of TT traffic achieving deterministic transmission based on the TAS, and 10\% to 70\% of traffic is SP. Therefore, the overall average traffic load on the network is 30\%-90\%. For each traffic load, we have used 20 randomly generated test cases. In order to fairly compare with ATS used individually, we still assume that all flows reshaped by ATS have the same priority and with no interference from other traffic types of lower priority (BE for example).

\begin{figure}[!t]
	\centering
	\includegraphics[width=0.29\textwidth]{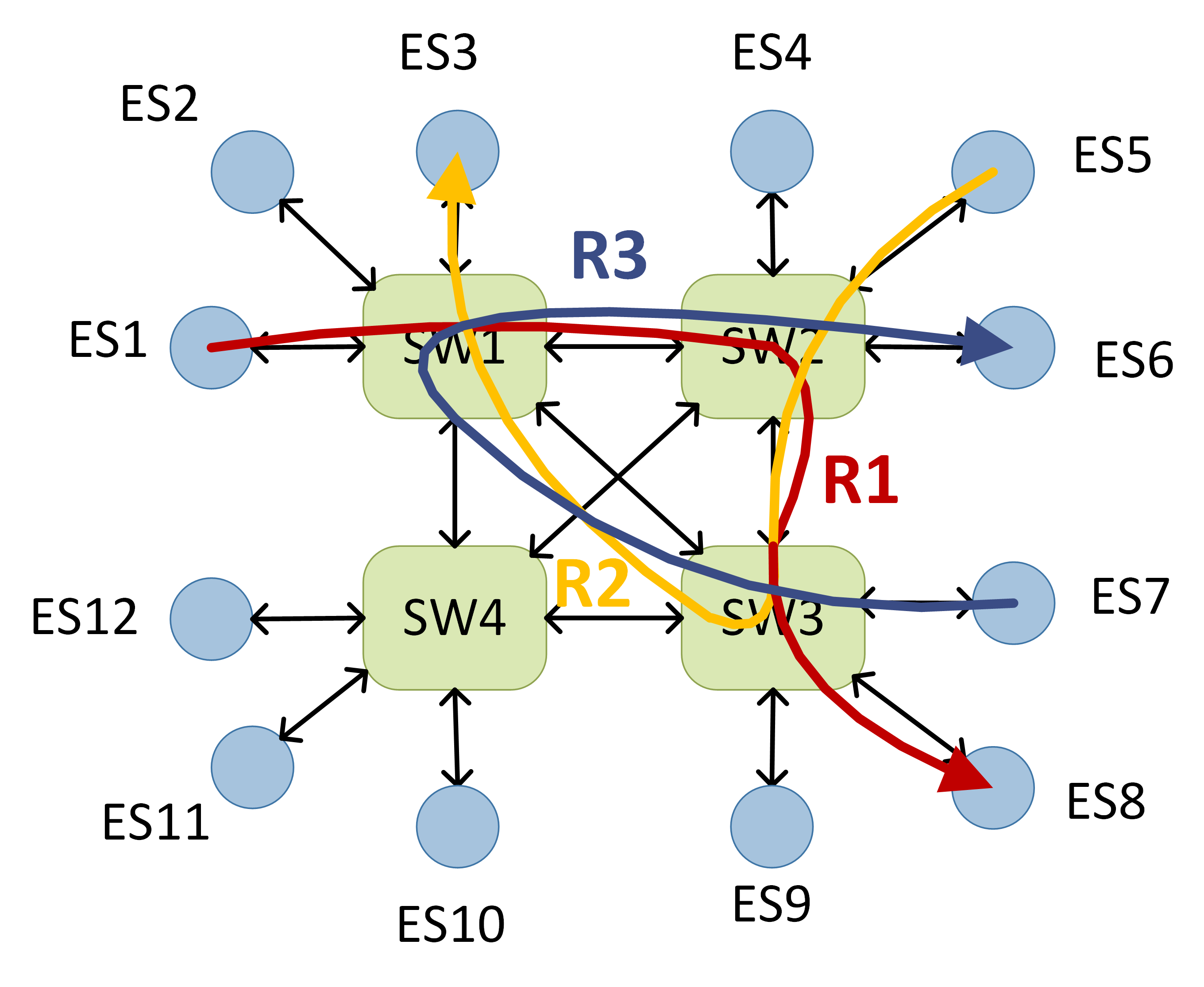}
	\caption{Topology with flows that have routes with cyclic dependencies}
	\label{fig:Cyclic}
\end{figure}

As can be seen from Fig.~\ref{fig:CmpTASATSSP}(a) and (b), similar to ATS used individually, with the increase of average traffic load, the performance of traffic shaped by ATS is gradually improved.
However, with the influence of TT traffic implemented by TAS, it is obvious that the optimization effect on latency bounds of ATS reshaping is better than that of the individual ATS traffic shaper. For the end-to-end latency bounds, the ATS traffic shaper used individually performs better only when the average traffic load is above 70\%. However, in the combined traffic shaper TAS+ATS+SP, as long as the average traffic load shaped by ATS reaches 20\% (overall average load is 40\%), the ATS traffic shaper is superior. Nevertheless, for the backlog bounds, ATS does not show its superiority in the combined architecture of TAS+ATS+SP either, due to the usage of shaped queues. The performance advantage of ATS in combination with TAS is similar to the performance advantage of reshaping for low priority traffic when ATS is used individually.
This is because TT traffic based on TAS has fixed transmission time slots and has the highest priority, and other traffic types cannot use the time slots allocated to TT traffic. The existence of TT traffic will greatly increase the possibility of mutual interference and backlog of other types of traffic. Then, the burst cascade of the flow on its route will be increased. Therefore, the time used by the ATS reshaping is lower than the waiting time caused by the bursty traffic.
Furthermore, we increase the average load of TT traffic to 30\%, while there is 10\% - 60\% of SP traffic, so the overall average traffic load in the network is 40\% to 90\%. Similarly, for each traffic load, there are 20 test cases generated randomly. Compared results of TAS+ATS+SP and TAS+SP are shown in Fig.~\ref{fig:CmpTASATSSP}(c) and (d), respectively.
By comparing with 20\% of the TT traffic load, it is found that with the increasing TT traffic load, the positive impact of ATS is not increasing significantly. 

\begin{table}[!t]
	\caption{Performance analysis of different traffic shapers on cyclic dependency topology}
	\label{tab:PerformanceCyclicDepen}
	\centering	
	\begin{tabular}{|c|c|c|c|c|}
		\cline{1-5}
		WCD &  \tabincell{c}{TAS+SP \\ ($\mu s$)} & \tabincell{c}{TAS+ATS+SP \\ ($\mu s$)} & \tabincell{c}{TAS+CBS \\ ($\mu s$)} & \tabincell{c}{TAS+ATS+CBS \\ ($\mu s$)}   \\ \hline\hline
		f1 & $\backslash$ & 2513.36 & $\backslash$ & 2702.53   \\	\hline
		f2 & $\backslash$ & 4103.04 & $\backslash$ & 4462.96   \\	\hline
		f3 & $\backslash$ & 2343.12 & $\backslash$ & 2662.75   \\	\hline
		f4 & $\backslash$ & 2917.16 & $\backslash$ & 3680.93   \\	\hline
		f5 & $\backslash$ & 3036.52 & $\backslash$ & 3609.21   \\ \hline
		f6 & $\backslash$ & 2858.64 & $\backslash$ & 3553.86   \\	\hline
		f7 & $\backslash$ & 2524.6 & $\backslash$ & 3079.79   \\	\hline
		f8 & $\backslash$ & 329.6 & $\backslash$ & 439.467   \\	\hline
		f9 & $\backslash$ & 4328.96 & $\backslash$ & 5048.18   \\	\hline
		f10 & $\backslash$ & 3133.76 & $\backslash$ & 3922.57   \\ \hline
		f11 & $\backslash$ & 4145.96 & $\backslash$ & 4817.71   \\	\hline
		f12 & $\backslash$ & 4764.4 & $\backslash$ & 5740.25   \\	\hline
		f13 & $\backslash$ & 4349.44 & $\backslash$ & 4725.44   \\	\hline
		f14 & $\backslash$ & 2343.12 & $\backslash$ & 2662.75   \\	\hline
		f15 & $\backslash$ & 2892.44 & $\backslash$ & 3590.99   \\ 	\hline
	\end{tabular}
\end{table}

\begin{figure*}[t!]
	\centering
	\subfigure[TAS+ATS+SP vs. TAS+SP: End-to-end latency bounds]{
		\includegraphics[width=0.48\textwidth]{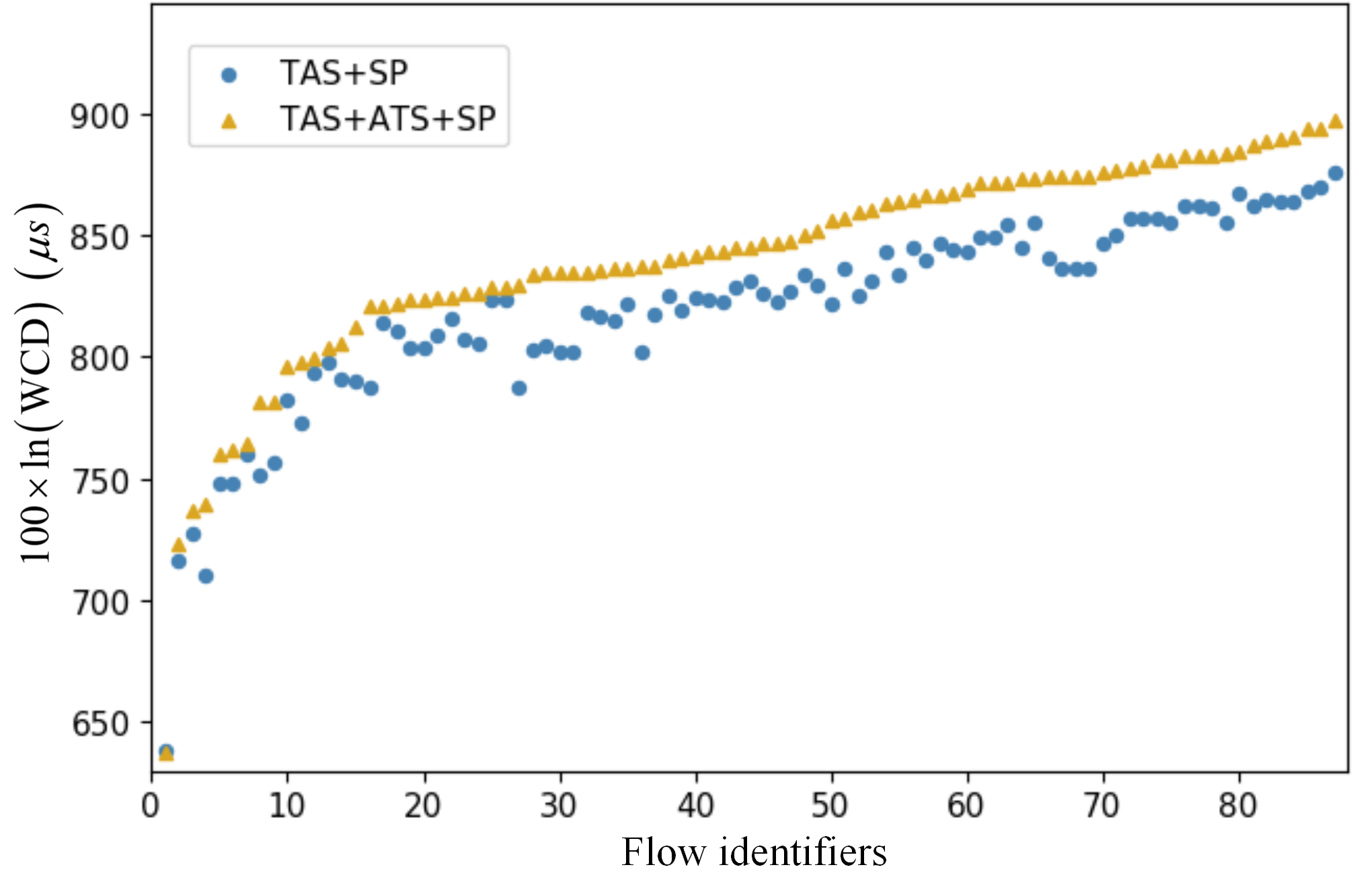}
	}
	\subfigure[TAS+ATS+SP vs. TAS+SP: Backlog bounds]{
		\includegraphics[width=0.48\textwidth]{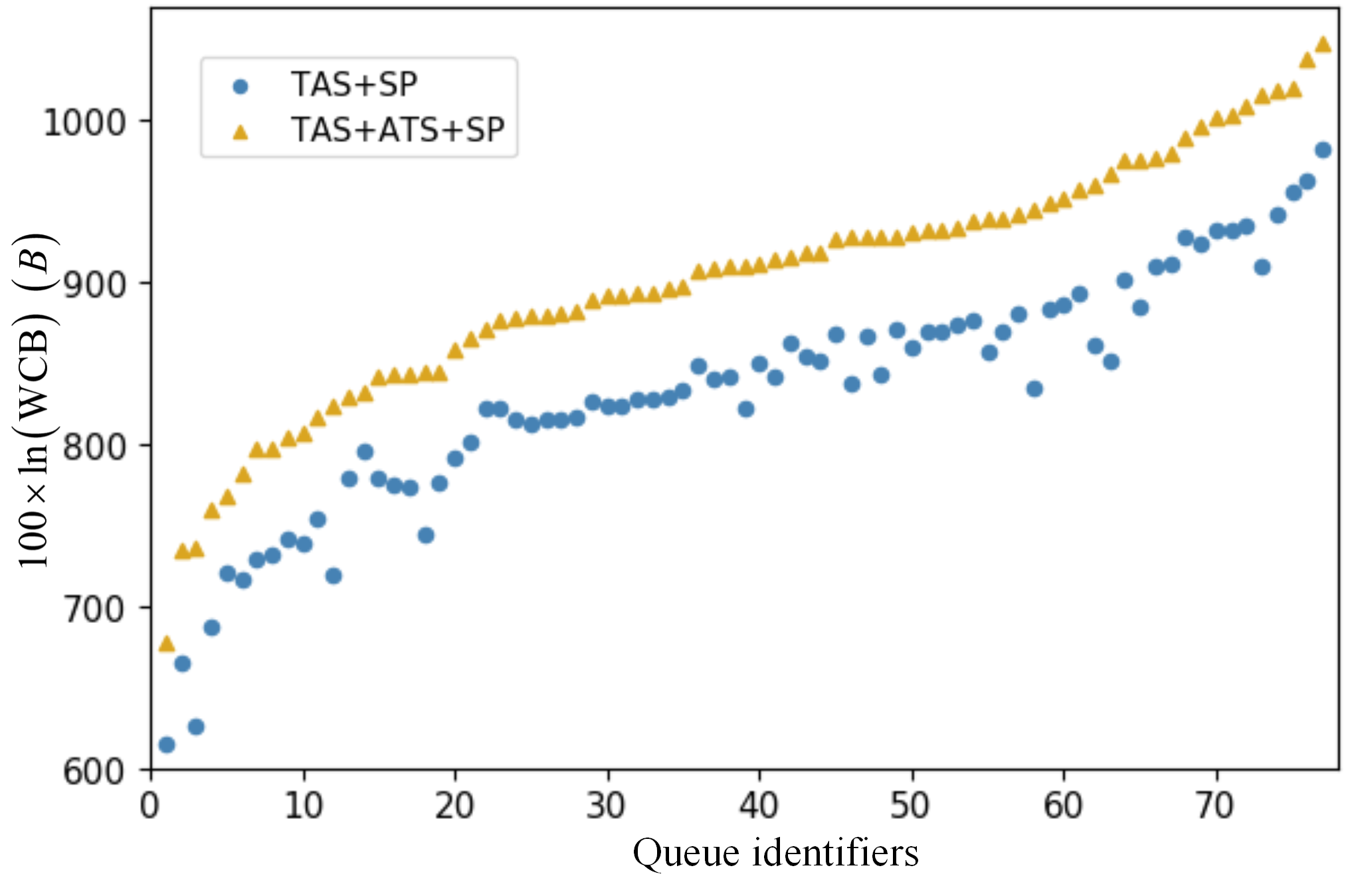}
	}
	\subfigure[TAS+ATS+CBS vs. TAS+CBS: End-to-end latency bounds]{
		\includegraphics[width=0.48\textwidth]{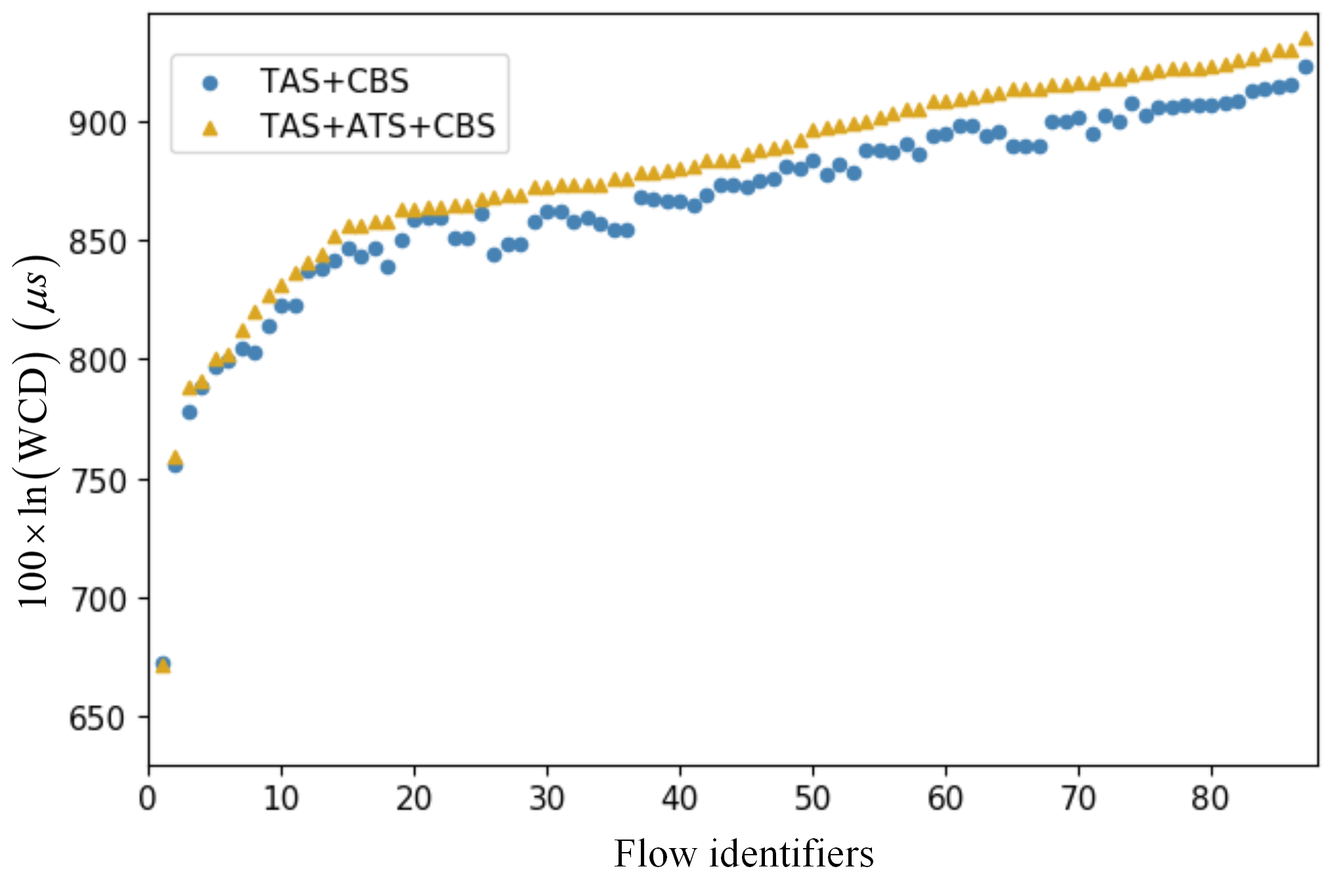}
	}
	\subfigure[TAS+ATS+CBS vs. TAS+CBS: Backlog bounds]{
		\includegraphics[width=0.48\textwidth]{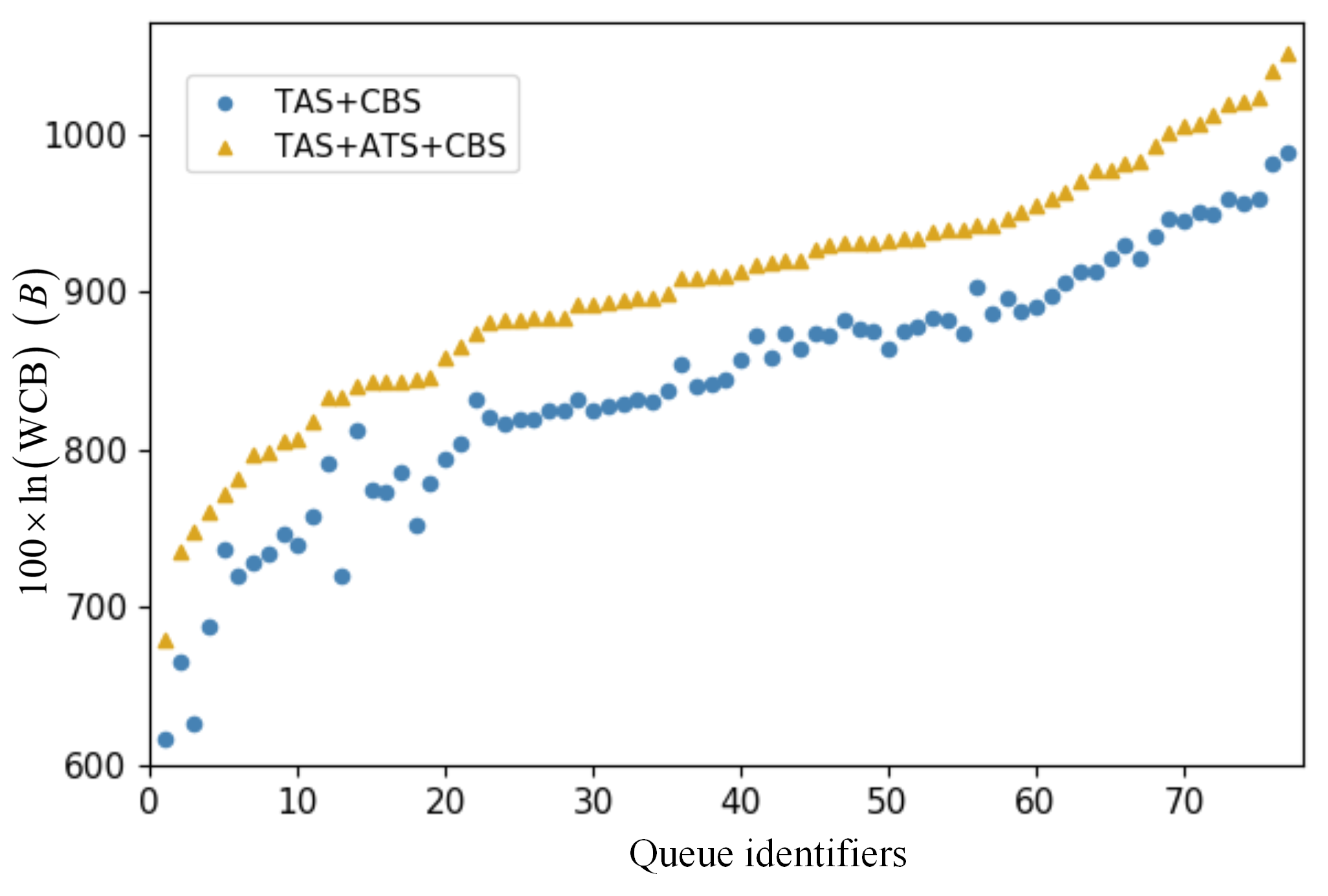}
	}
	\caption{\label{fig:CmpOrion1} Comparison of combined traffic shapers under the Orion CEV $\text{TC}_{\text{Orion}1}$ (TT - 1.5\%, SP/AVB with 1 priority)}
\end{figure*}

\begin{figure}[!b]
	\centering
	\includegraphics[width=0.47\textwidth]{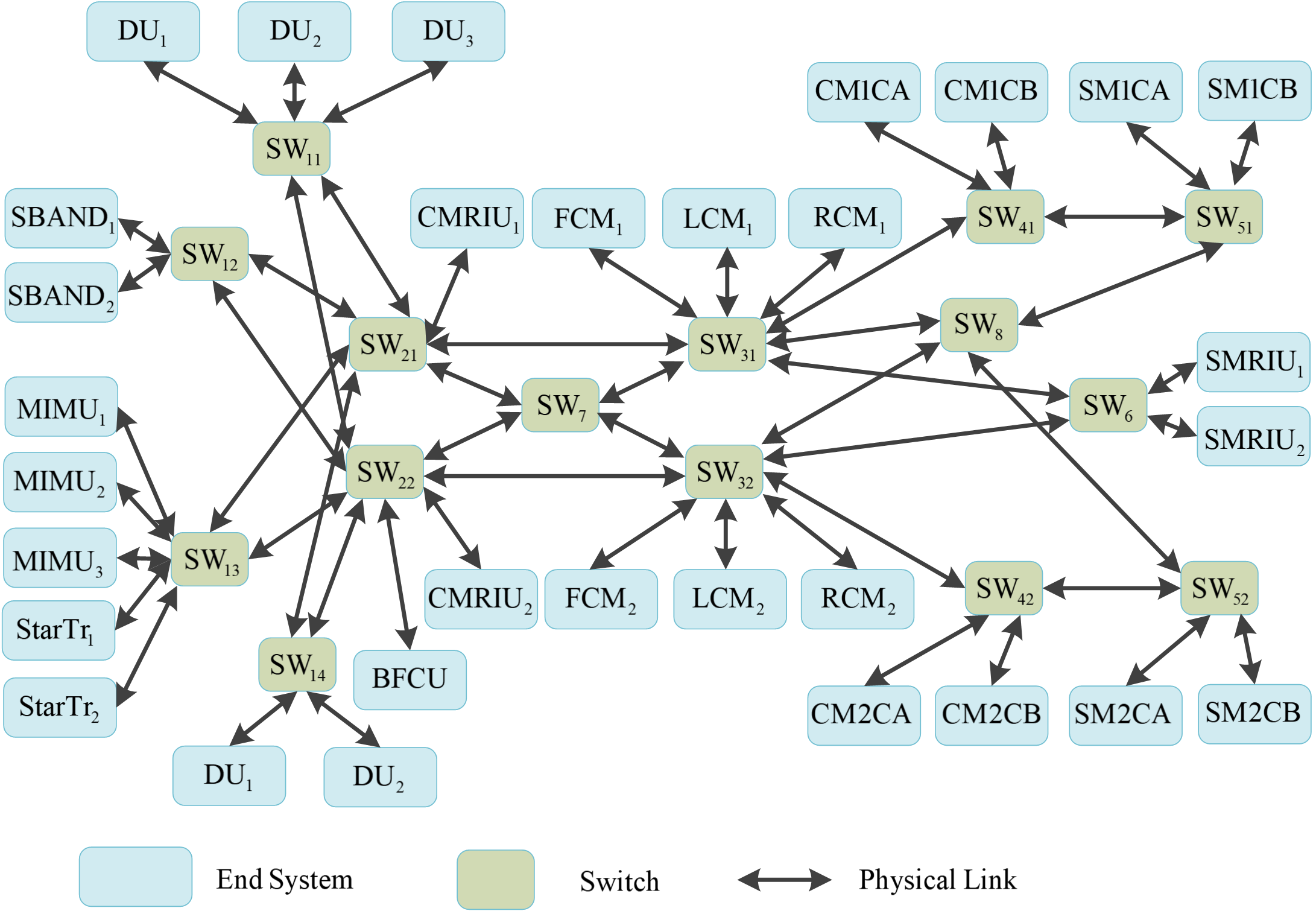}
	\caption{Network topology of Orion CEV}
	\label{fig:Orion_topo}
\end{figure}

\begin{figure*}[t!]
	\centering
	\subfigure[TAS+ATS+SP vs. TAS+SP: End-to-end latency bounds]{
		\includegraphics[width=0.48\textwidth]{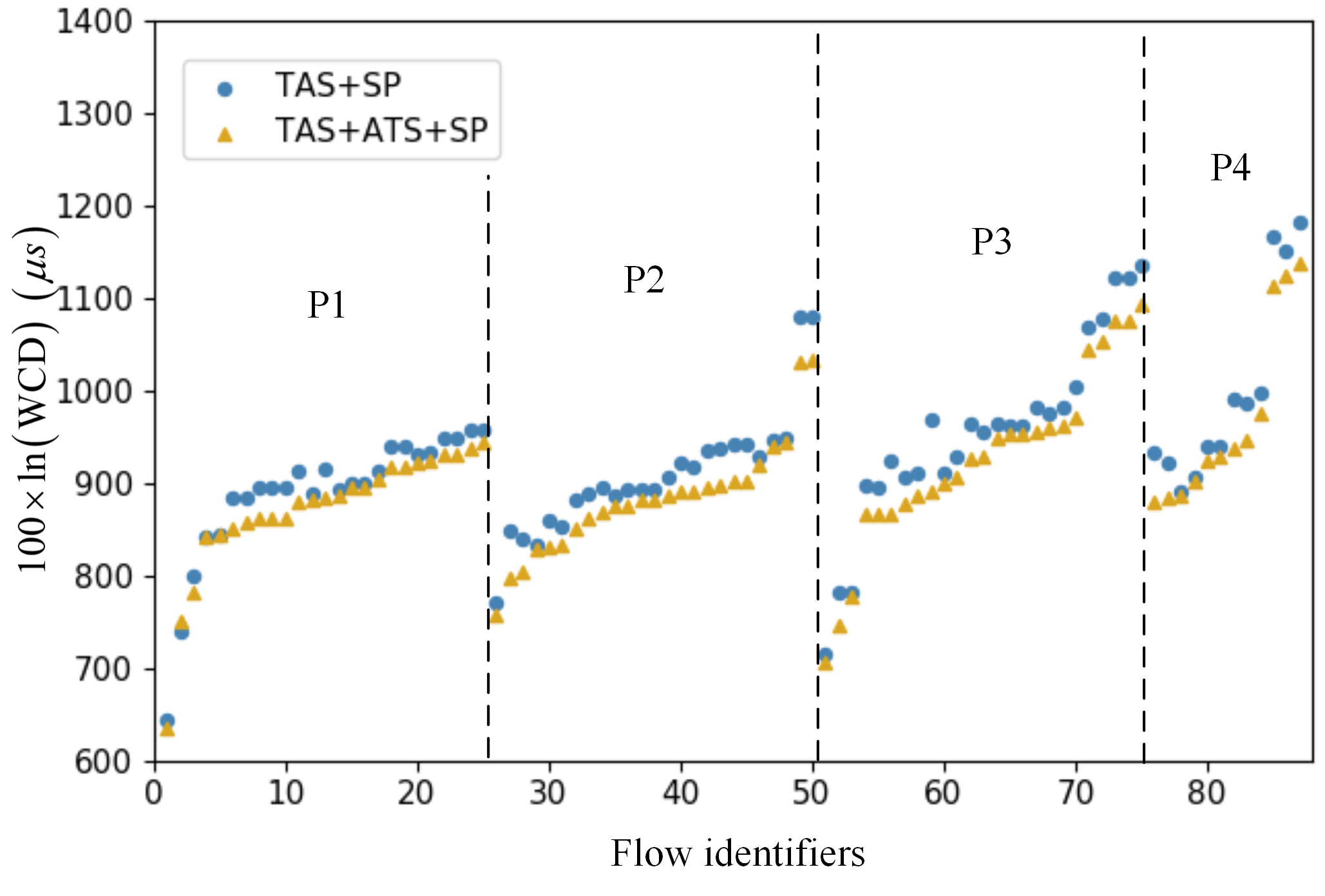}
	}
	\subfigure[TAS+ATS+SP vs. TAS+SP: Backlog bounds]{
		\includegraphics[width=0.48\textwidth]{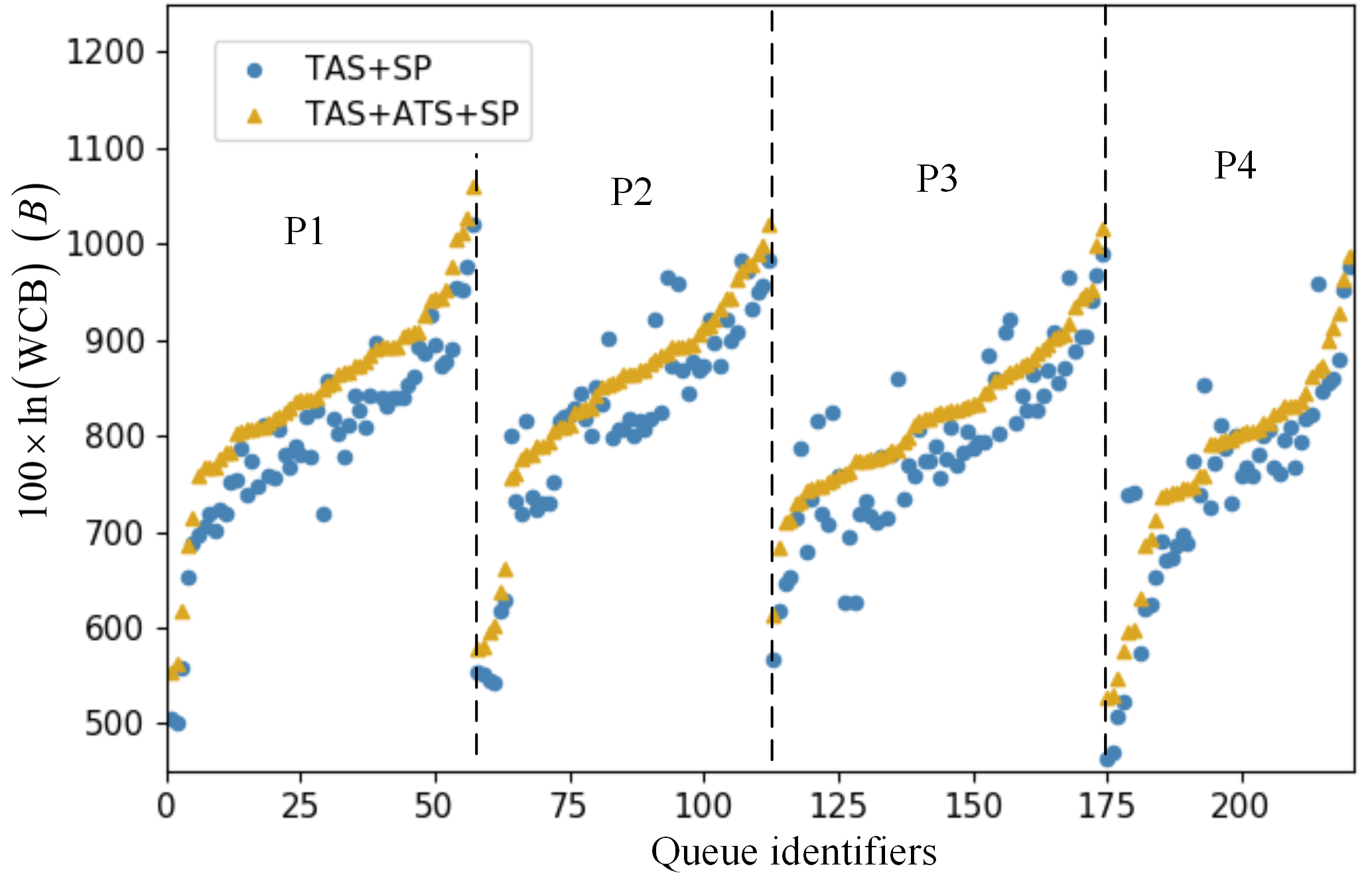}
	}
	\subfigure[TAS+ATS+CBS vs. TAS+CBS: End-to-end latency bounds]{
		\includegraphics[width=0.48\textwidth]{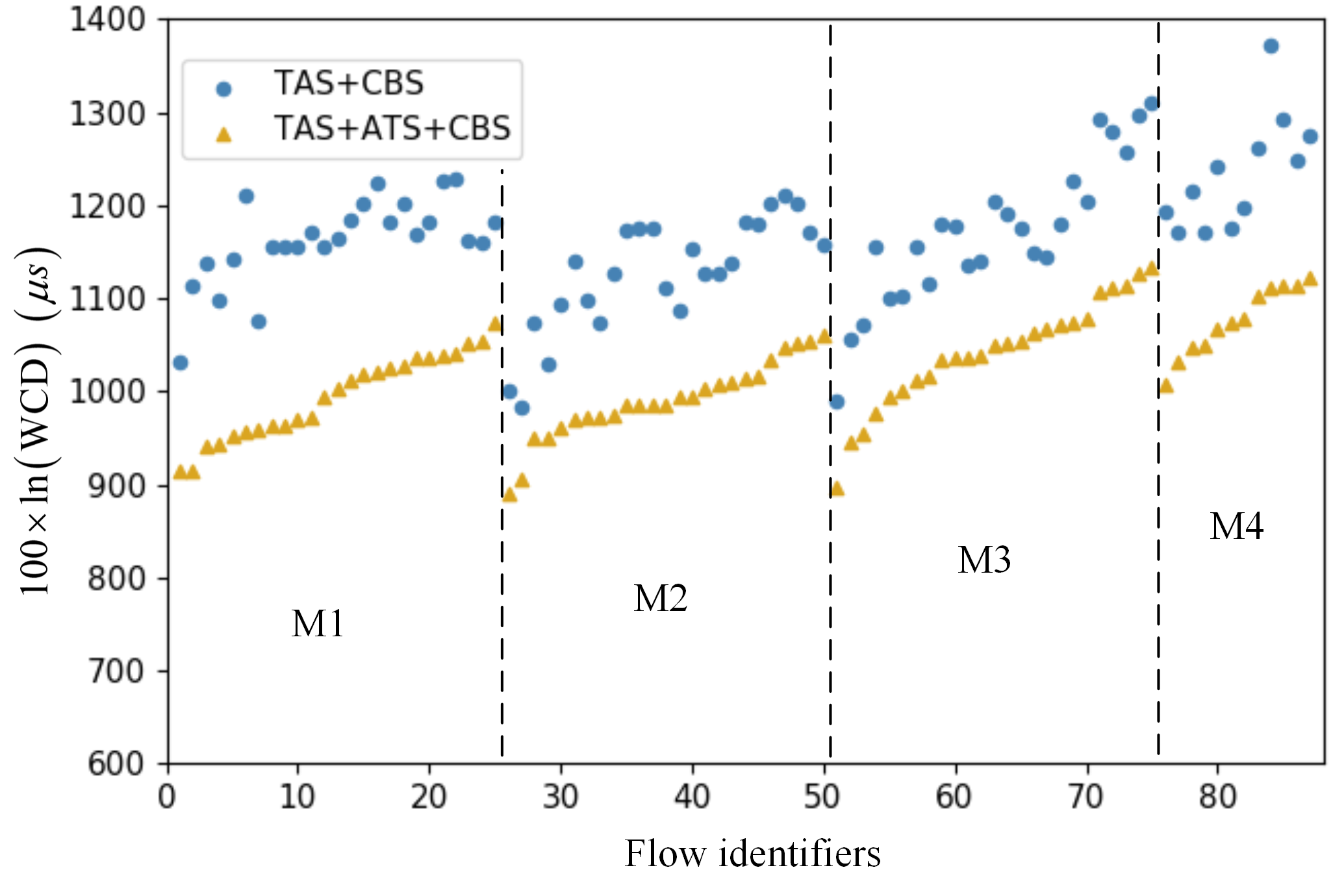}
	}
	\subfigure[TAS+ATS+CBS vs. TAS+CBS: Backlog bounds]{
		\includegraphics[width=0.48\textwidth]{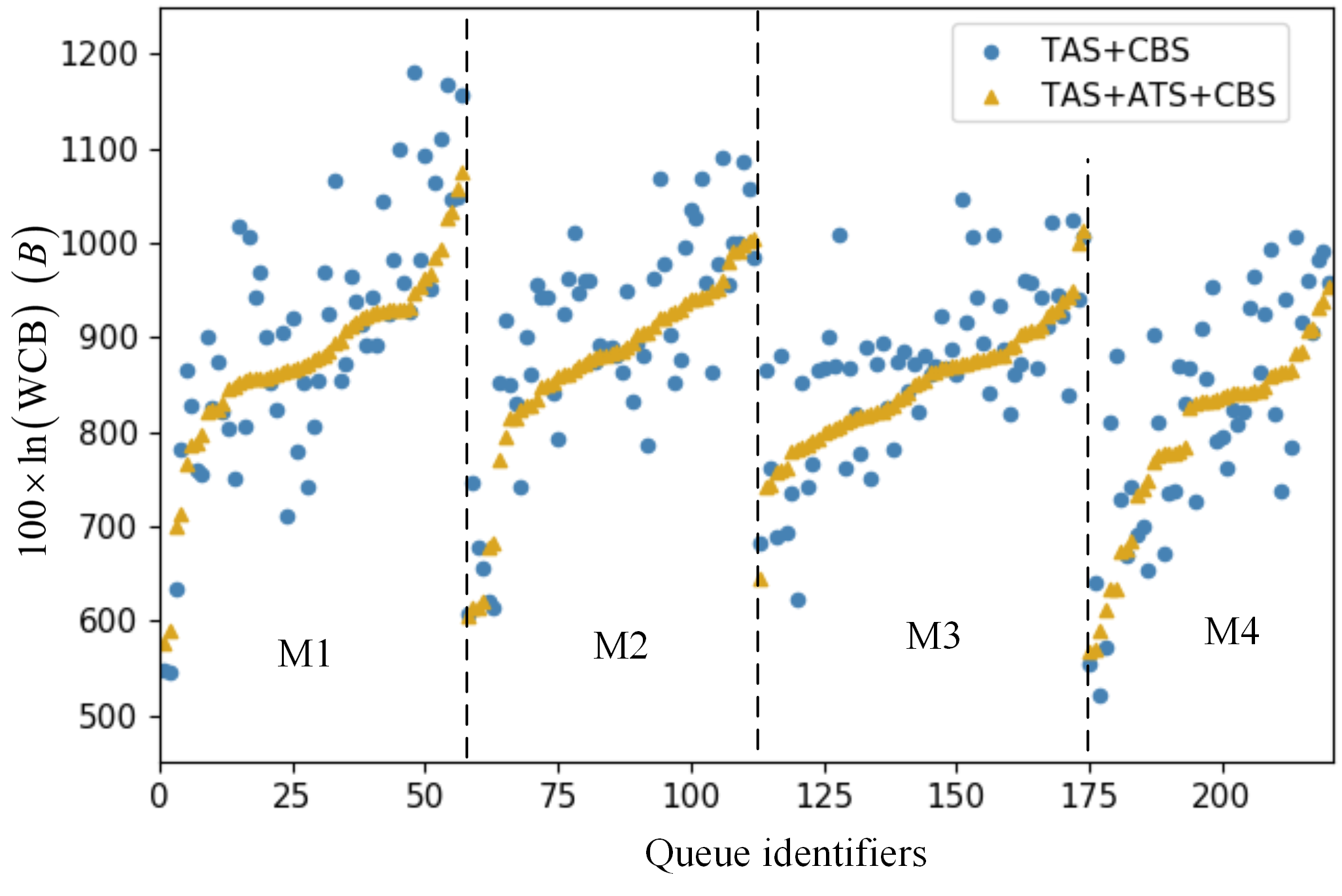}
	}
	\caption{\label{fig:CmpOrion2} Comparison of combined traffic shapers under the Orion CEV $\text{TC}_{\text{Orion}2}$ (TT - 15\%, SP/AVB with 4 priorities)}
\end{figure*}

Then, we will compare the performance of TAS+ATS+CBS and TAS+CBS without ATS. The TT traffic load is still 20\%. In addition, we consider the average load of AVB traffic is from 10\% to 40\%, as only 75\% of bandwidth is reserved for AVB traffic, which also includes the bandwidth occupied by TT traffic. In addition, it is also necessary to ensure that the maximum link load on the network does not exceed 75\%. Similarly, we have used 20 randomly generated test cases for each traffic load.
The comparison results are shown in Fig.~\ref{fig:CmpTASATSCBS}. Concerning the upper bounds of end-to-end latency, ATS also shows its superiority as long as the average load of AVB traffic is above 20\%, and concerning the backlog bounds, ATS performs worse than the backlog performance of the architecture without ATS, which is similar to the case with TAS+ATS+SP, as shown in Fig.~\ref{fig:CmpTASATSSP}.
But the difference is that with the AVB traffic load increasing, although the optimization of the delay performance with the ATS in the architecture TAS+ATS+CBS is also increasing, it is not as obvious as that in TAS+ATS+SP. For the upper bounds of backlog, the optimization effect of ATS does not increase with the traffic load. This may be because CBS is a bandwidth reservation service. CBS itself implements a fairer service by controlling credit, thereby reducing the long-term rate of arrival of flows. Therefore, the optimization effect of ATS itself in the TAS+ATS+CBS architecture is weakened.

All the synthetic test cases above are without cyclic dependencies. Next, we are interested in seeing the ability of ATS to remove the cyclic-dependency of flows in a topology. It is known that the classical NC approach, e.g., Total Flow Analysis, cannot be used directly for the analysis of cyclic dependencies~\cite{Amari17,Finzi19}. In the following, we will respectively test TAS+SP, TAS+ATS+SP, TAS+CBS, TAS+ATS+CBS on a MM topology with routes that have cyclic dependencies, which are shown in Fig.~\ref{fig:Cyclic}. As can be seen in the figure, R1=[[ES1,SW1], [SW1,SW2], [SW2,SW3], [SW3,ES8]], R2=[[ES5,SW2], [SW2,SW3], [SW3,SW1], [SW1,ES3]] and R3=[[ES7,SW3], [SW3,SW1], [SW1,SW2], [SW2,ES6]] are mutually cyclically dependent. The end-to-end latency bound (WCD) for each flow is shown in Table~\ref{tab:PerformanceCyclicDepen}. As can be seen from the results, for the architectures TAS+SP and TAS+CBS without ATS reshaping the flows in middle nodes, it is not possible to calculate WCDs using the classical NC approach. However, with the ATS reshaping function in the architectures TAS+ATS+SP and TAS+ATS+CBS, it is easy to break the cyclic dependency, and the WCDs for flows can be calculated using the classical NC method. It has been shown that cyclic dependencies can be eliminated by reshaping flows in the network~\cite{Thomas19}.
In conclusion from the above experiment, we can find that although from the perspective of worst-case latency and backlog, ATS does not always perform better than the other traffic shapers, ATS can always handle the cyclic dependency of flows. We can say that no matter for CBS, SP etc., the combination use of ATS before them will break the cyclic dependency and make the traditional network calculus analysis method feasible. Therefore, if there are cyclic dependencies between flows in the network, it would be suggested to use the ATS on node ports along the path of the cyclic-dependency flow to break the cycles to guarantee the deterministic transmission for the flows.

\subsubsection{Evaluation on the Realistic Test Case}
\label{sec:EvaRealTC}
In the last experiment, we use the realistic case of the Orion Crew Exploration Vehicle (CEV) from NASA~\cite{Tamas-Selicean14}. The test case topology is shown in Fig.~\ref{fig:Orion_topo}. The Orion CEV case has 31 ESes, 15 SWs, 188 dataflow routes connected by a physical link transmitting at 100 Mbps. 
In the last set of experiments, we are interested to see the effect of ATS on the three novel hybrid architectures of combined traffic shapers TAS+ATS+SP, TAS+ATS+CBS and ATS+CBS. We have run the NC-based performance analysis method for both combinations TAS+SP and TAS+ATS+SP (resp. TAS+CBS and TAS+ATS+CBS, ATS(H)+CBS(L) and CBS(H)+ATS(L)) on the Orion CEV case, and obtained for each combined traffic shaper the upper bounds of the maximum latency (WCD) for each flow and the upper bounds of the maximum backlog (WCB) for each priority queue in an egress port.

We first use the original traffic parameters~\cite{Paulitsch11} of the Orion CEV ($\text{TC}_{\text{Orion}1}$), including 99 TT flows and 87 Rate Constraint (RC) flows of the same priority. RC flows are considered as SP and AVB flows under respective combinations of traffic shapers in this paper. The idle slope for AVB traffic is set to 75\%. The average network load (resp. maximum link load) for TT traffic is around 1.5\% (resp. 5.5\%), and the overall traffic load on the network is 3.5\% on average and 10\% on maximum. The results are shown in Fig.~\ref{fig:CmpOrion1}, where the upper bounds are normalized to $100\times \ln(X)$ with $X=\{\text{WCD}, \text{WCB}\}$. The obtained results are sorted in increasing order by results. As can be seen from the figure, 
%even though the backlog upper bounds for most of queues in egress ports under TAS+ATS+SP/AVB perform better than TAS+SP/AVB, 
ATS does not show a positive impact on both the upper bounds of end-to-end latencies of the flows and the backlog upper bounds of egress ports under the architecture TAS+ATS+SP/AVB. This is because the average traffic load for both TT and SP/AVB is relatively low. The results for Orion CEV conform to the outcomes shown in Fig.~\ref{fig:CmpTASATSSP}(a), (b) and Fig.~\ref{fig:CmpTASATSCBS}.

Then we increase the traffic load in Orion CEV by raising the rate and keeping the frame size of the flow (called $\text{TC}_{\text{Orion}2}$). The average network load (resp. maximum link load) for TT traffic is increased to 15\% (resp. 54\%). The overall traffic load on the network is 25\% on average and 69\% on maximum. Moreover, we are interested in taking a look at ATS's effect on multiple priorities, and thus we classify the SP/AVB traffic into four priorities. There are 25 flows of priority $P_1$, 25 of priority $P_2$, 25 of priority $P_3$ and 12 of priority $P_4$. For the AVB traffic, due to the increased traffic load and the uneven load for each traffic type on each link, it is difficult to assign a fixed idleSlope for each AVB traffic class across the entire network. Thus, we calculate the idle slope of AVB Class $M_i$ for each egress port according to the actual bandwidth utilization~\cite[\S~8.6.8.2]{802.1Q}, i.e.,
\begin{equation*}\label{g:idSlCal}
	idSl_i=\text{operIdleSlope}(M_i)\cdot \frac{\text{OperCycleTime}}{\text{GateOpenTime}},
\end{equation*}
where $\text{operIdleSlope}(M_i)$~\cite[\S~34.3]{802.1Q} is the actual bandwidth that is currently reserved for the AVB class $M_i$ for each port, and $\frac{\text{OperCycleTime}}{\text{GateOpenTime}}$ is the fraction of effective time that the gate is open for AVB traffic. Similarly, the results are shown in Fig.~\ref{fig:CmpOrion2}. The obtained results are grouped by different priorities with vertical dotted lines and, respectively, sorted in increasing order by results within each priority. From the figure, we can find that with the increasing traffic load, the combination shaper TAS+ATS+SP (resp. TAS+ATS+CBS) outperforms TAS+SP (resp. TAS+CBS) from the latency upper bounds. For the backlog bounds, TAS+ATS+SP does not perform better than TAS+SP as conforming to results in Fig.~\ref{fig:CmpTASATSSP}. However, more than half of the backlog bounds under the TAS+ATS+CBS are better than under the TAS+CBS. This is due to the idle slope of CBS, which is related to the serviceability and is set according to the actual bandwidth of AVB traffic. The relative load of AVB is very large. 
%Therefore, the positive effect of ATS on CBS in combined TAS+ATS+CBS is more obvious than that on SP in the combined TAS+ATS+SP. 
What is more interesting is that, with the combination of ATS, the performance of SP and CBS gets close to each other, but CBS allows the bandwidth reservation for the traffic.

\begin{figure}[!t]
	\centering
	\includegraphics[width=0.48\textwidth]{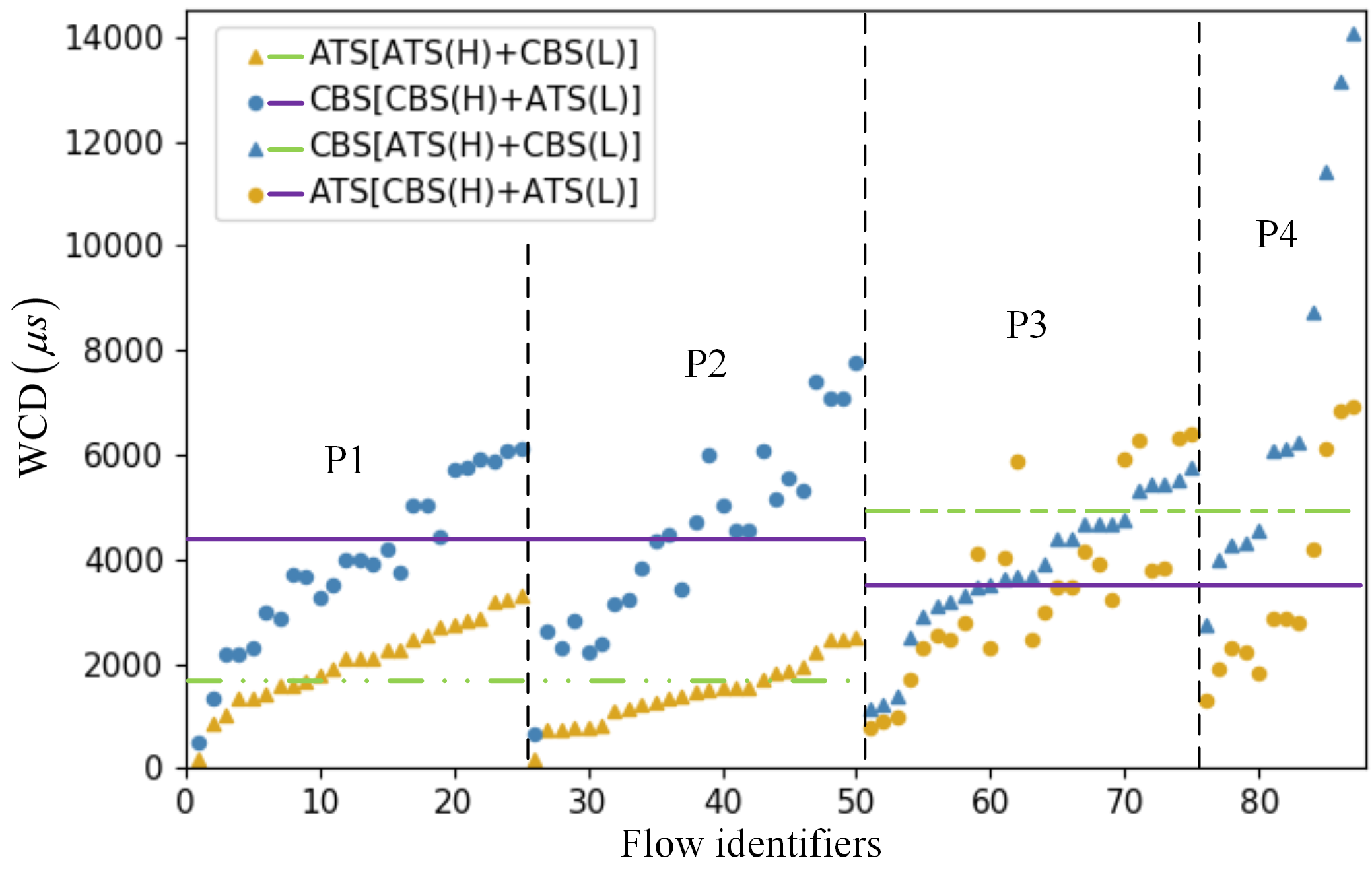}
	\caption{Comparison of ATS(H)+CBS(L) and CBS(H)+ATS(L) under the Orion CEV $\text{TC}_{\text{Orion}3}$}
	\label{fig:CmpOrion3}
\end{figure}

In the last experiment, we are interested in testing combinations ATS(H)+CBS(L) and CBS(H)+ATS(L) using the case adapted from Orion CEV, that has 25 flows of priority $P_1$ (the highest priority), 25 of priority $P_2$, 25 of priority $P_3$ and 12 of priority $P_4$ (called $\text{TC}_{\text{Orion}3}$). The average network load is 10\%. One case is ATS used for flows with priority $P_0$ and $P_1$, and CBS used for flows with priority $P_1$ and $P_2$. In the other case, CBS(H)+ATS(L) is the other way round. The idle slopes for each AVB Class are, respectively, 45\% and 30\%. The results are shown in Fig.~\ref{fig:CmpOrion3}, where the x-axis refers to the flow ID, and the y-axis presents the WCD in microseconds. It can be found that when both ATS and CBS are for high-priority or low-priority traffic, most ATS results are better than CBS. The average WCDs for ATS and CBS classes are given by horizontal lines. As can be seen from the figure, the average WCD of ATS is smaller than that of CBS under both the CBS(H)+ATS(L) (purple solid line) and ATS(H)+CBS(L) (green dotted line) architectures, which is conformed to the results shown in Fig.~\ref{fig:CmpATSCBS}(a), (b). Finally, the overall flows' average WCDs under the ATS(H)+CBS(L) and CBS(H)+ATS(L) are respectively $3091.7 \mu s$ and $2899.4 \mu s$. This is due to the relatively low overall traffic load of 10\%, as can also be seen in Fig.~\ref{fig:CmpATSCBS}(c).

\textbf{}

\section{Conclusion}
\label{sec:conclusion}
This paper has studied the qualitative performance comparison of the various individual traffic shapers and their possible combinations. The analysis is based on the Network Calculus (NC) approach. The paper summarized the existing NC-based analysis for ATS, CBS, SP individually used, and TAS+SP, TAS+CBS used in combination, and extended the NC-based analysis for some other traffic shapers used in combination which have not been discussed, including ATS and CBS used in combination for different priority queues, the novel architectures TAS+ATS+SP and TAS+ATS+CBS when ATS and CBS are used for the same queue. From plenty of test cases, we can conclude that SP and CBS have advantages in that SP is more beneficial to the transmission delay of high-priority traffic. While CBS can specify bandwidth reservations for each priority of traffic. In addition, due to the credit controlled by CBS, the long-term rate of traffic arrival is reduced, so the backlog upper bounds of AVB traffic are probably lower than SP traffic. Compared with SP and CBS, ATS has limited advantages for high-priority traffic. Only when the average traffic load of high-priority traffic in the network reaches around 80\%, ATS shows its superiority. The positive effect of ATS on low-priority traffic is more obvious. When the average traffic load of low-priority traffic on the entire network reaches about 20\% (overall load 40\%), the positive effect of ATS on the upper bound of delay performance begins to become prominent.
%In addition, when the average load of low-priority traffic reaches about 5\% (overall load 10\%), the effect of ATS on the upper bound of port backlog performance becomes positive.

For the combined traffic shapers, we first investigate the performance of ATS and CBS used for different queues. Whether to use ATS for high priority or CBS for high priority is not a definite conclusion, as each has its advantages.
Compared with all the above traffic shapers, TT traffic implemented with flow-based scheduling by TAS has the highest performance, with ultra-low latency, jitter and backlog. However, it is well known that TAS requires the synthesis of optimized GCLs, which do not scale to large networks with many flows. This problem can be mitigated by combining different traffic shapers in the same switch architecture, to reduce the number of flows handled by TAS. Moreover, the combined use of ATS with TAS will make ATS play a more active role, of which the effect is similar to the reshaping impact of ATS used individually on low priority traffic, and at the same time, TAS will maintain unchanged its advantages of ultra-low latency and jitter. Additionally, we have shown that even though, from the perspective of worst-case latency and backlog, ATS does not always perform better than the other traffic shapers, adding the ATS model before other traffic shapers (for example, SP and CBS) can always break the cyclic dependency of flows and make the traditional network calculus analysis method feasible.

\ifCLASSOPTIONcaptionsoff
\newpage
\fi



%\bibliographystyle{IEEEtran}

%\appendix[]
% or


\appendices

\section{Network Calculus Theory}
\label{sec:NC}
Network Calculus~\cite{LeBoudec01,Bouillard18} is a system theory proposed for analyzing performance guarantees in communication networks. By constructing arrival curve and service curve models, the maximum amount of flow data entered into network nodes and the minimum service offered by network nodes can be obtained. Network Calculus is build on min-plus algebra, which includes two basic operators on non-decreasing functions: $\mathcal{F}_\uparrow=\{f:\mathbb{R}_+\rightarrow\mathbb{R}|x_1<x_2\Rightarrow f(x_1)\leq f(x_2)\}$. One is the convolution operator $\otimes$,
\begin{equation*}\label{g:conv}
	(f\otimes g)(t)=\inf_{0\leq s\leq t}\{f(t-s)+g(s)\},
\end{equation*}
and the other is the deconvolution operator $\oslash$,
\begin{equation*}\label{g:deconv}
	(f\oslash g)(t)=\sup_{s\geq0}\{f(t+s)-g(s)\},
\end{equation*}
where $\inf$ means infimum and $\sup$ means supremum.

The arrival and service curves are defined by means of the min-plus convolution.
An arrival curve $\alpha(t)$ is a model constraining the arrival process $R(t)$ of a flow, where $R(t)$ represents the input cumulative function counting the total data bits of the flow that has arrived on the network node up to time $t$. We say that $R(t)$ is constrained by $\alpha(t)$ iff,
\begin{equation}\label{g:1}
	R(t)\leq\inf_{0\leq s\leq t}\left\{R(s)+\alpha(t-s)\right\}=(R\otimes\alpha)(t).
\end{equation}
\noindent
Note that an arrival curve $\alpha(t)$ should be a non-negative wide-increasing function. A typical example of an arrival curve is the ``leaky bucket'' constraint satisfying $\alpha(t)=b+r\cdot t$ for $t>0$ and $\alpha(0)=0$, with the maximum burst tolerance $b$ and long-term rate $r$ of the flow.

A service curve $\beta(t)$ models the processing capability of the available resource for the network node. Assume that $R^*(t)$ is the departure process, which is the output cumulative function that counts the total data bits of the flow departure from the network node up to time $t$. There are several definitions for the service curve. We say that the network node offers the min-plus minimal service curve $\beta(t)$ (considered in this paper) for the flow iff
\begin{equation}\label{g:2}
	R^{*}(t)\geq \inf_{0\leq s\leq t}\left\{R(s)+\beta(t-s)\right\}=(R\otimes \beta)(t),
\end{equation}
and offers the strict service curve $\beta(t)$ iff
\begin{equation}\label{g:strSC}
	R^{*}(t+\Delta t)-R^{*}(t)\geq \beta(\Delta t),
\end{equation}

\noindent
during any backlog period $[t,t+\Delta t)$. Note that a service curve $\beta(t)$ should be a non-negative wide-increasing function.

A shaping curve $\sigma(t)$ characterizes the maximum number of bits that are served during any period of time $\Delta t$, which means that the departure process $R^*(t)$ from the server is always constrained by the shaping curve. A server offers a shaping curve $\sigma(t)$ iff,
\begin{equation}\label{g:shaping curve definition}
	R^*(t+\Delta t)-R^*(t)\leq \sigma(\Delta t),
\end{equation}
i.e. $R^*(t)\leq R^*\otimes\sigma (t)$. Note that the shaping curve $\sigma(t)$ in Eq.~(\ref{g:shaping curve definition}) is non-greedy shaping, which is different from the definition of greedy shaper satisfying $R^*(t)=R\otimes\sigma (t)$. The shaping curve can be used to additionally constrain the output arrival curve by the minimum operation.

Three basic results of network calculus are given as follows. If the flow $R(t)$ constrained by the arrival curve $\alpha(t)$ traverses the network node offering the service curve $\beta(t)$, the latency experienced by the flow in the network node is bounded by the maximum horizontal deviation between the graphs of two curves $\alpha(t)$ and $\beta(t)$,
\begin{equation}\label{g:maxDh}
	D=\text{h}(\alpha,\beta)=\sup_{s\geq0}\left\{\inf\left\{\tau\geq0\mid\alpha(s)\leq\beta(s+\tau)\right\}\right\}.
\end{equation}
The backlog of the flow in the network node is bounded by the maximum vertical deviation between the graphs of two curves $\alpha(t)$ and $\beta(t)$,
\begin{equation}\label{g:maxBv}
	B=\text{v}(\alpha,\beta)=\sup_{s\geq0}\left\{\alpha(s)-\beta(s)\right\}.
\end{equation}
%The arrival curve of individual flow above can be extended to arrival curve of aggregate flows, and service curve should also be expanded to the service capability for aggregate flows. 
The output arrival curve for the output flow $R^*(t)$ is bounded by the arrival curve $\alpha^*(t)$,
\begin{equation}\label{g:outputArr1}
	\alpha^*(t)=\alpha\oslash\beta(t)=\sup_{s\geq 0}\left\{\alpha(t+s)-\beta(s)\right\},
\end{equation}
\noindent
With the effect of packetization, the output arrival curve in Eq.~(\ref{g:outputArr1}) is changed into,
\begin{equation}\label{g:outputArr3}
	\alpha^*(t)=\alpha\oslash(\beta-l^{\max})(t),
\end{equation}
which is proved by Corollary 8.3 in~\cite{Bouillard18}, where $l^{\max}$ is the maximum frame size of the flow.
With the known latency upper bound of the flow, the output arrival curve of the flow can also be given by,
\begin{equation}\label{g:outputArr2}
	\alpha^*(t)=\alpha(t) \oslash\delta_{D}(t),
\end{equation}
where $\delta_{D}(t)$ is the pure-delay function which equals to 0 if $t\leq D$ and $+\infty$ otherwise. The output arrival curve can also be regarded as the input arrival curve of the flow before reaching the next node. Note that in this paper, all the NC-based method is based on the Total Flow Analysis (TFA). For TFA, using the delay bound $D$ calculated from aggregate flows is always better than using the delay bound $D_f=\text{h}(\alpha_f,\beta_f)$ for the flow of interest $f$~\cite{Bouillard18}. Moreover, the calculation of $\beta_f$ when considering FIFO multiplexing depends on the choice of an efficient $\theta$, which is still an open research issue~\cite{Boyer20,Scheffler21},
\begin{equation}\label{g:indiSer}
	\beta_f(t)=\left[\beta-\sum_{j\neq f}\alpha_j\otimes\delta_{\theta}\right]^+\wedge \delta_{\theta}, \forall \theta\in\mathbb{R}_+.
\end{equation}


%\section{}

\section*{Acknowledgment}
The authors would like to thank Marc Boyer from ONERA for the valuable suggestions on the review version of this paper.

\begin{IEEEbiography}[{\includegraphics[width=1in,height=1.25in,clip,keepaspectratio]{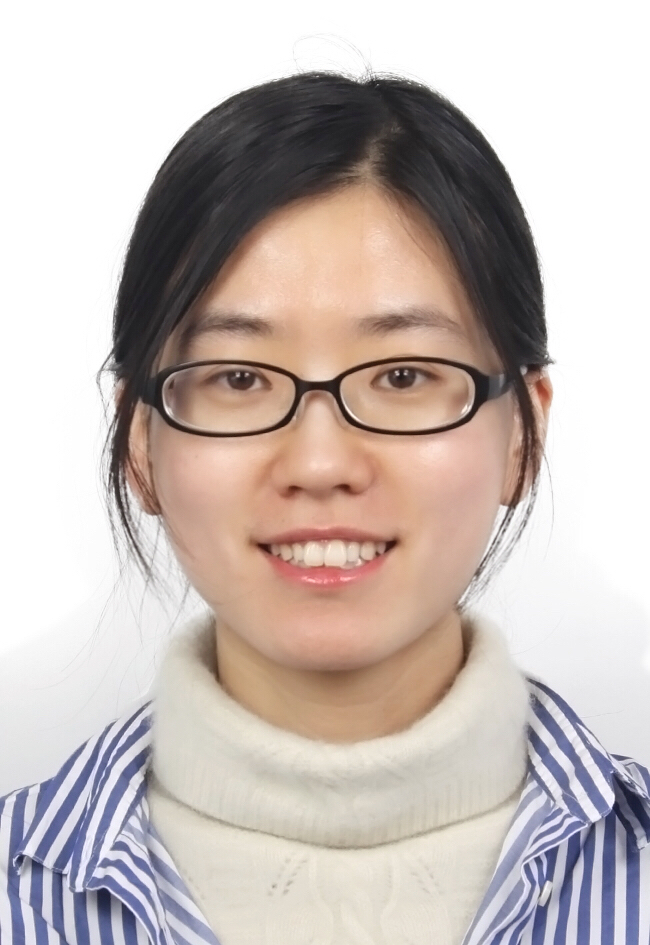}}]{Luxi Zhao}
	%%Biography text here.
	received the PhD in communication and information system from the Beihang University, Beijing, China, in 2017. She is currently an Associate Professor of Communication and Information System at department of Electronic and Information Engineering, Beihang University. She was a postdoc at DTU Compute, Technical University of Denmark (DTU) from 2017 to 2019. She has been a Marie-Curie research fellow at department of Electrical and Computer Engineering, Technical University of Munich (TUM). Her main research interest concerns worst-case analysis and performance evaluation of deterministic real-time and safety-critical networks.
\end{IEEEbiography}

\begin{IEEEbiography}[{\includegraphics[width=1in,height=1.25in,clip,keepaspectratio]{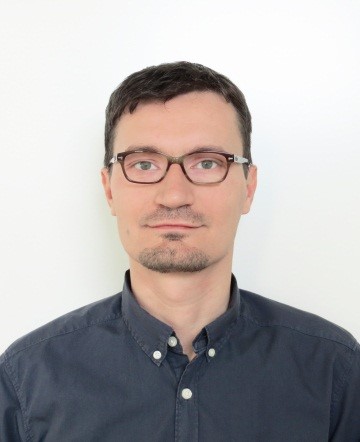}}]{Paul Pop}
	%%Biography text here.
	is a Professor of Cyber-Physical Systems at DTU Compute, Technical University of Denmark (DTU). He has received his Ph.D. degree in computer systems from Linköping University in 2003. His research is focused on developing methods and tools for the analysis and optimization of networked dependable cyber-physical systems. In this area, he has published over 130 peer-reviewed papers, three books, and seven book chapters. He has served as a technical program committee member on several conferences, such as DATE and ESWEEK. He has received the Best Paper Award at DATE 2005, RTIS 2007, CASES 2009, MECO 2013, DSD 2016, and ETFA 2020 . He is the Chairman of the IEEE Danish Chapter on Embedded Systems. He is the coordinator of the Nordic University Hub on Industrial IoT and of the European Training Network on Fog Computing for Robotics and Industrial Automation.
\end{IEEEbiography}

\begin{IEEEbiography}[{\includegraphics[width=1in,height=1.25in,clip,keepaspectratio]{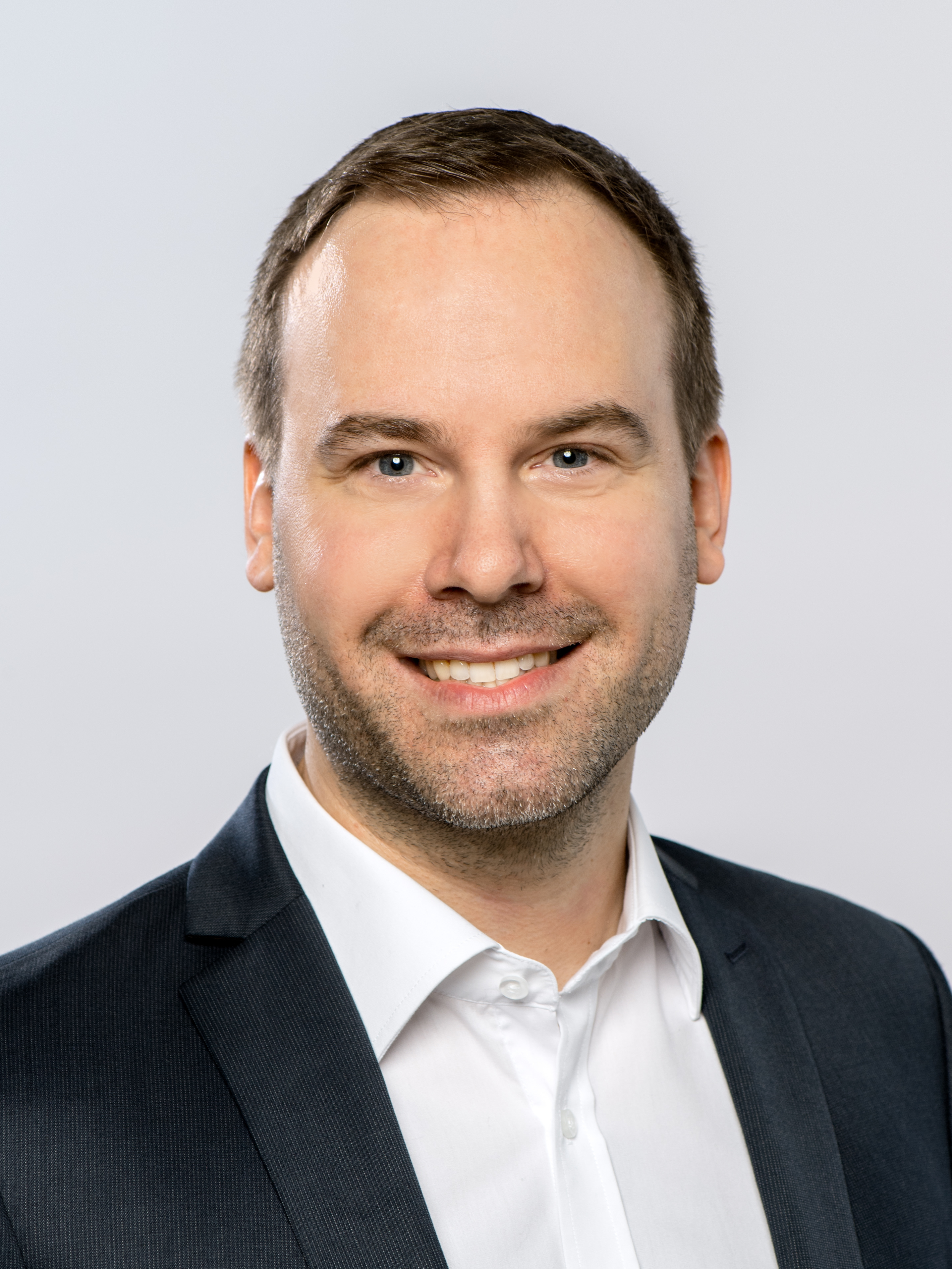}}]{Sebastian Steinhorst}
	%%Biography text here.
	(Senior Member, IEEE) received the M.Sc. (Dipl.-Inf.) and Ph.D. (Dr. phil. nat.) degrees in computer science from Goethe University, Frankfurt, Germany, in 2005 and 2011, respectively. He is currently an Associate Professor at the Technical University of Munich (TUM), Germany where he leads the Embedded Systems and Internet of Things Group, Department of Electrical and Computer Engineering. He was also a Co-Program PI with the Electrification Suite and Test Laboratory, Research Center TUMCREATE, Singapore. His research interests include design methodology and hardware/software architecture co-design of secure distributed embedded systems for use in IoT, automotive, and smart energy applications.

\end{IEEEbiography}

%\vfill


\end{document}